%% file: 2DFishnetsLong.tex
\documentclass[12pt,a4paper,sort&compress]{article}

\usepackage[T1]{fontenc}
\usepackage[latin1]{inputenc}
\usepackage[nosort]{cite}
\usepackage{color}
\usepackage{xfrac}
\usepackage{graphicx}
\usepackage{bbm}
\usepackage{amsmath}
\usepackage{amssymb}
\usepackage{physics}
\usepackage{mathtools}
\usepackage{subfig}
\usepackage{verbatim}
\usepackage{multirow}
\usepackage{tikz}
\usetikzlibrary{math}
\usepackage{amsthm}
\usepackage{fancyhdr}
\usepackage{wrapfig}
\usepackage{hyperref}
\usepackage{dsfont}%for identity operator \idop
\usepackage{delimset} % for brackets
\usepackage{shuffle} % for shuffle product
\usepackage{mathtools}% for underbracket
\usepackage{todonotes}%for Albrechts figure
\usepackage{bbold}

\input cyracc.def

\makeatletter \@addtoreset{equation}{section} \makeatother

\makeatletter
\let\old@startsection=\@startsection
\let\oldl@section=\l@section
\renewcommand{\@startsection}[6]{\old@startsection{#1}{#2}{#3}{#4}{#5}{#6\mathversion{bold}}}
\renewcommand{\l@section}[2]{\oldl@section{\mathversion{bold}#1}{#2}}
\makeatother

\makeatletter
\let\old@makecaption=\@makecaption
\def\@makecaption{\small\old@makecaption}
\makeatother

\let\oldPhi=\Phi
\let\oldPsi=\Psi
\let\oldGamma=\Gamma
\let\oldDelta=\Delta
\let\oldSigma=\Sigma
\let\oldTheta=\Theta
\let\oldPi=\Pi
\let\oldUpsilon=\Upsilon
\renewcommand{\Phi}{\mathnormal{\oldPhi}}
\renewcommand{\Psi}{\mathnormal{\oldPsi}}
\renewcommand{\Gamma}{\mathnormal{\oldGamma}}
\renewcommand{\Sigma}{\mathnormal{\oldSigma}}
\renewcommand{\Delta}{\mathnormal{\oldDelta}}
\renewcommand{\Theta}{\mathnormal{\oldTheta}}
\renewcommand{\Pi}{\mathnormal{\oldPi}}
\renewcommand{\Upsilon}{\mathnormal{\oldUpsilon}}

\newcommand{\superN}{\mathcal{N}}

\newcommand{\gen}[1]{\mathrm{#1}}
\newcommand{\levo}[1]{ \gen{\widehat #1}}

\newcommand{\Eval}{s} % Evaluation parameter
\makeatletter
\newlength{\apb@width}
\newcommand{\autoparbox}[2][c]{\settowidth{\apb@width}{#2}\parbox[#1]{\apb@width}{#2}}
\newcommand{\includegraphicsbox}[2][]{\autoparbox{\includegraphics[#1]{#2}}}
\makeatother
\ifx\genfrac\sdflkaj

\else

\fi
\newcommand{\alg}[1]{\mathfrak{#1}}

\newcommand{\Perm}{\textrm{Perm}_G}

\makeatletter
\def\mr@ignsp#1 {\ifx\:#1\@empty\else #1\expandafter\mr@ignsp\fi}%
\newcommand{\multiref}[1]{\begingroup%\let\protect\string%
\xdef\mr@no@sparg{\expandafter\mr@ignsp#1 \: }%
\def\mr@comma{}%
\@for\mr@refs:=\mr@no@sparg\do{\mr@comma\def\mr@comma{,}\ref{\mr@refs}}%
\endgroup}
\makeatother

\newcommand{\hypref}[2]{\ifx\href\asklfhas #2\else\href{#1}{#2}\fi}

\newcommand{\Figref}[1]{figure~\multiref{#1}}
\renewcommand{\eqref}[1]{(\multiref{#1})}

\ifx\href\asklfhas\newcommand{\href}[2]{#2}\fi

 \definecolor{darkred}{RGB}{180,30,30}

\DeclareMathOperator{\Aut}{Aut}
\DeclareMathOperator{\Sol}{Sol}
\DeclareMathOperator{\PFI}{\mathsf{PFI}}
\DeclareMathOperator{\YDI}{\mathsf{YDI}}

\newcommand{\ua}{\underline{a}}

\newcommand{\uPi}{\underline{\Pi}}
\newcommand{\ux}{\underline{x}}
\newcommand{\uz}{\underline{z}}

\newcommand{\rd}{\mathrm{d}}

\def\beq{\begin{equation}}
\def\eeq{\end{equation}}
\def\bsp#1\esp{\begin{split}#1\end{split}}

\newcommand{\cC}{\mathcal{C}}
\DeclareMathOperator{\K}{K}

\tikzset{
    cross/.pic = {
    \draw[rotate = 45] (-#1,0) -- (#1,0);
    \draw[rotate = 45] (0,-#1) -- (0, #1);
    }
}
\usetikzlibrary{shapes.geometric}

\begin{document}

\thispagestyle{empty}

\begin{flushright}\footnotesize
%\texttt{arXiv:xxxx.xxxx}\\
\texttt{BONN-TH-2024-04}\\
\texttt{TUM-HEP-1498/24}\\
\end{flushright}
\vspace{.2cm}

\begin{center}%
{\LARGE\textbf{\mathversion{bold}%
Geometry from Integrability:\\ Multi-Leg Fishnet Integrals
in Two Dimensions}\par}

\vspace{1cm}
{\textsc{Claude Duhr${}^a$, Albrecht Klemm${}^{a,b}$, Florian Loebbert${}^a$,\\ Christoph Nega${}^c$, Franziska Porkert${}^a$ }}
\vspace{8mm} \\
\textit{%
${}^a$Bethe Center for Theoretical Physics, Universit\"at Bonn, D-53115, Germany\\[2pt]
${}^b$Hausdorff Center for Mathematics, Universit\"at Bonn, D-53115, Germany\\[2pt]
${}^c$Physics Department, Technical University of Munich, D-85748 Garching, Germany.
}
\vspace{.5cm}

\texttt{\{cduhr,aoklemm,loebbert,fporkert\}@uni-bonn.de}\\[2pt]
\texttt{c.nega@tum.de}
%

%%%%%%%%
\par\vspace{10mm}

\textbf{Abstract} \vspace{5mm}

\begin{minipage}{12.8cm}
We generalise the geometric analysis of square fishnet integrals in two dimensions to the case of hexagonal fishnets with three-point vertices. Our results support the conjecture that fishnet Feynman integrals in two dimensions, together with their associated geometry, are completely fixed by their Yangian and permutation symmetries. As a new feature for the hexagonal fishnets, the star-triangle identity introduces an ambiguity in the graph representation of a given Feynman integral. This translates into a map between different geometric interpretations attached to a graph. We demonstrate explicitly how these fishnet integrals can be understood as Calabi-Yau varieties, whose Picard-Fuchs ideals are generated by the Yangian over the conformal algebra. In analogy to elliptic curves, which represent the simplest examples of fishnet integrals with four-point vertices, we find that the simplest examples of three-point fishnets correspond to Picard curves with natural generalisations at higher loop orders.

\end{minipage}
\end{center}

\newpage
\tableofcontents
\bigskip
\hrule

%%%%%%%%%%%%%%%%%%%%%%%%%%%%%%%%%%%%%%%%%%%%%%%%%%%%%%%%%%%%%%%%%%%%%%%%%%%%

\input{introduction}

\input{setup}

\input{geometry}

\input{yangian}

\input{examples}

%%%%%%%%%%%%%%%%%%%%%%%%%%%%%%%%%%%%%%%%%%%%%%%%%%%%%%%%%%%%%%%%%%%%%%%%%%%

\section{Conclusions and Outlook}
\label{sec:outlook}

In this paper, we have considered fishnet Feynman integrals in two spacetime dimensions, with particular focus on the interplay between their geometric interpretation and constraints from integrability. Our main results are the following:
\begin{itemize}
\item We have provided evidence that fishnet Feynman integrals in two dimensions are fully fixed by their Yangian and permutation symmetries. This implies that the symmetries of a given graph completely constrain the Feynman integral as a geometric object.
\item The star-triangle identity, which naturally acts on the considered three-point fishnets, relates different representations of a given Feynman graph and thus different geometric interpretations. This provides a very direct illustration of the insight that one cannot attach a unique geometry to a given Feynman integral.
\end{itemize}
There are various interesting directions in which the line of research subject to this paper should be extended. 

Firstly, in this paper, we have focussed on isotropic fishnets with the same rational propagator powers for all propagators. This choice implies the connection to Calabi-Yau geometry. It is natural to ask whether the Yangian and permutation symmetries still fully constrain fishnet Feynman integrals with parametric propagator powers, as originally considered by Zamolodchikov~\cite{Zamolodchikov:1980mb}. The Yangian symmetry of these graphs was recently established in ref.~\cite{Kazakov:2023nyu}.

Moreover, in this paper, we assumed the same value of the scaling dimension $\Delta$ entering into the holomorphic and anti-holomorphic copy of the Yangian. It would be natural to extend this analysis to the situation with different holomorphic and anti-holomorphic dimensions and to establish a geometric way to pair these into the result for the Feynman graph, thus generalising the CY intersection pairing employed above.

Finally, it would be very interesting to understand in how far the fruitful interplay between geometry and integrability extends to higher dimensional Feynman integrals. Both, Yangian symmetry and CY geometry have already proven useful for Feynman integrals in higher dimensions and combining them would thus be a natural next step.
%%%%%%%%%%%%%%%%%%%%%%%%%%%%%%%%%%%%%%%%%%%%%%%%%%%%%%%%%%%%%%%%%%%%%%%%%%%
%%%%%%%%%%%%%%%%%%%%%%%%%%%%%%%%%%%%%%%%%%%%%%%%%%%%%%%%%%%%%%%%%%%%%%%%%%%%

\section*{Acknowledgements} 

The work of FL was supported by funds of the Klaus Tschira Foundation gGmbH.
This work was co-funded by
the European Union (ERC Consolidator Grant LoCoMotive 101043686 (CD, FP) and ERC Starting Grant 949279
HighPHun (CN)). Views and opinions expressed are
however those of the author(s) only and do not necessarily reflect those of the European
Union or the European Research Council. Neither the European Union nor the granting
authority can be held responsible for them. AK wants to thank Kilian 
B\"ohnisch and Matt Kerr for discussions. Special thanks to Matt Kerr 
for reading section 3.5 and suggesting improvements.

%%%%%%%%%%%%%%%%%%%%%%%%%%%%%%%%%%%%%%%%%%%%%%%%%%%%%%%%%%%%%%%%%%%%%%%%%%%
%%%%%%%%%%%%%%%%%%%%%%%%%%%%%%%%%%%%%%%%%%%%%%%%%%%%%%%%%%%%%%%%%%%%%%%%%%%%

\appendix

\input{AppendixExamples}

\bibliographystyle{nb}
\bibliography{2DFishnetsLong}

\end{document}

%% file: introduction.tex
% !TEX root = 2DFishnetsLong.tex

\newpage
\section{Introduction and Summary}

The question of which underlying concepts fix the laws of nature has always been an appealing problem. While at the phenomenological level this question seems hard to answer, one can lift this challenge to a more abstract toy situation. A prime example of this idea is the Anti~de~Sitter/Conformal Field Theory (AdS/CFT) correspondence, where two highly symmetric but non-trivial theories are constrained by large amounts of symmetry. In its original formulation, the AdS/CFT conjecture relates $\superN=4$ maximally supersymmetric Yang-Mills (SYM) theory in four dimensions and IIB string theory on the background $\mathrm{AdS}_5\times \mathrm{S}^5$ \cite{Maldacena:1997re}. In particular, in the planar limit integrable structures emerge and imply strong constraints on both sides of this duality, cf.,~e.g., refs.~\cite{Beisert:2010jr,Bombardelli:2016rwb}.

The algebraic structure behind the integrability of planar $\superN=4$ SYM theory is associtated to Drinfeld's \emph{Yangian}~\cite{Drinfeld:1985rx,Drinfeld:1986in}. The Yangian algebra is well known to underly integrable models of rational type and tightly connected to solutions of the quantum Yang-Baxter equation, for reviews see refs.~\cite{Bernard:1992ya,MacKay:2004tc,Beisert:2010jq,Torrielli:2010kq,Loebbert:2016cdm}. In the context of the $\superN=4$ SYM theory, the Yangian was first identified by Dolan, Nappi and Witten as a symmetry of the theory's dilatation operator
\cite{Dolan:2003uh,Dolan:2004ps}. By now it has been found in many other situations. For example, it provides a symmetry of scattering amplitudes \cite{Drummond:2009fd}, smooth Wilson loops \cite{Muller:2013rta} or  the action \cite{Beisert:2017pnr} of $\superN=4$ SYM theory, and prominently features in the
AdS/CFT S-matrix \cite{Beisert:2006fmy}.

To further simplify the AdS/CFT setup and to detach it from supersymmetry, one can gamma-deform the original model \cite{Lunin:2005jy,Leigh:1995ep,Frolov:2005dj,Frolov:2005iq}, which, in addition to the coupling constant $g$ and the number of colors $N_\text{c}$, introduces three new parameters $\gamma_{1}$, $\gamma_2$ and $\gamma_3$ into the $\superN=4$ SYM theory.  Importantly, these additional parameters allow for the following double-scaling limit of the gamma-deformed model as introduced in ref.~\cite{Gurdogan:2015csr}: Taking $g\to 0$ and $\gamma_3\to i\infty$, with the new coupling constant $\xi^2\coloneqq g^2N_\text{c}e^{-i\gamma_3}$ held fixed, one finds a simple model of two complex scalars coupled by a chiral four-point vertex, see also refs.~\cite{Fokken:2013aea, Caetano:2016ydc, Loebbert:2020tje} for explicit Lagrangians. This bi-scalar model has been generalised to $D$ spacetime dimensions in ref.~\cite{Kazakov:2018qbr} with the Lagrangian
 \begin{equation}
\mathcal{L}_{\text{F}}^{\omega D}= N_\text{c}\tr \brk[s]*{ -X(-\partial_\mu \partial^{\mu})^{\omega} \bar{X}- Z (-\partial_\mu\partial^{\mu})^{\frac{D}{2}-\omega} \bar{Z}+\xi^2 XZ\bar{X}\bar{Z}}.
\label{eq:LagrD}
 \end{equation}
 Here $\omega$ denotes a parameter resulting in deformed propagator powers of the associated Feynman integrals. For $D=4$ and $\omega=1$ this theory corresponds to the above mentioned double-scaling limit of the gamma-deformed $\superN=4$ SYM theory.
It was shown that renormalisation requires additional double-trace couplings to be added to the above fishnet Lagrangian \cite{Fokken:2013aea,Sieg:2016vap}, which are irrelevant for the correlators considered in this paper. 
Notably, similar to $\superN=4$ SYM theory, the fishnet theory has a holographic dual, the so-called fishchain model introduced in refs.~\cite{Gromov:2019aku,Gromov:2019bsj}.

From a general quantum field theory perspective, it is an important insight that the above double-scaling limit draws a connection to the properties of individual Feynman integrals. Due to the chiral four-point vertex, the bi-scalar model is characterized by a limited number of Feynman graphs of fishnet structure, which in fact represent full correlation functions in the limit theory, cf.~refs.~\cite{Gurdogan:2015csr,Caetano:2016ydc,Chicherin:2017cns}. Though the number of Feynman graphs is small, the associated Feynman integrals are in general hard to compute. Here the relation to the planar AdS/CFT duality comes in handy, which results in integrable structures of these fishnet Feynman graphs. In particular, these fishnet integrals were shown to be invariant under a Yangian extension of the purely bosonic conformal algebra \cite{Chicherin:2017cns,Chicherin:2017frs,Loebbert:2020hxk,Kazakov:2023nyu}, which manifests the integrability of large classes of Feynman graphs.
As one may expect from the constraining power of integrability,%
\footnote{The Yangian typically appears in two-dimensional integrable models where it completely fixes the $S$-matrix and implies its factorization into two-to-two scattering, cf.~ref.~\cite{Loebbert:2018lxk}.} 
it has been demonstrated on a number of examples that the Yangian can be employed to bootstrap Feynman integrals from scratch \cite{Loebbert:2020hxk,Loebbert:2019vcj,Loebbert:2020glj,Corcoran:2020epz,Loebbert:2020aos,Duhr:2022pch} (see ref.~\cite{Loebbert:2022nfu} for a review). 
In this paper, we will take the Yangian perspective on fishnet Feynman integrals, but we emphasise that various associated integrability approaches have recently been pursued in this context, cf.~refs.~\cite{Chicherin:2017frs,Kazakov:2023nyu,Basso:2017jwq,
Gromov:2017cja,
Basso:2018agi,
Derkachov:2018rot,
Gromov:2018hut,
Derkachov:2019tzo,
Gurdogan:2020ppd,
Basso:2021omx,
Cavaglia:2021mft,
Aprile:2023gnh,
Kade:2023xet,
Alfimov:2023vev}.
Notably, integrable structures of fishnet integrals had already been observed by Zamolodchikov in 1980 \cite{Zamolodchikov:1980mb}, but were largely ignored for a long time.

While the presence of integrability implies new tools for the computation of Feynman integrals, an important question is what it means to successfully compute an integral. In fact, at higher loop orders this question is still to be settled, and developing the relevant theory of special functions is a very active area of research.
In particular, large efforts in the last decades have resulted in the fact that those Feynman integrals, which evaluate to polylogarithms or their elliptic generalisations are under relatively good control by now (see, e.g., refs.~\cite{Weinzierl:2022eaz,Bourjaily:2022bwx,Abreu:2022mfk,Badger:2023eqz} and references therein for recent reviews). 
These achievements can be considered as the tip of an iceberg of connections to geometry, which seem to provide the right framework for tackling more general classes of Feynman integrals. In particular, the connection to Calabi-Yau geometry recently turned out to be very fruitful~\cite{Duhr:2022pch,Brown:2010bw,Bloch:2014qca,MR3780269,Klemm:2019dbm,Bonisch:2020qmm,Bonisch:2021yfw,Duhr:2022dxb,Bourjaily:2018yfy,Bourjaily:2019hmc,Bourjaily:2018ycu,Vergu:2020uur,Pogel:2022ken,Pogel:2022vat,Duhr:2023eld}, since the latter has been developed over many decades in the context of string theory, see \cite{MR1177829,MR3965409} for reviews and further references, and pure mathematics,  see \cite{MR1963559} for a mathematical  introduction to Calabi-Yau geometries and  their periods \cite{MR3822913}.

In this paper, we will be interested in the case of two spacetime dimensions, where the family of fishnet Feynman integrals has a particularly simple structure, cf.~refs.~\cite{Kazakov:2018qbr,Duhr:2022pch,Derkachov:2018rot,Duhr:2023eld}. The results of the present paper build on ref.~\cite{Duhr:2022pch}, where we associated
a Calabi-Yau (CY) variety of complex dimension $\ell$, for short CY $\ell$-fold,  to each planar $\ell$-loop fishnet integral with vertices of valency $V=4$. These integrals can be understood as correlators of the above fishnet model from eq.~\eqref{eq:LagrD} with $D=2$ and $\omega=1/2$.
This relation to Calabi-Yau varieties opened a geometrical approach in which the special functions, which furnish the building blocks for fishnet integrals, are identified with periods of the CY $\ell$-fold. Concretely, a 
hermitian, monodromy invariant, real bilinear of these periods that geometrically is the calibrated volume 
or the quantum volume of the Calabi-Yau geometry, or its mirror, respectively, yields the fishnet integral \cite{Duhr:2022pch,Duhr:2023eld}. 

One of the main results of this paper is that the results of ref.~\cite{Duhr:2022pch} carry over to
 fishnet integrals for $V=3$. The latter can be understood as correlators in the $D=2$, $\omega_j=2/3$ version of the honeycomb fishnet theory of ref.~\cite{Kazakov:2022dbd} generalizing the model of ref.~\cite{Mamroud:2017uyz}:
\begin{equation}\label{eq:L_hex}
   \mathcal{L}_\text{hex}
=
N_\text{c}\tr \brk[s]*{ -X(-\partial_\mu \partial^{\mu})^{\omega_1} \bar{X}- Y (-\partial_\mu\partial^{\mu})^{\omega_2} \bar{Y}
- Z (-\partial_\mu\partial^{\mu})^{\omega_3} \bar{Z}
+\xi_1^2 \bar X Y Z +\xi_2^2 X\bar Y \bar Z
}\,.
\end{equation}
Here $\omega_{j=1,2,3}$ denote three parameters with $\omega_1+\omega_2+\omega_3=D$. The detailed investigation of isotropic ($\omega_1=\omega_2=\omega_3$) fishnet integrals with hexagon structure in two dimensions represents one of the main subjects of the present paper.
Moreover, we provide evidence that both square and hexagonal fishnet integrals in two spacetime dimensions are completely fixed by their Yangian and their permutation symmetries. 
Using this fact and the geometrical tools, we obtain analytical expressions 
that allow exact evaluation of multi-parameter Feynman integrals for many new examples.

An important consequence of our analysis is that for $V=3$, we cannot associate a unique geometry to the fishnet integral.
On the physics side, this can be traced back to star-triangle relations, which relate different conformally-invariant integrals.
In particular, we show that the (very singular) CY $\ell$-folds 
associated to the graphs with $V=3$ vertices are identified in this way 
with lower-dimensional Picard varieties (which are not CY geometries). 

The paper is structured as follows: In section~\ref{sec:setup} we introduce and review fishnet integrals in $D$ dimensions with a focus on their symmetries and relations induced via the star-triangle identity. In section~\ref{sec:fishnets_2D} we specify the case of integrals in $D=2$ dimensions, and we explain their interpretation in terms of Calabi-Yau geometry. We introduce the family of triangle track graphs and discuss singularities of the associated geometry. In section~\ref{sec:examples} we consider several explicit examples of $V=4$ train track and $V=3$ triangle track graphs. In section~\ref{sec:outlook} we finish with a brief conclusion and outlook.

%% file: setup.tex
% !TEX root = 2DFishnetsLong.tex

\section{Fishnet integrals in $D$ dimensions}
\label{sec:setup}

Before focussing on the specific case of $D=2$ dimensions in section~\ref{sec:fishnets_2D}, in this section we give an overview over $D$-dimensional fishnet integrals and their symmetries.

\subsection{Fishnet integrals}
\label{sec:fishnetintegrals}

\begin{figure}[t]
\centering
\resizebox{0.3\textwidth}{!}{
  \begin{tikzpicture}[line width=2pt]
 % \node (hex) at (0,0) [font=\Large, text width=1 cm]{hex};
  \foreach \i in {0,...,3} 
  \foreach \j in {0,...,1} {
   \foreach \a in {30,150,270} \draw (2*sin{60}+2*sin{60}*\i,1+2*\j+2*sin{30}*\j) -- +(\a:1);
  \foreach \a in {30,150,270} \draw (sin{60}+2*sin{60}*\i,1+2*\j+2*sin{30}*\j+1+sin{30}) -- +(\a:1);
  \foreach \a in {210,90,-30} \draw (sin{60}+2*sin{60}*\i,1+sin{30}+2*\j+2*sin{30}*\j) -- +(\a:1);
  \foreach \a in {210,90,-30} \draw (2*sin{60}+2*sin{60}*\i,2+2*sin{30}+2*\j+2*sin{30}*\j) -- +(\a:1);
}
\end{tikzpicture}}
\qquad
 \resizebox{0.28\textwidth}{!}{
\begin{tikzpicture}[line width =1.5pt]
%  \node (quad) at (0,0) [font=\small, text width=1 cm]{quad};
  \foreach \i in {0,...,3} 
  \foreach \j in {0,...,3} {
   \foreach \a in {0,90,180,270} \draw (1+\i,1+\j) -- +(\a:1);
}
\end{tikzpicture}}
\qquad
\resizebox{0.25\textwidth}{!}{
\begin{tikzpicture}[line width=1pt]
% \node (quart) at (0,0) [font=\Large, text width=1.3 cm]{trig};
\clip(.7,.3) rectangle (4.2,4.25);
  \foreach \i in {0,...,4} 
  \foreach \j in {0,...,2} {
   \foreach \a in {0,60,120,180,240,300} \draw (1+\i,1+2*sin{60}*\j) -- +(\a:1);
  %\foreach \a in {30,150,270} \draw (sin{60}+2*sin{60}*\i,1+2*\j+2*sin{30}*\j+1+sin{30}) -- +(\a:1);
  %\foreach \a in {210,90,-30} \draw (sin{60}+2*sin{60}*\i,1+sin{30}+2*\j+2*sin{30}*\j) -- +(\a:1);
  %\foreach \a in {210,90,-30} \draw (2*sin{60}+2*sin{60}*\i,2+2*sin{30}+2*\j+2*sin{30}*\j) -- +(\a:1);
  \draw (0,1.87) -- (6,1.87);
  \draw (0,3.6) -- (6,3.6);
}
\end{tikzpicture}}
\caption{
\label{Tilings}
The three regular tilings of the plane with vertices of valency $V=3,4,6$ respectively. 
}   
\end{figure}
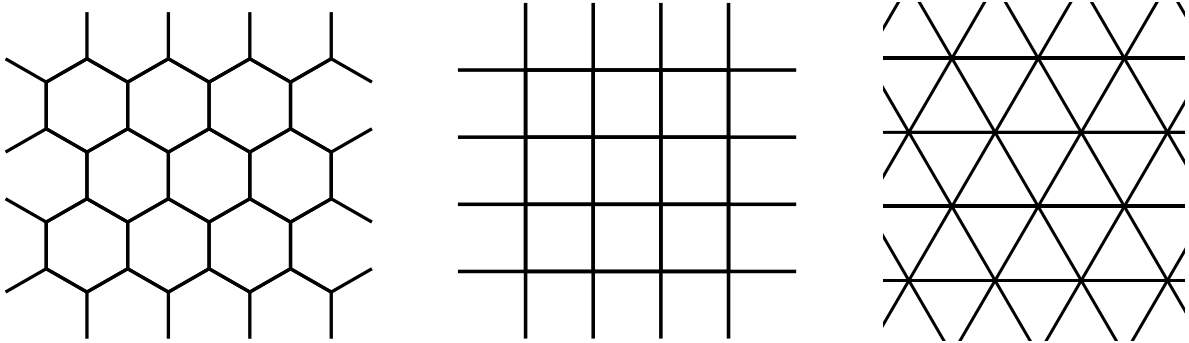

Scalar fields $\phi$ in $D$ spacetime dimensions have conformal dimension $\Delta_\phi=(D-2)/2$. To have  
a conformally invariant interaction $\phi^V$, its valency $V$ has to be $V=2D/(D-2)$. 
Positivity and integrality of $D$ and $V$ singles out the dimensions $D=3,4,6$ with 
valencies $V=6,4,3$, respectively. The planar Feynman graphs that are associated to these interacting theories can be cut out precisely of the three regular tilings of the plane, cf.~\Figref{Tilings}. The Feynman integrals associated to these three classes of graphs have astonishing structural features. It has been fruitful to deviate from the restriction to conformal dimensions of scalar 
fields and to study the structure of fishnet graphs in two-dimensional conformal field theories. 

In this paper, we investigate certain isotropic graphs in two dimensions where conformality requires that the conformal dimension of the fields, which maps to the propagator powers in the Feynman integrals, takes the value $\Delta_\phi=D/V=2/V$. Throughout this section, we keep the discussion general, so that it applies to any dimension $D$.
More precisely, the integrals that we are interested in are given by a class of Feynman integrals that can be defined as follows:
Consider a connected  region cut out along a closed oriented curve $\cC$ intersecting the edges of one of the regular  tilings from~\Figref{Tilings} (each edge is intersected at most once; see ref.~\cite{Chicherin:2017cns}).  In this way, we obtain a connected graph $G$ by considering only the edges of the tiling that intersect $\cC$ (the \emph{external edges}) or lie in its interior (the \emph{internal edges}). The vertices of $G$ are the vertices that lie in the interior of $\cC$ (the \emph{internal points}) and the intersection points between $\cC$ and the edges of the tiling (the \emph{external points}). The vertices are labelled by points in $\mathbb{R}^{D}$, where $D$ denotes the space-time dimension. We denote the labels of the internal and external points by $\xi_i$ and $\alpha_i$, respectively. We call such a graph (together with the labelling of the vertices) a \emph{fishnet graph}. 
Note that the curve $\cC$ defines a cyclic ordering on the external vertices. Unless specified otherwise, we always assume that the labels follow this ordering.

Every edge of $G$ connecting two vertices labelled by $a,b\in\mathbb{R}^D$ represents a propagator $[(a-b)^2]^{-D/V}$. We integrate over the positions of the internal vertices, and we refer to the number $\ell$ of internal vertices as the \emph{number of loops}%
\footnote{These are actual loops in a dual momentum space.}
of $G$. Every fishnet graph $G$ then defines a \emph{fishnet integral}
\beq\label{eq:fishnet_D}
I_G^{(D)}(\underline{\alpha}) = \int\Big[\prod_{i}\rd^D\xi_i\Big]\Big[\prod_{i,j}\frac{1}{[(\xi_i-\xi_j)^2]^{D/V}}\Big]\,\Big[\prod_{i,j}\frac{1}{[(\xi_i-\alpha_j)^2]^{D/V}}\Big]\,,
\eeq
where we introduced the vector of external labels $\underline{\alpha} = (\alpha_1,\ldots,\alpha_n)$. 

For square tilings ($V=4$), fishnet integrals can be understood as correlation functions in the
 bi-scalar fishnet theory defined via the Lagrangian in eq.~\eqref{eq:LagrD}, see refs.~\cite{Gurdogan:2015csr,Kazakov:2018qbr}.
Also for the triangular tiling ($V=6$) in $D=3$ dimensions an associated fishnet Lagrangian exists, which is understood as a limit of ABJM theory~\cite{Caetano:2016ydc}. In the case of the hexagonal lattice ($V=3$) in $D=6$, a scalar fishnet Lagrangian is known (see eq.~\eqref{eq:L_hex}), but so far no interpretation as a limit of some `mother theory' has been found~\cite{Mamroud:2017uyz}.

\subsubsection{Triangular tilings and star-triangle relation}
\label{eq:tri_tiling}

While it is possible to define isotropic fishnet integrals for all three tilings in \Figref{Tilings}, in the remainder of this paper we will mostly be concerned with fishnet integrals associated to the hexagonal and square tilings. The reason for this restriction will be discussed in section~\ref{sec:fishnets_2D}. Let us mention, however, that it is possible to relate some fishnet integrals for triangular and hexagonal tilings using the well-known star-triangle relation. Indeed, conformal symmetry implies that any three-point integral can be identified with a product of three propagators. In particular, for a single integration vertex with propagator powers obeying the condition $\alpha+\beta+\gamma=D$ one finds the well known {star-triangle} identity, which identifies an integration-star with a propagator-triangle,
\begin{align}
\int \frac{\dd^D \xi}{(\alpha_1-\xi)^{2\alpha}(\alpha_2-\xi)^{2\beta}(\alpha_3-\xi)^{2\gamma}}
&=
\frac{X_{\alpha\beta\gamma}}{(\alpha_{1}-\alpha_2)^{2\gamma'}(\alpha_{2}-\alpha_3)^{2\alpha'}(\alpha_{3}-\alpha_1)^{2\beta'}}\\
\includegraphicsbox{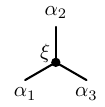}&=\,\,X_{\alpha\beta\gamma}\includegraphicsbox{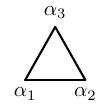}\, .
\label{eq:startriangle}
\end{align}
Here the overall constant factor reads
\begin{equation}
\label{eq:startrianglefactor}
    X_{\alpha\beta\gamma}=\pi^{D/2}\frac{\Gamma(\alpha')\Gamma(\beta')\Gamma(\gamma')}{\Gamma(\alpha)\Gamma(\beta)\Gamma(\gamma)} \, ,
\end{equation}
and we employ the notation $\alpha'=D/2-\alpha$, etc. We note that the star-triangle identity can be interpreted as a Yang-Baxter equation, which plays a distinguished role in the context of integrable models.

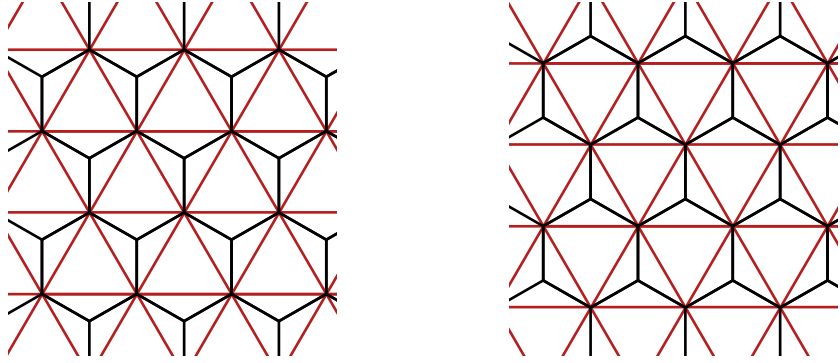
\begin{figure}[!t]
\begin{center}
 \definecolor{darkred}{RGB}{180,30,30}
\begin{tikzpicture}[scale=.9, line width=1pt]
\clip(0.2,.3) rectangle (5,5.5);

\begin{scope}[scale=1.385,draw=darkred,xshift=-4.97mm,yshift=-1.35mm]
  \foreach \i in {0,...,4} 
  \foreach \j in {0,...,2} {
   \foreach \a in {0,60,120,180,240,300} \draw (1+\i,1+2*sin{60}*\j) -- +(\a:1);
   }
  \draw (0,1.87) -- (6,1.87);
  \draw (0,3.6) -- (6,3.6);
  \end{scope}

\begin{scope}[scale=.8,draw=black]
  \foreach \i in {0,...,3} 
  \foreach \j in {0,...,1} {
   \foreach \a in {30,150,270} \draw (2*sin{60}+2*sin{60}*\i,1+2*\j+2*sin{30}*\j) -- +(\a:1);
  \foreach \a in {30,150,270} \draw (sin{60}+2*sin{60}*\i,1+2*\j+2*sin{30}*\j+1+sin{30}) -- +(\a:1);
  \foreach \a in {210,90,-30} \draw (sin{60}+2*sin{60}*\i,1+sin{30}+2*\j+2*sin{30}*\j) -- +(\a:1);
  \foreach \a in {210,90,-30} \draw (2*sin{60}+2*sin{60}*\i,2+2*sin{30}+2*\j+2*sin{30}*\j) -- +(\a:1);
}
\end{scope}
\end{tikzpicture}
%%%%%%%%%%%%%%%
\hspace{2cm}
\begin{tikzpicture}[scale=.9, line width=1pt]
\clip(0.2,-.7) rectangle (5,4.5);

\begin{scope}[scale=1.385,draw=darkred,xshift=-4.97mm,yshift=-1.35mm]
  \foreach \i in {0,...,4} 
  \foreach \j in {-1,...,2} {
   \foreach \a in {0,60,120,180,240,300} \draw (1+\i,1+2*sin{60}*\j) -- +(\a:1);
   }
  \draw (0,1.87) -- (6,1.87);
  \draw (0,3.6) -- (6,3.6);
  \draw (0,.14) -- (6,.14);
  \end{scope}

\begin{scope}[scale=.8,draw=black,xshift=0cm,yshift=-1cm]
  \foreach \i in {0,...,3} 
  \foreach \j in {0,...,1} {
   \foreach \a in {30,150,270} \draw (2*sin{60}+2*sin{60}*\i,1+2*\j+2*sin{30}*\j) -- +(\a:1);
  \foreach \a in {30,150,270} \draw (sin{60}+2*sin{60}*\i,1+2*\j+2*sin{30}*\j+1+sin{30}) -- +(\a:1);
  \foreach \a in {210,90,-30} \draw (sin{60}+2*sin{60}*\i,1+sin{30}+2*\j+2*sin{30}*\j) -- +(\a:1);
  \foreach \a in {210,90,-30} \draw (2*sin{60}+2*sin{60}*\i,2+2*sin{30}+2*\j+2*sin{30}*\j) -- +(\a:1);
}
\end{scope}
\end{tikzpicture}
\end{center}
\caption{Using the star-triangle relation in eq.~\protect\eqref{eq:startriangle}, the triangular fishnet (red) can be transformed into a hexagonal fishnet (black) consisting of elementary three-point vertices only. There are two different possibilities for this identification (left and right figure), distinguished by the subset of vertices to which the star triangle-identity is applied.}
\label{fig:tritohex}
\end{figure}

The above identity can be used to transform the triangular tiling into a hexagonal tiling (see \Figref{fig:tritohex}). On the level of fishnet graphs cut from a tiling, however, this duality only works for a restricted set of graphs. For instance, the following graph cut from a triangular tiling cannot be transformed into a graph that stems from a hexagonal tiling using the star-triangle identity:
\begin{equation}
\includegraphicsbox[scale=1]{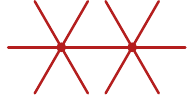} \quad.
\end{equation}
We can complete it, however, by a coincidence limit of two pairs of external points into a graph that has a dual three-point representation:
\begin{equation}\label{eq:hex_to_tri_incidence}
\includegraphicsbox[scale=1]{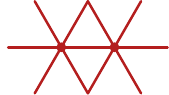}
\quad\to\quad
\includegraphicsbox[scale=1]{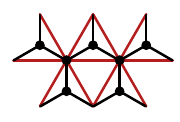}
\quad.
\end{equation}
Hence, even though we restrict in the following to fishnet integrals for the hexagonal and square tilings, (some of) our results can be extended to fishnet integrals for triangular tilings whenever they are related to a hexagonal tiling via a sequence of star-triangle relations.

%%%%%%%%%%%%%%%%%%%%%%%%%%%%%%%%%%%%%%%%%%%%%%%%%%%%%%
%%%%%%%%%%%%%%%%%%%%%%%%%%%%%%%%%%%%%%%%%%%%%%%%%%%%%%
%%%%%%%%%%%%%%%%%%%%%%%%%%%%%%%%%%%%%%%%%%%%%%%%%%%%%%

\subsection{Symmetries of fishnet integrals}
\label{sec:symmetries}
In this section we discuss symmetries of fishnet integrals, i.e., transformations that leave the integrals $I_G^{(D)}(\underline{\alpha})$ invariant. We discuss both infinitesimal and discrete symmetry transformations. 

\subsubsection{Yangian symmetry}\label{sec:yangian}

The Yangian algebra $Y(\mathfrak{g})$ of a Lie algebra $\mathfrak{g}$ can be defined in different equivalent ways, (see, e.g., ref.~\cite{Loebbert:2016cdm}).
In its so-called \emph{first realization}, the Yangian is generated by level-zero generators $\gen{J}^a$ (the generators of the Lie algebra $\mathfrak{g}$) and level-one generators $\levo{J}^a$. These generators obey the commutation relations
\begin{align}
\comm{\gen{J}^a}{\gen{J}^b  } &= f^{ab} {} _c \gen{J}^c \,, &
\big [\gen{J}^a,\levo{J}^b  \big] &= f^{ab} {} _c \levo{J}^c\,,
\label{eq:YangComm}
\end{align}
where $f^{ab}{} _c$ are the structure constants of $\mathfrak{g}$.
The action of the generators on the external labels $\alpha_i$ is given by
\begin{align}
\label{eq:yangian_gens}
\gen{J}^a &= \sum_{j=1}^n \gen{J}_{j}^a\,, &
 \levo{J}^a &=\frac{1}{2} f^a{}_{bc}\sum_{j<k} \gen{J}_{j}^c \gen{J}_{k}^b+ \sum_{j=1}^n \Eval_j \gen{J}_{j}^a\,,
\end{align}
where $\gen{J}_j^a$ is a differential operator in $\alpha_j$. The $\Eval_j$ are constants called \emph{evaluation parameters}, and they parametrise an external automorphism of the Yangian. Their explicit values for fishnet integrals are discussed below. In addition, the Yangian generators have to obey the so-called Serre relations, a quantum extension of the Jacobi identity for Lie algebras, which constrains the representation, see,~e.g.,~refs.~\cite{Loebbert:2016cdm,Miczajka:2022jjv,Dokmetzoglou:2022mfd} for the conformal representation considered below.

In the context of the fishnet integrals studied here, we are interested in the Yangian over the conformal algebra $\mathfrak{so}(1,D+1)$ in $D$ Euclidean spacetime dimensions. We consider a differential representation of the conformal algebra $\mathfrak{so}(1,D+1)$ with generators acting on the external labels $\alpha_j\in\mathbb{R}^D$:
\begin{align}
\gen{P}_j^\mu &= -i \partial_{\alpha_{j}}^\mu\,, 
& \gen{K}_{j}^\mu &= -2i\alpha_j^\mu \brk*{\alpha_j^\nu  \partial_{\alpha_{j},\nu}  + \Delta_j} +i \alpha_j^2 \partial_{\alpha_{j}}^\mu\,,\notag\\
\gen{L}_j^{\mu\nu} &=i\brk1{\alpha_j^\mu \partial_{\alpha_{j}}^\nu - \alpha_j^\nu\partial_{\alpha_{j}}^\mu}\,, &
\gen{D}_j &= -i \brk*{\alpha_j^\mu \partial_{\alpha_{j},\mu} + \Delta_j}\,.
\label{eq:Ddimconfgens}
\end{align}
Here the scaling dimension $\Delta_j$ represents the conformal dimension of the external leg $j$ of the fishnet graph.
If all external labels $\alpha_j$ are distinct, the fishnet integrals in eq.~\eqref{eq:fishnet_D} are invariant under the generators of the Yangian over the conformal algebra~\cite{Chicherin:2017frs,Loebbert:2020hxk,Kazakov:2023nyu}. The Yangian-invariance of fishnet integrals is then expressed by the fact that the latter are annihilated by level-zero and level-one generators:
\beq
\label{eq:invariance}
\gen{J}^aI_G^{(D)} = \levo{J}^aI_G^{(D)} =0\,.
\eeq
Since the generators act via differential operators, eq.~\eqref{eq:invariance} determines a set of partial differential equations satisfied by the fishnet integrals. 
%\cite{Chicherin:2017cns,Chicherin:2017frs}. 
Note that the structure of the conformal algebra allows one to deduce full Yangian invariance from invariance under the level-zero Lie algebra and under one additional level-one generator, e.g., the level-one momentum generator $\levo{P}^\mu$. Moreover, since the Yangian $Y(\mathfrak{so}(1,D+1))$ contains the conformal algebra $\mathfrak{so}(1,D+1)$ as a subalgebra, the fishnet integrals are, in particular, conformally invariant with conformal weight $\Delta_j$ at the external point $\alpha_j$. It follows that we can write a fishnet integral associated with a given Feynman graph $G$ in the form
\beq\label{eq:IFPhi_D}
I_G^{(D)}(\underline{\alpha}) = \mathcal{F}_G^{(D)}(\underline{\alpha})\,\phi_G^{(D)}(\underline{\chi})\,,
\eeq
where $ \mathcal{F}_G^{(D)}(\underline{\alpha})$ is an algebraic function carrying the conformal weight and $\phi_G^{(D)}(\underline{\chi})$ only depends on conformal cross ratios:
\beq
\label{defcrossratios}
\chi_{ijkl} \coloneqq \frac{\alpha_{ij}^2\alpha_{kl}^2}{\alpha_{ik}^2\alpha_{jl}^2}\,,\qquad \alpha_{ij} \coloneqq \alpha_i-\alpha_j\,.
\eeq

%%%%%%%%%%%%%%%

\begin{figure}[t]
\begin{center}
\includegraphicsbox{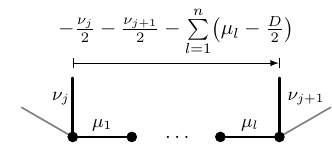}
\end{center}
\caption{Illustration of the rule to associate evaluation parameters to a given Feynman graph. The evaluation parameters $s_j$ and $s_{j+1}$, which are associated to the external legs $j$ and $j+1$, respectively, are related by a term depending on the powers of external ($\nu$) and internal ($\mu$) propagators which lie on the propagator path that connects these external points.}
\label{fig:evalrule}
\end{figure}

\paragraph{Evaluation parameters.}
In order to fully specify the representation by which the Yangian $Y(\mathfrak{so}(1,D+1))$ acts on fishnet integrals, we need to specify the values of the conformal dimensions $\Delta_j$ and the evaluation parameter $\Eval_j$ for each external leg of a fishnet graph $G$. 
For isotropic fishnet graphs, the scaling dimensions are fixed to $\Delta_j=\frac{D}{V}$.
The choice of evaluation parameters, which guarantees the Yangian invariance is determined by the following rule~\cite{Chicherin:2017cns,Loebbert:2020hxk,Kazakov:2023nyu}: We pick an arbitrary external leg to be leg 1, and we fix the associated evaluation parameter to $\Eval_1=0$.%
\footnote{We may shift all evaluation parameters by the same arbitrary constant, due to the level-zero conformal symmetry of the integral.}
Starting with leg 1, we then move clockwise from leg to leg along the boundary of the graph. Depending on the powers of the propagators which connect two neighbouring legs, we add the following term when passing from evaluation parameter $s_j$ to $s_{j+1}$ (cf.~\Figref{fig:evalrule}):
\begin{equation}
   s_{j+1}-s_j=-\frac{\nu_j}{2}-\frac{\nu_{j+1}}{2}-\sum_{l=1}^n\brk*{\mu_l-\frac{D}{2}}. 
\end{equation}
Here the $\nu_j$ denote the powers of external propagators, while the $\mu_k$ denote internal propagator powers on the boundary of the graph. The number of internal propagators connecting two neighbouring external legs may be zero or non-vanishing.
Using this rule, the topology of the graph determines the full set of evaluation parameters of the associated representation of the Yangian. 

\paragraph{Two-point Yangian symmetries.}
One-loop integrals, i.e., integrals with all external legs attached to the same integration vertex, are invariant under any transposition of two external points (provided the respective propagator powers transform as well). In fact, this permutation symmetry can be used to prove that the above level-one Yangian invariance of the full $n$-point one-loop graph is equivalent to the invariance under the level-one generator annihilating any pair $j,k$ of the external legs with propagator powers $\nu_j$ and $\nu_k$ \cite{Loebbert:2020glj}:
\begin{equation}
\levo{J}^a_{jk}=\frac{1}{2} f^a{}_{bc} \gen{J}_{j}^c \gen{J}_{k}^b+  \tilde\Eval_j \gen{J}_{j}^a+  \tilde\Eval_k \gen{J}_{k}^a\,,
\qquad
\tilde\Eval_j=-\frac{\nu_k
}{2} \, , 
\quad \tilde\Eval_k=\frac{\nu_j}{2}\, .
    \label{eq:twositesyms}
\end{equation}
Since this is a local property of the product of propagators, it directly generalises to higher-loop integrals, which are invariant under the above two-site density of the Yangian generator, if the legs $j$ and $k$ are attached to the same integration vertex, here labelled $0$:
\begin{equation}
\label{eq:two-site-density}
\levo{J}^a_{jk}\frac{1}{x_{j0}^{2\nu_j}x_{k0}^{2\nu_k}}=0\,.
\end{equation}
These two-point Yangian symmetries thus furnish additional differential equations for a given Feynman integral. In the subsequent sections we will merely employ the two-point level-one momentum symmetry. 

\subsubsection{Permutation symmetries}
\label{sec:hidden}
Besides the Yangian symmetry, which acts on the fishnet integrals via differential operators, there is also a set of discrete symmetries acting on the fishnet integrals via a permutation of the external points. We denote by $\Perm$ this subgroup of the group of the permutations of the external labels that leaves the integral invariant:
\beq\label{eq:I_G_perm}
I_G^{(D)}(\sigma\cdot \underline{\alpha}) = I_G^{(D)}(\underline{\alpha}) \,, \textrm{   for all } \sigma \in \Perm\,.
\eeq
 In particular, every automorphism of $G$ (i.e., every permutation of its edges and vertices that preserves the incidence relation) acts as a permutation on the labels $\alpha_j$ of the external points,
 and the fishnet integral is invariant under such a permutation. Hence, we see that the group $\Aut(G)$ of automorphisms of $G$ is always a subgroup of $\Perm$. 

\paragraph{Star-triangle relations and hidden symmetries.}

The star-triangle relation \eqref{eq:startriangle} allows one to relate certain conformally-invariant integrals to each other. In particular, we may use the star-triangle relation to relate certain fishnet integrals to other Feynman integrals, with different loop orders and different propagator powers. Note that we have to exclude star-triangle relations which involve a star with exactly one external vertex. Indeed, those relations will result in a graph with an external vertex of valency at least 2, and Yangian invariance for such graphs is not established.%
\footnote{Note that graphs for which an external valency-2 vertex can be turned into a valency-1 vertex are very special since the propagator powers have to obey the star-triangle condition. In general, an external valency-2 vertex can not be eliminated in this way.}
We will call a star-triangle relation that does not involve a star with exactly one external vertex \emph{admissible}.

It can happen that the integral resulting from the application of an admissible star-triangle relation manifests a higher degree of symmetry than the original fishnet integral.
For example, if the graph $G'$ is obtained from a fishnet graph $G$, it may happen that $G'$ has a larger group of automorphisms than $G$. Since $I_G^{(D)}(\underline{\alpha})=I_{G'}^{(D)}(\underline{\alpha})$ (possibly up to some overall rational factor), these additional automorphisms give rise to additional permutation symmetries of $I_G^{(D)}(\underline{\alpha})$. As a consequence, the group $\Perm$ can be larger than the group of automorphisms $\Aut(G)$.

\begin{figure}[!t]
\begin{center}
 \includegraphicsbox[scale=1]{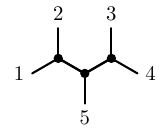} 
 \quad $\to$ \quad
    \includegraphicsbox[scale=1]{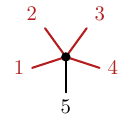} 
    \caption{\label{fig:3-loop_to_pentagon}A three-loop fishnet integral before and after applying twice of the star-triangle relation. Black and red edges correspond to propagators raised to the powers $D/6$ and $D/3$, respectively.}
\end{center}
\end{figure}

As an example, consider the three-loop fishnet graph $G$ shown in~\Figref{fig:3-loop_to_pentagon}. We have $\Aut(G)=\mathbb{Z}_2^3$. Applying the star-triangle relation to the vertices on either end, we obtain a one-loop five-point graph $G'$, with $\Aut(G') = S_4$ (which corresponds to permuting the four red legs~in~\Figref{fig:3-loop_to_pentagon}). It follows that $I_G^{(D)}(\underline{\alpha})$ enjoys a permutation symmetry that is larger than $\Aut(G)$:
\beq
\Aut(G) = \mathbb{Z}_2^3 \subset \Perm=S_4\,.
\eeq

\paragraph{Permutation symmetries and Yangian symmetry.}
The expression for the level-one generator in eq.~\eqref{eq:yangian_gens} depends on an order of the external points $\alpha_j$. At this point there is an ambiguity. If $\gen{J}^a$ and $\levo{J}^a$ are the generators corresponding to a given ordering of the external points and $\sigma\in\Perm$, then
\beq
 \gen{J}^a_{\sigma} \coloneqq \sigma\gen{J}^a\sigma^{-1} \textrm{~~~and~~~}  \levo{J}^a_{\sigma} \coloneqq \sigma\levo{J}^a\sigma^{-1}
 \eeq
 define another representation corresponding to a different ordering, that also annihilates the integral.
 From eq.~\eqref{eq:yangian_gens} we immediately see that $\gen{J}^a_{\sigma}=\gen{J}^a$ for all $\sigma\in \Perm$. Moreover, Yangian invariance follows from conformal invariance plus invariance under the level-one momentum operator $\levo{P}^\mu$, and thus we only need to consider the action of the permutations on the latter operator. For a non-trivial permutation, the differential operator $\levo{P}^{\mu}_{\sigma}$ will in general be different from $\levo{P}^{\mu}$. In other words, there is a family of level-one momentum operators $\Perm$ that annihilate $I_G^{(D)}$:
 \beq
\levo{P}^\mu_{\sigma}I_G^{(D)} =0\,, \textrm{   for all } \sigma \in \Perm\,.
\eeq
For example, the fishnet integral corresponding to the graph in \Figref{fig:3-loop_to_pentagon} is annihilated by $\levo{P}^{\mu}_{\sigma}$ for every permutation $\sigma\in S_4$ that exchanges the labels of the legs in red in the right-hand graph.

Similar conclusions can be drawn for the differential operators which correspond to the two-site densities $\levo{J}^a_{ij}$, but here we can find even more hidden symmetries. This is most easily explained on the example in~\Figref{fig:3-loop_to_pentagon}. Clearly, the integral is annihilated by all two-site densities $\levo{J}^a_{ij}$, where $i$ and $j$ is any pair of red legs in the right-most graph, in agreement with the permutation symmetries from $\Perm$. However, since eq.~\eqref{eq:two-site-density} holds for arbitrary propagator powers, the integral for the graphs in~\Figref{fig:3-loop_to_pentagon} is annihilated by the two-site densities for \emph{any} pair of legs $i$ and $j$ attached to the same vertex. 
Hence, we see that for the two-site densities, there is an even larger set of hidden symmetries.

\subsubsection{Summary}
From the preceding discussions, we conclude that the symmetry algebra of a fishnet integral $I_G^{(D)}(\underline{\alpha})$ has the form of a semi-direct product $\Perm\ltimes \mathcal{Y}_G$. Here $\Perm$ denotes the subgroup of the permutations of the external labels that leaves the integral invariant (cf. eq.~\eqref{eq:I_G_perm}) and $\mathcal{Y}_G$ is the algebra generated by 
\begin{itemize}
\item the Yangian $Y(\mathfrak{so}(1,D+1)$,
\item the two-site densities $\levo{J}^a_{jk}$ where $\alpha_j$ and $\alpha_k$ are attached to the same internal vertex of any graph obtained from $G$ through application of an admissible star-triangle relation. 
\end{itemize}
A permutation $\sigma\in\Perm$ acts on $\mathcal{Y}_G$ by sending $\gen{J}\in \mathcal{Y}_G$ to $\sigma \gen{J}\sigma^{-1}$. To summarise, fishnet integrals enjoy a rather large degree of symmetry. It is an interesting question if the fishnet integrals are in fact uniquely fixed by this symmetry algebra. While in general dimension $D$ the answer is not known, we provide a conjectural answer in $D=2$ in the next sections.

%%%%%%%%%%%%%%%%%%%%%%%%%%%%%%%%%%%%%%%%%%%%%%%%%%%%%%
%%%%%%%%%%%%%%%%%%%%%%%%%%%%%%%%%%%%%%%%%%%%%%%%%%%%%%
%%%%%%%%%%%%%%%%%%%%%%%%%%%%%%%%%%%%%%%%%%%%%%%%%%%%%%

%%%%%%%%%%%%%%%%%%%%%%%%%%%%%%%%%%%%%%%%%%%%%
%%%%%%%%%%%%%%%%%%%%%%%%%%%%%%%%%%%%%%%%%%%%%

\section{Fishnet integrals in 2 dimensions and Calabi-Yau motives}
\label{sec:fishnets_2D}

\subsection{Fishnet integrals in 2 dimensions}

While so far the discussion was valid for arbitrary space-time dimensions $D$, in ref.~\cite{Duhr:2022pch} we have focused on fishnet integrals for the square tiling in $D=2$ dimensions. In two Euclidean dimensions, it is convenient to package the labels $\alpha_j,\xi_j\in\mathbb{R}^2$ into complex variables:
\beq
a_j \coloneqq \alpha_j^1 + i\alpha_j^2 \textrm{~~~and~~~}x_j \coloneqq \xi_j^1 + i\xi_j^2\,.
\eeq
It is easy to see that the fishnet integral from eq.~\eqref{eq:fishnet_D} can be cast in the form
\beq
\label{32}
I_G(\ua) = \int\left(\prod_{j=1}^{\ell}\frac{\rd x_j\wedge\rd \overline{x}_j}{-2i}\right)\frac{1}{|P_G(\ux,\ua)|^{4/V}}\,,
\eeq
with $\ua= (a_1,\ldots,a_n)$, $\ux = (x_1,\ldots,x_\ell)$, and we dropped the dependence of the integral on the space-time dimension for readability.
The integrand is defined by the polynomial
\beq\label{eq:P_G_def}
P_G(\ux,\ua) = \Big[\prod_{i,j}(x_i-x_j)\Big]\,\Big[\prod_{i,j}(x_i-a_j)\Big]\,,
\eeq
where the product ranges depend on the graph topology.

The conformal algebra $\alg{so}(1,3)$ in 2 dimensions splits into a holomorphic and anti-ho\-lo\-mor\-phic copy of $\alg{sl}(2,\mathbb{R})$.
The Yangian likewise splits into holomorphic and antiholomorphic parts:
\beq\label{eq:split_Y}
Y(\mathfrak{so}(1,3)) = Y(\mathfrak{sl}(2,\mathbb{R})) \oplus \overline{Y(\mathfrak{sl}(2,\mathbb{R}))}\,.
\eeq 
The holomorphic generators of $\mathfrak{sl}(2,\mathbb{R})$ can be expressed in terms of the conformal generators in two spacetime dimensions from eq.~\eqref{eq:Ddimconfgens}:
\begin{align}\label{eq:P2D}
\mathbb{P}&=-\frac{i}{2}\brk*{\gen{P}_1+i\gen{P}_0}=-i\partial_a\, ,
% =\gen{P}^\text{1D}_{\alpha\to a}(\Delta/2)\,,
\\
\mathbb{K}&=+\frac{i}{2}\brk*{\gen{K}_1-i\gen{K}_0}=-ia(a\partial_a+\Delta)\, ,
% =\gen{K}^\text{1D}_{\alpha\to a}(\Delta/2)\,,
\\
\mathbb{D}&=+\frac{i}{2}\brk*{\gen{L}_{01}-i\gen{D}}=-i\brk*{a\partial_a+\Delta/2}\, .
% =\gen{D}^\text{1D}_{\alpha\to a}(\Delta/2)\,.
\end{align}
Similar expressions hold for the antiholomorphic versions, and we have employed the abbreviation
\begin{equation}
\partial_a=\frac{1}{2}\brk*{\partial_{\alpha_0}-i\partial_{\alpha_1}}\,.
\end{equation}
% \begin{align}
% \gen{P}_\pm=\half \brk*{\gen{P}_0\pm \gen{P}_1}\,,
% \qquad
% \gen{K}_\pm=\half\brk*{\gen{K}_0\mp \gen{K}_1}\,,
% \qquad
% \gen{D}_\pm=\half\brk*{\gen{D}\mp \gen{L}_{01}}\,.
% \end{align}
The above generators obey the $\mathfrak{sl}(2,\mathbb{R})$ commutation relations
\begin{align}
\comm{\mathbb{D}}{\mathbb{P}}=i \mathbb{P}\,,
\qquad
\comm{\mathbb{D}}{\mathbb{K}}=-i \mathbb{K}\,,
\qquad
\comm{\mathbb{K}}{\mathbb{P}}=2i \mathbb{D}\,.
\end{align}
This decomposition of the conformal algebra carries over to the level-one generators $\levo{J}^a$. In particular, the level-one momentum generator for the Yangian $Y\big(\alg{sl}(2,\mathbb{R})\big)$ takes the simple form
\begin{equation}\label{eq:Phat_2D}
\widehat{\mathbb{P}}
% =\sum_{j<k}\levo{P}_{jk}
=\frac{i}{2}\sum_{j<k}\brk*{\mathbb{P}_j \mathbb{D}_k-\mathbb{P}_k \mathbb{D}_j}+\sum_j \Eval_j \mathbb{P}_j \,,
\end{equation}
which can be understood as a sum over a two-site density
\begin{equation}
    \widehat{\mathbb{P}}_{jk}=\frac{i}{2}\brk*{\mathbb{P}_j \mathbb{D}_k-\mathbb{P}_k \mathbb{D}_j}+\tilde\Eval_j \mathbb{P}_j+\tilde\Eval_k \mathbb{P}_k\, .
\end{equation}

The previous discussion implies that the symmetry algebra of a fishnet graph $G$ in two dimensions (cf.~section~\ref{sec:symmetries}) can be separated into a holomorphic and an antiholomorphic part. Indeed, we can write 
\beq
\mathcal{Y}_G = Y_G \oplus \overline{Y}_G\,,
\eeq
where $Y_G$ is generated by the Yangian $Y(\mathfrak{sl}(2,\mathbb{R}))$ and the holomorphic two-site densities (and their permutations, cf.~section~\ref{sec:hidden}), and $\overline{Y}_G$ is its complex conjugate. The permutation symmetries from $\Perm$ naturally act separately on the holomorphic and antiholomorphic variables. Hence, we obtain:
\beq
\Perm\ltimes \mathcal{Y}_G = (\Perm\ltimes {Y}_G )\oplus(\Perm\ltimes \overline{Y}_G )\,.
\eeq
This symmetry algebra allows one to attach a differential ideal (i.e., a right-ideal of differential operators) to every fishnet graph in two dimensions. We define the \emph{Yangian differential ideal of $G$}, denoted by $\YDI(G)$, as the differential ideal of holomorphic differential operators generated by 
\begin{itemize}
\item the Yangian generators from $Y(\mathfrak{sl}(2,\mathbb{R}))$ and all its $\Perm$ permutations,
\item the holomorphic two-site densities, and all its permutations, obtained from admissible star-triangle relations.
\end{itemize}

Conformal invariance implies that we can write the integral in a form very similar to eq.~\eqref{eq:IFPhi_D},
\beq\label{eq:I-Fphi}
I_G(\ua) = \big|F_G(\ua)\big|^2\,\phi_G(\uz) =(-1)^{\frac{\ell(\ell-1)}{2}}\left(-{2i}\right)^{-\ell}\big|F_G(\ua)\big|^2 \int\overline{\Omega}\wedge\Omega\,, 
\eeq
where $F_G(\ua)$ is a holomorphic rational function that carries the conformal weights, and $\phi_G(\uz)$ is a function only of holomorphic conformal cross ratios:
\beq
\label{defzcrossratios}
\chi_{i,j,k,l} \coloneqq \frac{a_{ij}^2a_{kl}^2}{a_{ik}^2a_{jl}^2}\,,\qquad a_{ij} \coloneqq a_i-a_j\,,
\eeq
and $\uz$ denotes a vector of independent cross ratios.
We also introduced the $(\ell,0)$-form
\beq
\Omega = \frac{1}{F_G(\ua)}\,\frac{\rd x_1\wedge\ldots\wedge \rd x_{\ell}}{P_G(\ux,\ua)^{2/V}}\,.
\label{eq:Omega}
\eeq
Note that $\Omega$ is conformally invariant and only depends on the conformal cross ratios $\uz$.

In ref.~\cite{Duhr:2023bku}, two of the authors have shown that, in the case of non-integer propagator powers, Feynman integrals in two Euclidean dimensions can be expressed as single-valued analogues of Aomoto-Gel'fand hypergeometric functions. This condition is always satisfied for the fishnet integrals considered here, because the propagator powers are $2/V$ with $V=3,4$. It follows that we can write the integral in the form
\beq\label{eq:phi_SV}
\phi_G(\uz) = (-i)^\ell\,\uPi_G(\uz)^{\dagger}\Sigma_G\uPi_G(\uz)\,,
\eeq
where $\uPi_G(\uz)$ are ordinary/holomorphic Aomoto-Gel'fand hypergeometric functions, which are twisted periods of some twisted cohomology group:
\beq\label{eq:period_vector}
\uPi_G(\uz) = \Big(\int_{\Gamma_0}\!\!\!\Omega,\ldots,\int_{\Gamma_{b_\ell-1}}\!\!\!\!\!\!\Omega\Big)^T\,.
\eeq
Here $b_\ell$ denotes the dimension of the relevant twisted cohomology group, and the $\Gamma_i$ form a basis of the corresponding twisted homology group. $\Sigma_G$ is (the inverse of) the intersection matrix of the twisted cycles $\Gamma_i$. Equation~\eqref{eq:phi_SV} implies that we can compute $\phi_G$ from the knowledge of the twisted periods $\uPi_G(\uz)$ and the intersection matrix $\Sigma_G$. While there are methods how to do that in principle (cf. ref.~\cite{AomotoKita} for a review), this can be rather difficult in practise. In the case of fishnet graphs, however, we can apply more powerful techniques by using knowledge about the underlying geometry.

%%%%%%%%%%%%%%%%%%%%
\subsection{Fishnet integrals and Calabi-Yau varieties}
\label{sec:fishnet_CYs}

%%%%%%%%%%%%%%%%%%%%

In ref.~\cite{Duhr:2022pch} we argued that fishnet integrals for square tilings in $D=2$ dimensions are closely related to Calabi-Yau (CY) geometries. In  ref.~\cite{Duhr:2023eld} we have generalized this analysis as reviewed below, such that it also applies to fishnet integrals for hexagonal tilings. We can then leverage methods developed for CY varieties for the computation of fishnet integrals. This was done explicitly in ref.~\cite{Duhr:2022pch} for train track integrals for square tilings of low loop order, and extended to four-point fishnet integrals for square tilings in ref.~\cite{Duhr:2023eld}. One of the main results of this paper is to use these methods to obtain explicit expressions for fishnet integrals associated to hexagonal tilings of low loop order. As explained in ref.~\cite{Duhr:2023eld}, we are currently unable to prove that the geometry associated to triangular tilings is CY, and so our results are currently restricted to square and hexagonal tilings. However, as explained in section~\ref{eq:tri_tiling}, some fishnet integrals for the triangular tiling are related to the hexagonal tiling via the star-triangle relations, and (some of) our results can be extended to those cases. We also stress that (some of) the results of this subsection are valid for fishnet graphs where some of the external labels are identified. Before we discuss concrete examples in subsequent sections, we review some of the general concepts of CY varieties for fishnet integrals introduced in refs.~\cite{Duhr:2022pch,Duhr:2023eld}

Loosely speaking, a CY $\ell$-fold is a complex $\ell$-dimensional K\"ahler manifold $M$ that admits a unique holomorphic $(\ell,0)$-form $\Omega$. The \emph{periods} of $M$ are obtained by integrating $\Omega$ over a basis of cycles $\Gamma_i$ that span the middle homology $H_\ell(M,\mathcal{R})$, and $b_{\ell}\coloneqq\dim H_\ell(M,\mathcal{R})$. Here $\mathcal{R}$ is a ring, typically $\mathcal{R}=\mathbb{Z}$ or some extension of $\mathbb{Z}$ in the case the CY is generically singular. The CY variety associated to a fishnet graph $G$ in two dimensions can be constructed as follows: To each internal vertex of $G$ 
we associate a $\mathbb{P}^1$  with homogeneous coordinates $[x_i:u_i]$, $i=1,\ldots, l$ over which we want to  integrate with the measure 
\begin{equation}  
\mathrm d \mu_i=u_i \mathrm dx_i-x_i \mathrm du_i \, . 
\end{equation} 
To every edge connecting two internal vertices we assign a factor $(u_j x_i-x_j u_i)$, and to every edge connecting an internal vertex to the external vertex labelled by $a_j$ a factor $(x_i-a_j u_i)$. In this way we can  associate  
an (in general  singular) $\ell$-dimensional CY variety $M_G$ as the ($d\coloneqq6-V$)-fold cover 
\begin{equation} 
 W={y^d}- P([\underline{x}:\underline{u}];\underline{a})=0
\label{eq:defsingCY} 
\end{equation} 
over the base $B=(\mathbb{P}^1)^\ell$ branched at (cf.~eq.~\eqref{eq:P_G_def})
\begin{equation}  
P([\underline{x}:\underline{w}];\underline{a})=\prod_{ij}(u_j x_i-x_j u_i)  \prod_{ij}(x_i-a_j u_i)=0 \,.
\label{eq:defsingbranch} 
\end{equation}   
Note that for square and hexagonal tilings, we have $d=2$ and $d=3$ respectively. 
The orders of the covering automorphism exchanging the sheets will play a 
crucial role when studying the geometry.  

Equation~\eqref{eq:defsingCY} defines a  
CY manifold, because the canonical 
class of the base is given by
\begin{equation}  
K_B=2 \bigoplus_{i=1}^{\ell}  H_i \,,
\end{equation}   
with  $H_i$ the hyperplane class of the $i^{\textrm{th}}$ $\mathbb{P}^1$,
and the CY condition  ensuring  $K_{M_\ell}=-c_1(T_{M_\ell})=0$  reads:
\begin{equation}  
\frac{d}{d-1} K_B=[P([\underline{x}:\underline{u}];\underline{a})]=V \bigoplus_{i=1}^{\ell}  H_i =\frac{V}{2}\,K_B\,.
\label{eq:KB} 
\end{equation} 
This constraint is fulfilled precisely for $d=6-V$ and $V=3,4$, which corresponds to the hexagonal and the square tilings, 
respectively. This implies
that the CY $(\ell,0)$-form
\begin{equation}
\tilde \Omega=
\frac{\prod_{i=1}^\ell {\rm d} \mu_i}{\frac{\partial W}{\partial y}}
=\frac{\prod_{i=1}^\ell {\rm d} \mu_i}{P^\frac{d-1}{d}([\underline{x}:\underline{u}];a)}
\end{equation}
is well defined under scaling of the projective coordinates. 
It agrees up to K\"ahler gauge transformations by $F_G(a)$ with eq.~\eqref{eq:Omega} written in affine coordinates, and hence the fishnet integral in eq.~\eqref{eq:I-Fphi} is the period bilinear of the Calabi-Yau 
$\ell$-variety defined in eq.~\eqref{eq:defsingCY}. Note that for the second equal sign in eq.~\eqref{eq:KB}, we used the fact 
that all valencies and propagator weights are equal, which 
will not be true for non-isotropic graphs.

%% file: geometry.tex
Due to the special factorised form of the branch locus in eq.~\eqref{eq:defsingbranch}, our CY $\ell$-folds 
defined as triple or double coverings by eq.~\eqref{eq:defsingCY} are singular. 
However, they are easily deformable to a smooth hypersurface in projective toric ambient 
spaces $\mathbb{P}_{\triangle^*}$ and $\mathbb{P}_{\triangle}$  specified by a pair of reflexive lattice polyhedra $(\triangle,\triangle^*)$. To define the latter, 
let $e_i$, $i=1,\ldots, \ell+1$, be the unit vectors  in $\mathbb{R}^{\ell+1}$ spanning a 
lattice $\mathbb{Z}^{\ell+1}$. Then 
\begin{equation} 
\begin{array}{rl}
\triangle &=\displaystyle{{\rm Conv} \left\{ (d-1)e_1-\sum_{k=2}^{\ell+1} e_k,\  -e_1+\sum_{k=2}^{\ell+1}\left[\frac{V}{2}(1\pm 1) -1\right] e_k\right\}} \ ,  \\ [5 mm]
\triangle^*&=\displaystyle{{\rm Conv}\left\{-\left\lfloor\frac{4}{d}\right\rfloor e_1-e_i, i=2,\ldots, \ell+1, \ e_i, i=1,\ldots, \ell+1\right\}}  \ . 
\end{array}      
\label{eq:polyhedra} 
\end{equation}  
Here $\triangle$ is the Newton polytope  of $W$  with the generic deformations of $P$ compatible with eq.~\eqref{eq:KB}. We call the corresponding Newton polynomial $W_{\triangle}$.  
The polytope $\triangle$ has $1+2^\ell$ vertices  specified by the uncorrelated sign choices in the first line of eq.~\eqref{eq:polyhedra}.  The fact that the polar polyhedron 
$\triangle^*$ is a lattice polyhedron is obvious from the integrality of its $2\ell+1$ vertices. This is another way to see that the hypersurfaces defined by
$W_{\triangle}=0$ in $\mathbb{P}_{\triangle^*}$  and  $W_{\triangle^*}=0$ in $\mathbb{P}_{\triangle}$ are mirror pairs of CY manifolds. We will illustrate the deformation and resolution of singularities on examples in section~\ref{sec:blow-up}.

Having identified the geometry associated to fishnet integrals in two dimensions for square and hexagonal tilings with CY manifolds, we may use the arsenal developed for the computation of CY periods to evaluate the $\uPi_G(\uz)$. Indeed, it is possible to compute a basis of periods as the solutions of an ideal of differential operators, the so-called \emph{Picard-Fuchs differential ideal} of $M_G$, denoted in the following by $\PFI(M_G)$. While there are general techniques to compute the PFI of a given family of CY varieties, this can still be a monumental task, especially for families parametrised by more than one modulus (which is typically the case for fishnet integrals). In ref.~\cite{Duhr:2022pch} we conjectured that for a fishnet graph associated with a square tiling, $\PFI(M_G)$ is generated precisely by (the holomorphic part of) the Yangian generators and two-point symmetries together with all permutation symmetries corresponding to the automorphisms of $G$. The conjecture of ref.~\cite{Duhr:2022pch} was motivated by investigating square fishnet graphs with up to four loops. Based on our results for hexagonal tilings (see section~\ref{sec:examples}), we observe that this conjecture extends to hexagonal tilings, albeit only if hidden permutation symmetries are taken into account (cf. the discussion in section~\ref{sec:symmetries}). We therefore have the following

\begin{quote}
{\bf Conjecture:} \emph{The Picard-Fuchs ideal $\PFI(M_G)$ for the CY varieties $M_G$ attached to an isotropic%
\footnote{Here isotropic means that all propagator powers are equal.} fishnet graph $G$ in two dimensions associated to square and hexagonal tilings is equal to the Yangian differential ideal $\YDI(G)$. As a consequence, these fishnet integrals are completely determined by their symmetry algebra \emph{Perm}$_G\ltimes{Y}_G$.}
\end{quote}

Let us conclude this section with two comments. First, the argument why fishnet graphs associated to square and hexagonal tilings are associated to CY geometries holds for arbitrary values of the external points, including cases where some of the external points are identified. However, since Yangian-invariance is only established for fishnet integrals with generic external points (i.e., all external points are distinct), our conjecture relating the PFI and the Yangian differential ideal only holds for graphs with generic external points. Second, we have already mentioned that we are currently unable to extend our results to general fishnet integrals for triangular tilings. In cases where we can apply a sequence of star-triangle relations to relate fishnet graphs for the triangular and hexagonal tilings, also the graph for the triangular tiling will be associated to a CY
 geometry. However, since this application of the star-triangle relation will generically give rise to fishnet graphs with $V=3$ and with identified external points (cf.~eq.~\eqref{eq:hex_to_tri_incidence}), our conjecture about the PFI does not apply to triangular tilings.

\subsection{Triangle tracks and Picard varieties } 
\label{sec:CYMinPicardvarieties}

In this section, we have a closer look at the geometries that arise from hexagonal tilings, and we relate them to varieties that have appeared in the mathematics literature.

 In the context of square fishnet graphs, a particularly interesting subclass of diagrams are the $\ell$-loop traintrack graphs $G_{1,\ell}$, owing their name to the dual (green line) representation in momentum space,%
 \footnote{Here the momenta $p_j$ relate to the $x$-variables via $p_j=x_j-x_{j+1}$ (not via Fourier transform).} cf.~ref.~\cite{Bourjaily:2018ycu}:
\begin{equation}
 \includegraphicsbox[scale=1]{FigTraintrackDual} \, .	
\end{equation}
The natural generalisation in the context of three-point fishnet integrals, which we denote \emph{triangle track} graphs  $Z_\ell$,
takes the form%
\footnote{This is the new track pattern currently implemented by \emph{Deutsche Bahn} all over Germany, which explains the frequent delays on the road towards efficiency, see e.g.\ this \href{https://www.theguardian.com/business/2023/oct/14/its-the-same-daily-misery-germanys-terrible-trains-are-no-joke-for-a-nation-built-on-efficiency}{link}.}
\begin{equation}
 \includegraphicsbox[scale=1]{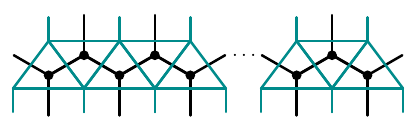} \, .	
\end{equation}
%
% In the previous section we have seen that there is a fundamental difference between the CY varieties obtained from the square and hexagonal tilings. 
%
In order to relate the associated Feynman integrals to notions of geometry, we will review some mathematical concepts in the following.

A \emph{Picard curve} is a Riemann 
surface (of complex dimension 1) given by a triple covering of $B=\mathbb{P}^1$  defined in affine 
coordinates by the equation\footnote{For an early reference, see, e.g., refs.~\cite{MR1504014,MR1554595}. See also ref.~\cite{MR1350073} for a review restricted to ${\rm deg}_x\widetilde{P}=4$.} 
\begin{equation}
y^3=\widetilde{P}(x,\underline{a}) \ .
\label{eq:PicardCurves} 
\end{equation}
Here the polynomial $\widetilde{P}(x,\underline{a})$ has degree
 ${\rm deg}_x(\widetilde{P})>3$ in $x$.
The genus $g$, as calculated by the Hurwitz formula for the generic deformation 
of $\widetilde{P}(x,\underline{a})$, is greater than 1, $g>1$, and the parameters $\underline{a}$ denote the complex moduli of the curve. To be more concrete, let us explain how we can compute the genus of a complex curve $C$. The genus $g=(2-\chi(C))/2$ is given in terms of its Euler number $\chi(C)$ determined by the Riemann-Hurwitz formula 
\begin{equation}
\chi(C)= d \chi(C')-\sum_{q\in C} (\nu(q)-1)\ .   
\end{equation}
Here $d$ denotes the generic number of sheets in the multicover 
of $C$ over $C'$, $q$ are the branch points and $\nu(q)$ is the 
branching order at $q$. For example, hyperelliptic curves are 
double covers of $\mathbb{P}^1$ ($\chi(\mathbb{P}^1)=2$) branched at 
$2n$ points with branching order $2$ and have therefore genus $g=n-1$.
Picard curves have been studied first 
in ref.~\cite{MR1504014} already with the understanding that their periods have interesting modular properties \cite{MR1554595}.

\begin{figure}[!t]
\begin{center}
$Z_2$:\quad \includegraphicsbox{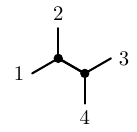}
\quad$\to$\quad
\includegraphicsbox{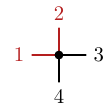}
\end{center}

\caption{The two-loop graph $Z_2$ can be identified with a box graph (modulo an overall rational factor), when using the star-triangle identity. Black  and red propagators are raised to powers $1/3$ and $2/3$, respectively. }
\label{fig:2zigzag}
\end{figure}

To see how Picard curves arise from CY varieties attached to fishnet graphs, let us start by considering the two-loop graph $Z_2$.
After the application of the star-triangle relation, the integral reduces to a one-loop four-point integral, albeit with two different powers for the propagators (see \Figref{fig:2zigzag}).  
The geometry associated to a one-loop integral is a complex curve (complex dimension $\ell=1$). In other words, 
application of the star-triangle relation changes the geometry from 
a (singular) {K3} surface to a (singular) Picard curve of genus 2 given by 
\begin{equation} 
y^3=(x-a_1)(x-a_2)(x-a_3)^2 (x-a_4)^2\ .
\label{eq:Picard1}
\end{equation}

Similarly, we can consider the two-parameter family of singular 
three-dimensional CY varieties for the triangle track graph $Z_3$ (see \Figref{fig:3zigzag}). Application of the star-triangle relation allows us to express this three-loop integral as a one-loop five-point function, whose geometry is again a Picard curve, but this time with two independent moduli:
\begin{figure}[t]
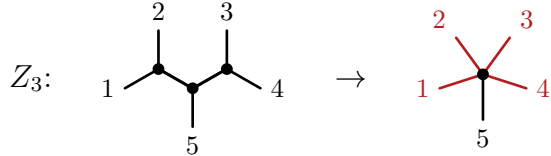

\begin{center}
$Z_3$:\quad \includegraphicsbox{Fig3LoopZigZag}
\quad$\to$\quad
\includegraphicsbox{Fig3LoopZigZagToPentagon}
\end{center}

\caption{The three-loop triangle track graph can be related to a one-loop five-point function via the star-triangle identity. Black  and red propagators are raised to powers $1/3$ and $2/3$, respectively. }
\label{fig:3zigzag}
\end{figure}
\begin{equation} 
y^3=(x-a_1)(x-a_2)(x-a_3) (x- a_4)(x-a_5)^2\ .
\label{eq:Picard2}
\end{equation}
We can compute the genus, and we find $g=3$. Both the Picard curves in eqs.~\eqref{eq:Picard1} and~\eqref{eq:Picard2} have been discussed in ref.~\cite{CFOZ}, where it was in particular found that the periods of the curve can be expressed in terms of so-called Picard modular forms. We will discuss the evaluation of the periods in section~\ref{sec:examples}.

At higher loops we define analogously a 
\emph{Picard variety} as the triple covering of $(\mathbb{P}^1)^r$ 
given by the affine equation 
\begin{equation}
y^3=\widetilde{P}_G(\underline{x},\underline{a}) \, ,
\label{eq:PicardVarities} 
\end{equation}
where we now understand the $r$ coordinates $\underline{x}=(x_1,\ldots, x_r)$ 
as inhomogeneous coordinates of the $(\mathbb{P}^1)^r$. Note that the polynomial $\widetilde{P}_G$ is different from the polynomial $P_G$ that defines the CY variety in eq.~\eqref{eq:P_G_def}, but it can easily be obtained from the latter via the star-triangle relation. Note that
every application of the star-triangle relation lowers the first Chern
class of the variety, by lowering the canonical  class of the base $K_B$
and increasing the class of the branch locus. Even though
eq.~\eqref{eq:PicardVarities} defines a singular variety, its first Chern class is well defined by considering the manifold that corresponds to 
the smooth generic deformation of eq.~\eqref{eq:PicardVarities}. 
Since we start with the Calabi-Yau variety the first Chern class 
of the Picard variety is always negative and bounded by 
the CY variety that is obtained by applying the star-triangle relation backwards.

\subsection{Varieties vs.\ motives}
\label{sec:motives}
Let us now discuss the consequences of identifying the geometries of triangle track graphs with Picard curves (or Picard varieties at higher loops). In section~\ref{sec:fishnet_CYs} we have shown that we can attach to every $\ell$-loop fishnet graph associated with a hexagonal tiling a family of (singular) CY $\ell$-folds. We can explicitly describe this family of CY $\ell$-folds as the triple cover over the base $B=\big(\mathbb{P}^1\big)^{\ell}$ given in affine coordinates $\ux=(x_1,\ldots,x_\ell)$ of $B$ by
\beq
y^3 = P_G(\ux,\ua)\,.
\label{eq:cubics}
\eeq
Alternatively, we may apply the star-triangle relation $(\ell-r)$ times to attach to a triangle track graph a family of $r$-dimensional Picard varieties, which can be explicitly described as the triple-cover over the base $\big(\mathbb{P}^1\big)^{r}$ given in affine coordinates $\ux=(x_1,\ldots,x_r)$ of $\big(\mathbb{P}^1\big)^{r}$ by
\beq
y^3 = \widetilde{P}_G(\ux,\ua)\,.
\eeq
Examples of the different CY  and Picard varieties obtained for low loop orders for triangle track graphs can be found in 
table~\ref{Table:topdataone}. 

\begin{table}[h!]
{{ 
\begin{center}
	\begin{tabular}{|c|c|c|}
		\hline
$\ell$ & \small CY variety  & \small Picard variety  \\ \hline

\multirow{2}{*}{$1$} & \small Elliptic curve (CY one-fold) &\multirow{2}{*}{--}\\[0.5ex]
&$y^3=(x_1-a_1)(x_1-a_2)(x_1-a_3)$ &  \\[1ex]
\hline
\multirow{3}{*}{$2$} & \small singular K3 surface (CY two-variety) &  \small Picard curve ${\cal C}^{(2)}_1$ of genus $g=2$ \\[0.5ex]
& $y^3=(x_1-a_1)(x_1-a_2)(x_1-x_2)$ &  $ y^3=(x_1-a_1)(x_1-a_2)(x_1-a_3)^2 $  \\
&$(x_2-a_3)(x_2-a_4)$ &  $(x_1-a_4)^2 $\\[1ex]
\hline
\multirow{3}{*}{$3$} & \small CY three-variety &  \small Picard two-variety ${\cal P}_2$  \\[0.5ex]
& $y^3=(x_1-a_1)(x_1-a_2)(x_1-x_2)(x_2-a_5)$ & $ y^3=(x_1-a_1)^2(x_1-a_2)^2(x_1-a_5)$  \\
& $(x_2-x_3)(x_3-a_3)(x_3-a_4) $&  $(x_1-x_2)(x_2-a_5)  $\\
& &  $(x_2- a_3)^2(x_1-a_4)^2 $\\[1ex]
\multirow{3}{*}{} &  &  \small Picard curve ${\cal C}_1^{(3)}$ of genus $g=3$ \\[0.5ex]
&  & $ y^3=(x_1-a_1)(x_1-a_2)(x_1-a_3) $  \\
&  & $(x_1- a_4) (x_1-a_5)^2$  
\\[1ex]
\hline
\multirow{3}{*}{$4$} & \small CY four-variety &  \small Picard two-variety $\tilde {\cal P}_2$\\[0.5ex]
& $y^3=(x_1-a_1)(x_1-a_2)(x_1-x_2)(x_2-a_3)$ & $y^3=(x_1-a_1)(x_1-a_2)(x_1-a_3)^2$\\	
&$(x_2-x_3)(x_3-a_4) (x_3-x_4)$& 
$\phantom{---}(x_1-x_2)^2 (x_2-a_4)^2 (x_2-a_5)$\\
&$(x_4-a_5)(x_4-a_6) $& 
$(x_2-a_6) $\\[1ex]
\hline
	 \end{tabular}	
	 
\end{center}}}
\vskip - 5mm 
\caption{The CY and Picard varieties attached to triangle track graphs at low loop order. For example, for $\ell=3$ we consider the two 
inequivalent ways to reduce the three loop triangle track graph depicted in figure~\ref{fig:3-loop_to_K3_to_pentagon}, leading to a Picard two-variety  and a Picard curve  of genus 3. For the CY four-variety we list only one possible reduction.}
  \label{Table:CYPicard}
\end{table}

Our analysis shows that it is in general not possible to assign a unique variety or geometry to a given Feynman integral, but the triangle track graphs provide an infinite family of graphs to which we can associate geometries that differ substantially in their topological properties, and even their dimension! This is in line with the genus-drop recently observed for the two-loop non-planar box integral in ref.~\cite{Marzucca:2023gto} (albeit in that case the dimension remained the same, and only the topological properties changed). Note that the number of different varieties we can assign to a fishnet graph is not limited to two, because we can apply the star-triangle relation in different ways, and each application may lead to a different variety (e.g., the dimensions may differ). For example we can apply the star triangle relation to the middle internal vertex of the three-loop triangle track graph. In this way, we can associate to the triangle track graph a two-dimensional Picard variety. Using the first 
equality in eq.~\eqref{eq:KB} for $d=3$ and the representation 
of the Picard variety in Table~\ref{Table:CYPicard}, we see that 
$\frac{3}{2} K_B=3(H_1+H_2)<[\tilde P_G([\underline{x},\underline{u}];a]=6 (H_1+H_2)$. Hence this variety 
has negative first Chern class, as expected.

\begin{figure}[!t]
\begin{center}
 \includegraphicsbox[scale=1]{Fig3LoopZigZag} 
 \quad \qquad \quad
    \includegraphicsbox[scale=1]{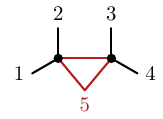} 
  \quad \qquad \quad
    \includegraphicsbox[scale=1]{Fig3LoopZigZagToPentagon} 
    \caption{\label{fig:3-loop_to_K3_to_pentagon} Different applications of the star-triangle relation may lead to different varieties attached to the three-loop triangle track graph: A CY 3-fold (left), a Picard  
    two-variety (middle) or a genus-3 Riemann curve (right).}
\end{center}
\end{figure}

In ref.~\cite{Bonisch:2021yfw} it was observed that 
different CY ($\ell-1$)-varieties yield the periods that correspond to the $\ell$-loop Banana graph 
integrals in two dimensions. On the one hand, these
periods are periods of singular $(\ell-1)$-dimensional hypersurfaces in toric ambient spaces, defined like in eq.~\eqref{eq:polyhedra} from the graph polynomial. On the 
other hand, they can be realised  as periods of a complete intersection CY of two constraints in $(\mathbb{P}^1)^{\ell+1}$. These spaces have the same dimension but different topology. Therefore it was argued that instead of the geometry one should focus on the unique familiy of motives associated to a Feynman integral from whose data we can efficiently evaluate the latter. The examples here,  as well as similar examples in ref.~\cite{Marzucca:2023gto} for curves of different 
genera, show again that it is generically not possible to assign a unique geometry to a given Feynman integral. In our case even the dimension of the geometric
representation is not fixed, as one can see in Table~\ref{Table:CYPicard} for 
the  different geometric representations of the $\ell$-loop triangle track graphs. Again, geometric representations of the
same motive in  different dimension were observed before in ref.~\cite{MR4257883}, where the  elliptic motive for the double 
box integral was identified in a singular five dimensional hypersurface. The star-triangle relation, that is easy 
to understand on the Feynman integral side, produces 
infinite series of such examples. A family of CY motives 
is defined by the periods as solutions of a Gauss-Manin 
system, or equivalently a Picard-Fuchs differential ideal, together with a monodromy-invariant integral intersection  pairing, represented by $\Sigma_G$ in eq.~\eqref{eq:phi_SV}, that defines the hermitian pairing, given in eq.~\eqref{eq:phi_SV} as well as a holomorphic pairing ${\underline \Pi}_G^T\Sigma_G {\underline \Pi}_G$ fulfilling Griffith transversality. Moreover we require the monodromy group to be defined over the integers or more general a ring ${\cal R}$ that is an 
algebraic extension of the integers, i.e. the monodromy group is a subgroup of ${O}(\Sigma,{\cal R})$. Indeed this data is all
present in the CY manifolds $M_G$ associated to graphs 
that come from the square lattice with $V=4$ with ${\cal R}=\mathbb{Z}$ 
and was used intensively  in refs.~\cite{Duhr:2022pch,Duhr:2023eld,Duhr:2023bku}.

%% file: yangian.tex
% !TEX root = 2DFishnetsLong.tex

%With the intersection pairing $\Sigma = \left(\begin{smallmatrix} 0 & 1 \\ -1 & 0\end{smallmatrix}\right)$ we find
%\beq\bsp
%I_{G_{1,1}}^{(2)}(\ua) \,&= -i\,\uPi_{G_{1,1}}(\ua)^{\dagger}\Sigma\uPi_{G_{1,1},}(\ua)\\
%&\, = \frac{4}{\sqrt{|a_{31}a_{42}|^2}}\,\left[\K(1-\overline{z})\K(z) + \K(\overline{z})\K(1-z)\right]\,.
%\esp\eeq

%We use conformal symmetry to set $_1=0$, $x_2=z$, $x_3=1$ and $x_4=\infty$ such that
%\begin{equation}
%\phi(z)=\int_{-\infty}^\infty \dd x_a\frac{1}{\sqrt{|x_a||x_{a}-z||x_{a}-1|}}.
%\end{equation}
%Assuming $0<z<1$ and using some identities for elliptic $K$, we find the following result by direct integration in Mathematica: 
%\begin{equation}
%f(z):=\phi^{0<z<1}(z)=4\brk[s]*{K(z)+K(1-z)}.
%\end{equation}
%In order to evaluate the result in the other kinematic regions, we 
%can use the above permutation symmetries of the box integral which imply the identities:
%\begin{align}
%&3\leftrightarrow 4: & &\phi(z)=\frac{1}{\sqrt{1-z}}\,\phi(\sfrac{z}{z-1}),
%\\
%&1\leftrightarrow 3:& &\phi(z)=\phi(1-z),
%\\
%&2 \leftrightarrow 3: & & \phi(z)=\frac{1}{\sqrt{z}}\phi(\sfrac{1}{z}).
%\end{align}
%Note that permutations change the ordering of the external kinematics and thus relate different regions of the function $\phi$:
%\begin{align}
%z<0:& & &\phi(z)=\frac{1}{\sqrt{1-z}}\,f(\sfrac{z}{z-1}),
%\\
%0<z<1:& &&\phi(z)=f(z), 
%\\
%1<z:& & &\phi(z)=\frac{1}{\sqrt{z}}f(\sfrac{1}{z}).
%\end{align}
%Modulo some identities for elliptic $K$, this agrees with direct integration in Mathematica.

%% file: examples.tex
% !TEX root = 2DFishnetsLong.tex
\subsection{Resolution of singularities and splitting of the Hodge structure}
\label{sec:blow-up}

As already mentioned, the CY varieties associated to fishnet graphs are typically singular, but we can either deform the singularities, yielding
a description of a smooth Calabi-Yau manifold as hypersurface $M_{\ell,\rm def}$ 
in a toric ambient space defined by the pair of reflexive polyhedra in eq.~\eqref{eq:polyhedra}, or we can resolve the singularities. 
In the following, we discuss the resolution of the singularities at low loop orders. This also clarifies some subtle differences between fishnet graphs for square and hexagonal tilings.

First, the topological types of the smooth deformations $M_{\ell,\rm def}$ of the singular Calabi-Yau spaces defined as double or triple coverings of $B=(\mathbb{P}^1)^l$ according to eqs.~\eqref{eq:defsingCY} and ~\eqref{eq:defsingbranch}
for $\ell\le 4$ are listed in Table~\ref{Table:topdataone}.
\begin{table}[h!]
{{ 
\begin{center}
	\begin{tabular}{|c|c|c|c|r||c|c|c|r|}
		\hline
$\ell$ &    \multicolumn{3}{c|}{ hexagonal tiling $(V,d)=(3,3)$} &$\chi$& \multicolumn{3}{c|}{ square tiling $(V,d)=(4,2)$}&$\chi$  \\ \hline
1 &  \multicolumn{3}{c|}{ $\Gamma(3)$ modulus at c.m. pt }   & $0$ &  \multicolumn{3}{c|}{ $\Gamma(2)$ one free modulus }& $0$ \\ \hline 
2 & $h_{11}^t=18$&$h^a_{11}=2$&&$24$& $h^t_{11}=18$&$h_{11}^a=2$\!\!\!\!&&$24$ \\ \hline
3 &$h_{21}=81$&$h_{11}=3$&&$ -156$&$h_{21}=115$&$h_{11}=3$\!\!\!\!&& $-224$ \\ \hline
4 & $h_{31}=324$&$h_{11}=4$&$h_{21}=0$\!\! & $ 2016$& $h_{31}=612$&$h_{11}=4$\!\!\!\!&$h_{21}=0$\!\! & $3744$\\ \hline
	 \end{tabular}	
\end{center}}}
\vskip - 5mm 
\caption{Euler number $\chi$ and dimensions $h_{p,q}$ of the non-trivial 
Hodge groups for the deformed CY manifolds associated to graphs with up 
to four vertices for the hexagonal and the quadratic tilings, respectively. For the elliptic curves, we indicate the modular group. For the K3 surface, the superscripts $t$ and $a$ stand for transcendental and algebraic elements of the homology.}
  \label{Table:topdataone}
\end{table}
Note that the different singular Calabi-Yau spaces that correspond to graphs with the same number $\ell$ of vertices, but with different graph topology, are all singular 
limits of the same $M_{\ell, {\rm def}}$. The K\"ahler resolution of these different singular Calabi-Yau geometries, however, 
depends on the specifics of the singular degeneration and is much more involved. We therefore only illustrate it for a few examples of train track and triangle track graphs.
 
At one loop, $\ell=1$, the CY manifold is a smooth variety:  the quadratic and the cubic covers define smooth 
elliptic curves (see Table~\ref{Table:topdataone}), for which the evaluation of the periods is straightforward.
There is, however, one important difference between the one-loop integrals for the square and hexagonal tilings:  
For the square tiling, the corresponding family of elliptic curves is the Legendre curve 
\beq 
y^2=x(x-1)(x-z),
\eeq 
which depends on one parameter $z$. In the hexagonal case, the elliptic curve is at a special point of complex multiplication (CM) with an order three automorphism. Using the conformal transformation to fix the three points $(a_1,a_2,a_3)$ in the left-hand side graph in eq.~\eqref{eq:startriangle} to $(1,\alpha,\alpha^2)$ (with $\alpha=\exp(2 \pi i/3)$) and homogenizing eq.~\eqref{eq:defsingCY} with $d=3$, which gives eq.~\eqref{eq:cubics} in the $(Y:X:Z)$ homogeneous coordinates of $\mathbb{P}^2$, we get the CM elliptic curve ${\cal C}^{(1)}_1$:   
\beq 
Y^3=X^3-Z^3\ .
\label{eq:orbifold}
\eeq
Let $K=\mathbb{Q}[\alpha]$ and ${\underline K}=H^1_K({\cal C}_1^{(1)})/\mathbb{Q}$ be the Hodge structure of the CM elliptic curve ${\cal C}^{(1)}_1$. 
The monodromies of the deformation family of the elliptic 
curve with $3 \psi XYZ=3 z^{-1/3}XYZ$ and the periods at the orbifold points $\psi=0$ are given in eqs.~(6.9) and (6.22) of ref.~\cite{Aganagic:2006wq}. This particular CM Hodge structure
will play a key r\^ole in the geometric interpretation of the star-triangle relation. First  note (explicitly from the analysis of ref.~\cite{Aganagic:2006wq})  that the Hodge 
structure ${\underline K}$  splits as
\beq
{\underline K}=H^1_{K}({\cal C}_1^{(1)})/\mathbb{Q}= H_+({\cal C}^{(1)}_1) \oplus H_-({\cal C }^{(1)}_1)\ ,\eeq
i.e., it splits into an invariant and an anti-invariant part under complex conjugation.

\begin{figure}[t]
\begin{center}
\begin{tabular}{c c c}
% \centering
 % \resizebox{0.22\textwidth}{!}{

\raisebox{1cm}{
\begin{tikzpicture}[line width=1.2pt,scale=.9]
 \tikzmath{\s1 = 1; \ys=3;}
\node (G1) at (1,\ys) [font=\small, text width=1 cm]{$G_{2,1}$};
  \foreach \i in {1,...,2} 
  \foreach \j in {1,...,1} {
   \foreach \a in {0,90,180,270} \draw (1+\i,\ys+\j) -- +(\a:\s1);
   \node [{below right}] at (1+\i,\ys+\j) {$x_\i$};
  \draw[fill=black!100] (1+\i,\ys+\j)  circle (1.5pt); }
   \foreach \k in {2,...,3} \node [{above}] at  (\k,\ys+1+1){$a_\k$};
   \foreach \k in {5,...,6} \node [{below}] at  (8-\k,\ys){$a_\k$};
   \node [{left}] at  (1,\ys+1){$a_1$};
   \node [{right}] at  (4,\ys+1){$a_4$};
\end{tikzpicture}
}

&
\raisebox{2.3cm}{\quad$\to$\quad}
&

% }
 % \resizebox{0.25\textwidth}{!}{%

\begin{tikzpicture}[>=latex, square/.style={regular polygon,regular polygon sides=4}]
\coordinate (0) at (0,0);
\coordinate (x2) at (0,4);
\coordinate (x1) at (4,0);
\coordinate (a1) at (.5,0);
\coordinate (a6) at (1,0);
\coordinate (a2) at (1.5,0);
\coordinate (e1) at (.5,4);
\coordinate (e6) at (1,4);
\coordinate (e2) at (1.5,4);
\coordinate (a3) at (0,2);
\coordinate (a4) at (0,2.5);
\coordinate (a5) at (0,3);
\coordinate (e3) at (4,2);
\coordinate (e4) at (4,2.5);
\coordinate (e5) at (4,3);
\coordinate (ad) at (0,0);
\coordinate (ed) at (4,4);
\draw[->, thick] (0) to (x1);
\draw[->, thick] (0) to (x2);
\node (x1) at  (x1) [below=0 of x1] {$x_1$};
\node (x2) at  (x1) [left=0 of x2] {$x_2$};
\draw[-] (a1) to (e1);
\node (a1) at  (a1) [ below=0 of a1] {$a_1$};
\draw[-] (a6) to (e6);
\node (a6) at  (a6) [below=0 of a6] {$a_2$};
\draw[-] (a2) to (e2);
\node (a2) at  (a2) [ below=0 of a2] {$a_6$};
\draw[-] (a3) to (e3);
\node (a3) at  (a3) [ left=0 of a3] {$a_3$};
\draw[-] (a4) to (e4);
\node (a4) at  (a4) [ left=0 of a4] {$a_4$};
\draw[-] (a5) to (e5);
\node (a5) at  (a5) [ left=0 of a5] {$a_5$};
\draw[-] (ad) to (ed);
\node [{left} ] at (4,4){$x_1=x_2$};

\node at (.5,2) [square,draw,fill,scale=0.3pt, rotate=45] (s1) {};
\node at (.5,2.5) [square,draw,fill,scale=0.3pt, rotate=45] (s1) {};
\node at (.5,3) [square,draw,fill,scale=0.3pt, rotate=45] (s1) {};
\node at (1,2) [square,draw,fill,scale=0.3pt, rotate=45] (s1) {};
\node at (1,2.5) [square,draw,fill,scale=0.3pt, rotate=45] (s1) {};
\node at (1,3) [square,draw,fill,scale=0.3pt, rotate=45] (s1) {};
\node at (1.5,2) [square,draw,fill,scale=0.3pt, rotate=45] (s1) {};
\node at (1.5,2.5) [square,draw,fill,scale=0.3pt, rotate=45] (s1) {};
\node at (1.5,3) [square,draw,fill,scale=0.3pt, rotate=45] (s1) {};
\node at (.5,0.5) [square,draw,fill,scale=0.3pt, rotate=45] (s1) {};
\node at (1,1) [square,draw,fill,scale=0.3pt, rotate=45] (s1) {};
\node at (1.5,1.5) [square,draw,fill,scale=0.3pt, rotate=45] (s1) {};
\node at (2,2) [square,draw,fill,scale=0.3pt, rotate=45] (s1) {};
\node at (2.5,2.5) [square,draw,fill,scale=0.3pt, rotate=45] (s1) {};
\node at (3,3) [square,draw,fill,scale=0.3pt, rotate=45] (s1) {};

\end{tikzpicture}

\\

 \raisebox{1cm}{
 % \resizebox{0.18\textwidth}{!}{
  \begin{tikzpicture}[line width=1.2pt,scale=.9]
   \tikzmath{\nv = 3;}
 \node (hex) at (0,0) [font=\small, text width=1 cm]{$Z_2$};
  \foreach \i in {0,...,0} 
  \foreach \j in {0,...,0} {
   \foreach \a in {30,150,270} \draw (2*sin{60}+2*sin{60}*\i,1+2*\j+2*sin{30}*\j) -- +(\a:1);
    \draw[fill=black!100] (2*sin{60}+2*sin{60}*\i,1+2*\j+2*sin{30}*\j)  circle (1.5pt); 
   \foreach \a in {210,90,-30} \draw (sin{60}+2*sin{60}*\i,1+sin{30}+2*\j+2*sin{30}*\j) -- +(\a:1);
   \draw[fill=black!100] (sin{60}+2*sin{60}*\i,1+sin{30}+2*\j+2*sin{30}*\j)  circle (1.5pt); 
}
\foreach \k in {1,...,1}  \node [{above}] at (2*sin{60}*\k,1+2*0+2*sin{30}*0){$x_{2}$};
\foreach \k in {1,...,1}  \node [{below}] at (sin{60}*\k,1+sin{30}){$x_{1}$};
\foreach \k in {2,...,2} \node [{above}] at  (-3*sin{60}+2*sin{60}*\k,1+sin{30}+2*0+2*sin{30}){$a_\k$};
\foreach \k in {4,...,4} \node [{below}] at  (10*sin{60}-2*sin{60}*\k,0){$a_\k$};
\node [{left}] at  (0,1){$a_1$};
\node [{right}] at  (2+sin{60},1+sin{30}){$a_3$};
\end{tikzpicture}
}

&
\raisebox{2.5cm}{\quad$\to$\quad}
&

 % \resizebox{0.25\textwidth}{!}{%
\begin{tikzpicture}[>=latex, square/.style={regular polygon,regular polygon sides=4}]
\coordinate (0) at (0,0);
\coordinate (x2) at (0,4);
\coordinate (x1) at (4,0);
\coordinate (a1) at (.5,0);
\coordinate (a2) at (1,0);
\coordinate (e1) at (.5,4);
\coordinate (e2) at (1,4);
\coordinate (a3) at (0,2.5);
\coordinate (a4) at (0,3);
\coordinate (e3) at (4,2.5);
\coordinate (e4) at (4,3);
\coordinate (ad) at (0,0);
\coordinate (ed) at (4,4);
\draw[->, thick] (0) to (x1);
\draw[->, thick] (0) to (x2);
\node (x1) at  (x1) [below=0 of x1] {$x_1$};
\node (x2) at  (x1) [left=0 of x2] {$x_2$};
\draw[-] (a1) to (e1);
\node (a1) at  (a1) [ below=0 of a1] {$a_1$};
\draw[-] (a2) to (e2);
\node (a2) at  (a2) [ below=0 of a2] {$a_2$};
\draw[-] (a3) to (e3);
\node (a3) at  (a3) [ left=0 of a3] {$a_3$};
\draw[-] (a4) to (e4);
\node (a4) at  (a4) [ left=0 of a4] {$a_4$};
\draw[-] (ad) to (ed);
\node [{left} ] at (4,4){$x_1=x_2$};

\node at (.5,2.5) [square,draw,fill,scale=0.3pt, rotate=45] (s1) {};
\node at (.5,3) [square,draw,fill,scale=0.3pt, rotate=45] (s1) {};
\node at (1,2.5) [square,draw,fill,scale=0.3pt, rotate=45] (s1) {};
\node at (1,3) [square,draw,fill,scale=0.3pt, rotate=45] (s1) {};
\node at (.5,.5) [square,draw,fill,scale=0.3pt, rotate=45] (s1) {};
\node at (1,1) [square,draw,fill,scale=0.3pt, rotate=45] (s1) {};
\node at (2.5,2.5) [square,draw,fill,scale=0.3pt, rotate=45] (s1) {};
\node at (3,3) [square,draw,fill,scale=0.3pt, rotate=45] (s1) {};

\end{tikzpicture}
\end{tabular}
% }
\caption{
\label{fignodalcuspk3}
Comparison of the two-loop train track $G_{1,2}$ and triangle track $Z_2$ graphs. On the right hand side, we illustrate the singularities of the associated K3 surface, denoted $M_{G_{1,2}}$ and $M_{Z_2}$. 
Note that the coordinates $a_i$ can be set to $0,1,\infty$ by a diagonal $\textrm{PSL}(2,\mathbb{C})$ acting on the 
projective plane in which the $a_i$ lie.
}
\end{center}
\end{figure}
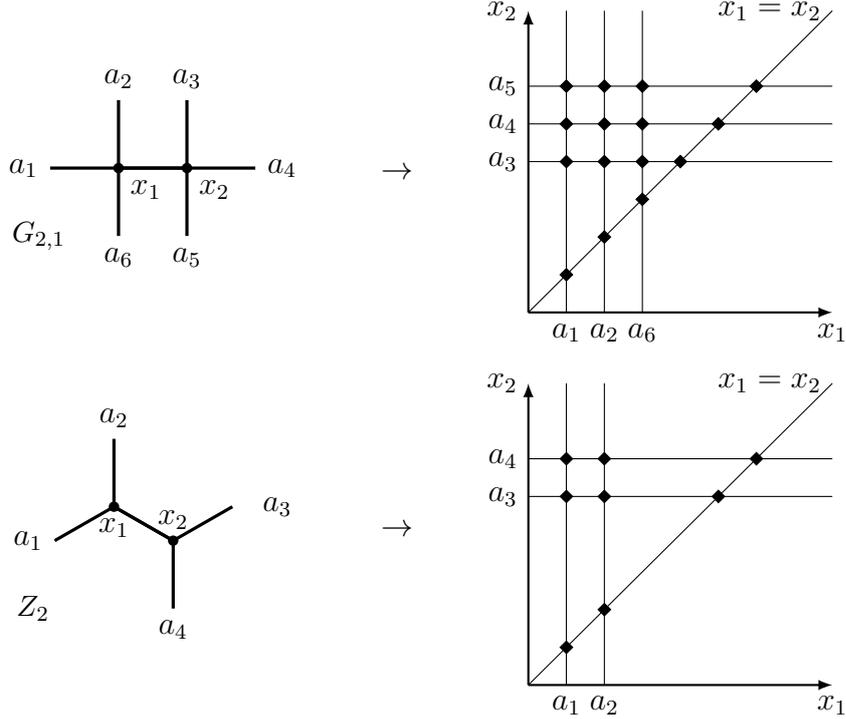

In the two-loop case, $\ell=2$, which corresponds to a K3 surface, we can see how the difference between the square and hexagonal tiling extends to higher-dimensional 
geometries. In both cases, the smooth K3 surface defined by the toric variety specified by eq.~\eqref{eq:polyhedra}  has Picard 
rank ${\rm rk(Pic)}=2$ coming from $B=(\mathbb{P}^1)^2$. For the square tiling, 
the geometry  acquires fifteen  nodes at the intersection diamonds in \Figref{fignodalcuspk3}, while for the  hexagonal tiling, it acquires eight cusp singularities.
The local situation near the singularities is simply   
\begin{equation} 
\textrm{square}: \quad y^2=\epsilon^2 \qquad \qquad \textrm{hexagonal}: \quad y^3=  \epsilon^2 \ .   
\end{equation} 
Hence, for the square tiling the fifteen nodes in the upper right 
diagram of figure \ref{fignodalcuspk3} correspond to $A_1$ singularities.
The latter can be resolved by a blow up of one $\mathbb{P}^1$ at each node, so that the total Picard rank increases to ${\rm rk(Pic)}=17=2+(9+6)$, and we get a smooth K3 surface for which the twenty-dimensional $H_2(\textrm{K3})$ splits into 
$17$ algebraic and $3$ transcendental 2-cycles. Here the $2$ comes from the two $\mathbb{P}^1$'s in the base $B=\mathbb{P}^1 \times \mathbb{P}^1$. By the Tian-Todorov theorem, the  latter correspond to the $3$ generic complex structure deformations of $M_{G_{1,2}}$, which in turn are identified with the physical complex deformations of the two-loop graph.  

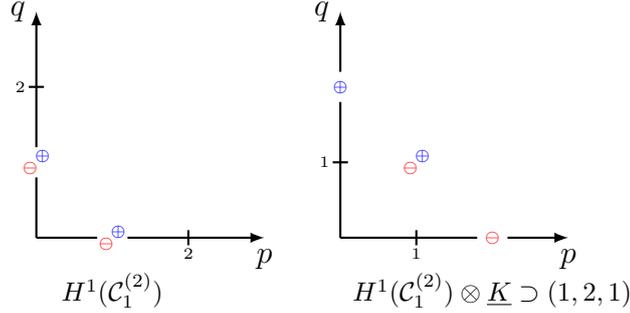
\begin{figure}[!t]
\begin{center}
\begin{tikzpicture}[>=latex, square/.style={regular polygon,regular polygon sides=4}]
\coordinate (0) at (0,0);
\coordinate (y1) at (0,1);
\coordinate (y2) at (0,2);
\coordinate (y3) at (0,3);
\coordinate (x1) at (1,0);
\coordinate (x2) at (2,0);
\coordinate (x3) at (3,0);
\draw[-, thick] (0) to (0.8,0);
\draw[->, thick] (1.2,0) to (x3);
\draw[-, thick] (0) to (0,0.8);
\draw[->, thick] (0,1.2) to (y3);
\draw[-,thick](2,-.1) to (2,.1); 
\node (x2) at (x2) [below=0 of x2]{\tiny $2$};   
\node (y2) at (y2) [left=0 of y2]{\tiny $2$};   
\draw[-,thick](-.1,2) to (.1,2); 
\node (x3) at  (x3) [below=0 of x3] {$p$};
\node (y3) at  (y3) [left=0 of y3] {$q$};
\node (s1) at  (y1) [xshift=.08cm, yshift=.08cm, blue] {\tiny $ \oplus$};
\node (s2) at  (y1) [xshift=-.08cm, yshift=-.08cm, red] {\tiny $ \ominus$};
\node (s1) at  (x1) [xshift=.08cm, yshift=.08cm, blue] {\tiny $ \oplus$};
\node (s2) at  (x1) [xshift=-.08cm, yshift=-.08cm, red] {\tiny $ \ominus$};
\node (l2) at (1,-.7)  [xshift=.00cm, yshift=.00cm] {\footnotesize $H^1({\cal C}_1^{(2)})$};
%-------------------------
\coordinate (0) at (4,0);
\coordinate (y1) at (4,1);
\coordinate (y2) at (4,2);
\coordinate (y3) at (4,3);
\coordinate (x1) at (5,0);
\coordinate (x2) at (6,0);
\coordinate (x3) at (7,0);
\draw[-, thick] (0) to (5.8,0);
\draw[->, thick] (6.2,0) to (x3);
\draw[-, thick] (0) to (4,1.8);
\draw[->, thick] (4,2.2) to (y3);
\draw[-,thick](5,-.1) to (5,.1); 
\node (x1) at (x1) [below=0 of x1]{\tiny $1$};   
\node (y1) at (y1) [left=0 of y1]{\tiny $1$};   
\draw[-,thick](3.9,1) to (4.1,1); 
\node (x3) at  (x3) [below=0 of x3] {$p$};
\node (y3) at  (y3) [left=0 of y3] {$q$};
\node (s1) at  (5,1) [xshift=.08cm, yshift=.08cm, blue] {\tiny $ \oplus$};
\node (s2) at  (5,1) [xshift=-.08cm, yshift=-.08cm, red] {\tiny $ \ominus$};
\node (s1) at  (y2) [xshift=.0cm, yshift=.0cm, blue] {\tiny $ \oplus$};
\node (s2) at  (x2) [xshift=-.0cm, yshift=-.0cm, red] {\tiny $ \ominus$};
\node (l2) at (6,-.7)  [xshift=.00cm, yshift=.00cm] {\footnotesize $H^1({\cal C}_1^{(2)})\otimes {\underline K} \supset (1,2,1)$};
\end{tikzpicture}
\end{center}
\caption{Relations of the Hodge structure $H^{q,p}$ of the genus two Picard curve ${\cal C}_1^{(2)}$ and the K3 in the second row $\ell=2$ of Table  \ref{Table:CYPicard} over 
${K} =\mathbb{Z}[\alpha]$. }
\label{rel:Hodgel=2}
\end{figure}

In the hexagonal case,  we get $A_2$ singularities at the eight cusps. See the 
lower diagram in \Figref{fignodalcuspk3}. The resolution of the latter 
introduces two exceptional  $\mathbb{P}^1$'s at each cusp singularity. 
They are all independent, raising the Picard group to ${\rm rk(Pic)}=18=2+2\cdot(4+4)$. 
Hence, the twenty-dimensional $H_2(M_{Z_2})$ has two transcendental cycles. This suggests that it has two independent complex structure deformations. 
The corresponding integral, however, has only one complex physical parameter. In fact, the $\mathbb{Z}_3$ symmetries of $M_{Z_2}$ enforces the manifold still to be in a special one-dimensional slice in the generically two-dimensional complex moduli space of the resolution. This can be understood systematically as follows: Under the action of the covering $S_3$ group in eq.~\eqref{eq:cubics} that permutes the covering sheets, the middle cohomology $H^\ell_K(M_{Z_2})$ splits over $K$ into the invariant and anti-invariant part under complex conjugation as 
\beq
H^{\ell}_K(M_{{Z_2}},)=H^{\ell}_+(M_{{Z_2}})\oplus H^{\ell}_-(M_{Z_2}).
\eeq
In particular, the corresponding two-dimensional transcendental lattice of $M_{Z_2}$ splits and the rank 4 K3 motive  splits over  
$K$ into two equivalent rank 2 motives (see section~\ref{sec:motives} for a review of CY motives) 
\begin{equation} 
\begin{array}{rll} 
&  & (1,1)_+ \\ [ -2mm]
(1,2,1) \rightarrow && \\ [-2 mm]
  & (1,1)_- &   \ .\\
  \end{array}
  \end{equation} 
The $(1,1)_\pm$ Hodge structures are  the ones of the genus two Picard  curve ${\cal C}^{(2)}_1$, whose rational Hodge structure of the latter splits over $K$ into 
\beq 
H^1_K({\cal C}^{(2)}_1)=H^1_+({\cal C}^{(2)}_1) \oplus   H^1_-({\cal C}^{(2)}_1)\ .
\eeq 
The two solutions that correspond to the period integrals of either  
$H^1({\cal C}^{(2)}_1)_\pm$ are discussed in detail after eq.~\eqref{var2zigzag}.
We see here at the point of maximal unipotent monodromy one holomorphic
and one logarithmic period. This monodromy weight filtration
of the degeneration is related to the Hodge filtration $(1,1)_\pm$, 
as familiar in mirror symmetry. The Hodge structure of the  K3 variety is obtained by tensoring it with the Hodge structure of the complex multiplication curve 
\beq
H^1({\cal C}_1^{(2)})\otimes {\underline K} \supset (1,2,1) \sim H^2(M_{Z_2}, K)^\text{trans}
\eeq
induces the shift in the Hodge structure as depicted on the right-hand side of figure  \ref{rel:Hodgel=2}. Each tensoring with $\underline K$ 
corresponds to one application of the star-triangle relation.   

\begin{figure}[!t]
\begin{center}
\begin{tikzpicture}[>=latex, square/.style={regular polygon,regular polygon sides=4}]
\coordinate (0) at (0,0);
\coordinate (y1) at (0,1);
\coordinate (y2) at (0,2);
\coordinate (y3) at (0,3);
\coordinate (y4) at (0,4);
\coordinate (x1) at (1,0);
\coordinate (x2) at (2,0);
\coordinate (x3) at (3,0);
\coordinate (x4) at (4,0);
\draw[-, thick] (0) to (0.8,0);
\draw[->, thick] (1.2,0) to (x4);
\draw[-, thick] (0) to (0,0.8);
\draw[->, thick] (0,1.2) to (y4);
\draw[-,thick](2,-.1) to (2,.1); 
\node (x2) at (x2) [below=0 of x2]{\tiny $2$};
\draw[-,thick](-.1,2) to (.1,2);    
\node (y2) at (y2) [left=0 of y2]{\tiny $2$};   
\draw[-,thick](3,-.1) to (3,.1); 
\node (x3) at (x3) [below=0 of x3]{\tiny $3$};
\draw[-,thick](-.1,3) to (.1,3); 
\node (y3) at (y3) [left=0 of y3]{\tiny $3$};  
\node (x4) at  (x4) [below=0 of x4] {$p$};
\node (y4) at  (y4) [left=0 of y4] {$q$};
\node (s1) at  (y1) [xshift=.03cm, yshift=.08cm, blue] {\tiny $ \oplus$};
\node (ss1) at  (y1) [xshift=.16cm, yshift=.-.04cm, blue] {\tiny $ \oplus$};
\node (s2) at  (y1) [xshift=-.02cm, yshift=-.1cm, red] {\tiny $ \ominus$};
\node (s1) at  (x1) [xshift=.08cm, yshift=.08cm, blue] {\tiny $ \oplus$};
\node (s2) at  (x1) [xshift=-.06cm, yshift=-.06cm, red] {\tiny $ \ominus$};
\node (ss2) at  (x1) [xshift=.11cm, yshift=-.11cm, red] {\tiny $ \ominus$};
\node (l2) at (1,-.7)  [xshift=.00cm, yshift=.00cm] {\footnotesize $H^1({\cal C}_1^{(3)})$};
%---------------------
\coordinate (0) at (5,0);
\coordinate (y1) at (5,1);
\coordinate (y2) at (5,2);
\coordinate (y3) at (5,3);
\coordinate (y4) at (5,4);
\coordinate (x1) at (6,0);
\coordinate (x2) at (7,0);
\coordinate (x3) at (8,0);
\coordinate (x4) at (9,0);
\draw[-, thick] (0) to (6.8,0);
\draw[->, thick] (7.2,0) to (x4);
\draw[-, thick] (0) to (5,1.8);
\draw[->, thick] (5,2.2) to (y4); 
\node (x1) at (x1) [below=0 of x1]{\tiny $1$};   
\draw[-,thick](4.9,1) to (5.1,1); 
\node (y1) at (y1) [left=0 of y1]{\tiny $1$};   
\draw[-,thick](6,-.1) to (6,.1); 
\draw[-,thick](8,-.1) to (8,.1); 
\node (x3) at (x3) [below=0 of x3]{\tiny $3$};   
\draw[-,thick](4.9,3) to (5.1,3); 
\node (y3) at (y3) [left=0 of y3]{\tiny $3$};
\node (x4) at  (x4) [below=0 of x4] {$p$};
\node (y4) at  (y4) [left=0 of y4] {$q$};
\node (s1) at  (5,2) [xshift=.08cm, yshift=.08cm, blue] {\tiny $ \oplus$};
\node (s2) at  (5,2) [xshift=-.08cm, yshift=-.08cm, blue] {\tiny $ \oplus$};
\node (s1) at  (6,1) [xshift=.08cm, yshift=.08cm, blue] {\tiny $ \oplus$};
\node (s2) at  (6,1) [xshift=-.08cm, yshift=-.08cm, red] {\tiny $ \ominus$};
\node (s1) at  (7,0) [xshift=.08cm, yshift=.08cm, red] {\tiny $ \ominus$};
\node (s2) at  (7,0) [xshift=-.08cm, yshift=-.08cm, red] {\tiny $ \ominus$};
\node (l3) at (7,-.7) [xshift=.00cm, yshift=.00cm] {\footnotesize $H^1({\cal C}_1^{(3)})\otimes {\underline K} \simeq H^2({\cal P}_2)\otimes {\underline K} \supset (1,2,2,1)$};
%-------------------------
\coordinate (0) at (10,0);
\coordinate (y1) at (10,1);
\coordinate (y2) at (10,2);
\coordinate (y3) at (10,3);
\coordinate (y4) at (10,4);
\coordinate (x1) at (11,0);
\coordinate (x2) at (12,0);
\coordinate (x3) at (13,0);
\coordinate (x4) at (14,0);
\draw[-, thick] (0) to (12.8,0);
\draw[->, thick] (13.2,0) to (x4);
\draw[-, thick] (0) to (10,2.8);
\draw[->, thick] (10,3.2) to (y4);
\draw[-,thick](11,-.1) to (11,.1); 
\node (x1) at (x1) [below=0 of x1]{\tiny $1$};   
\draw[-,thick](9.9,1) to (10.1,1); 
\node (y1) at (y1) [left=0 of y1]{\tiny $1$};   
\draw[-,thick](12,-.1) to (12,.1); 
\node (x2) at (x2) [below=0 of x2]{\tiny $2$};   
\draw[-,thick](9.9,2) to (10.1,2); 
\node (y2) at (y2) [left=0 of y2]{\tiny $2$};   
\node (x4) at  (x4) [below=0 of x4] {$p$};
\node (y4) at  (y4) [left=0 of y4] {$q$};
\node (s1) at  (12,1) [xshift=.08cm, yshift=.08cm, blue] {\tiny $ \oplus$};
\node (s2) at  (12,1) [xshift=-.08cm, yshift=-.08cm, blue] {\tiny $ \oplus$};
\node (s1) at  (11,2) [xshift=.08cm, yshift=.08cm, red] {\tiny $ \ominus$};
\node (s2) at  (11,2) [xshift=-.08cm, yshift=-.08cm, red] {\tiny $ \ominus$};
\node (s1) at  (y3) [xshift=.0cm, yshift=.0cm, red] {\tiny $ \ominus$};
\node (ss2) at  (x3) [xshift=-.0cm, yshift=-.0cm, blue] {\tiny $ \oplus$};
\node (l2) at (12.9,-.7)  [xshift=.00cm, yshift=.00cm] {\footnotesize  $H^1({\cal C}_1^{(3)})\otimes {\underline K} \otimes {\underline K} \supset (1,2,2,1)$};
\end{tikzpicture}
\end{center}
\caption{Relations of the Hodge structure $H^{q,p}$  of the genus three Picard curve ${\cal C}_1^{(3)}$, the Picard two-variety $\mathcal{P}_2$ and the Calabi-Yau three-variety  in the third 
row $\ell=3$ of Table  \ref{Table:CYPicard} over ${K} =\mathbb{Z}[\alpha]$. }
\label{rel:Hodgel=3}
\end{figure}
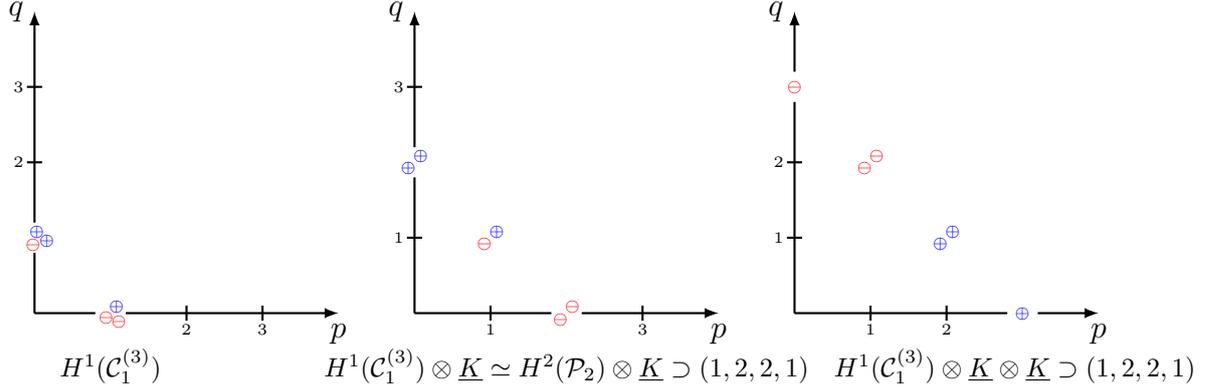

We can give as further example the Calabi-Yau three variety  with two parameters  that corresponds to the $\ell=3$ triangle track graph. Here, 
we get by two applications of the star-triangle relation a genus three Picard curve whose solutions correspond to the Appell hypergeometric function $F_1$ with the solutions discussed in eq.~\eqref{frob3zigzag}. Its solution structure  
corresponds precisely to the splitting of the Hodge structures that 
is outlined in figure  \ref{rel:Hodgel=3}, where we also depict 
the conjectured Hodge structure of the Picard two variety. Note that 
on the right-hand side of figure \ref{rel:Hodgel=3} the cohomology 
group $H^{2,1}(M_{Z_3})$  (which corresponds to the complex structure deformation of $M_{Z_3}$) does not split into even and odd parts with respect 
to complex conjugation. Hence, this geometry has two complex structure deformations, consistent with the physical parameter count.  

\begin{figure}[!t]
\begin{center}
\begin{tikzpicture}[>=latex, square/.style={regular polygon,regular polygon sides=4}]
\coordinate (0) at (0,0);
\coordinate (y1) at (0,.5);
\coordinate (y2) at (0,1);
\coordinate (y3) at (0,1.5);
\coordinate (y4) at (0,2);
\coordinate (x1) at (.5,0);
\coordinate (x2) at (1,0);
\coordinate (x3) at (1.5,0);
\coordinate (x4) at (2,0);
\draw[-, thick] (0) to (1.4,0);
\draw[->, thick] (1.6,0) to (x4);
\draw[-, thick] (0) to (0,1.4);
\draw[->, thick] (0,1.6) to (y4);
\draw[-,thick](.5,-.1) to (.5,.1); 
\node (x1) at (x1) [below=0 of x1]{\tiny $1$};   
\draw[-,thick](-.1,.5) to (0.1,.5); 
\node (y1) at (y1) [left=0 of y1]{\tiny $1$};   
\draw[-,thick](1,-.1) to (1,.1); 
\node (x2) at (x2) [below=0 of x2]{\tiny $2$};   
\draw[-,thick](-.1,1) to (.1,1); 
\node (y2) at (y2) [left=0 of y2]{\tiny $2$};   
\node (x4) at  (x4) [below=0 of x4] {$p$};
\node (y4) at  (y4) [left=0 of y4] {$q$};
\node (s1) at  (1,.5) [xshift=.07cm, yshift=.07cm, blue] {\tiny $ \oplus$};
\node (s2) at  (1,.5) [xshift=-.07cm, yshift=-.07cm, red] {\tiny $\ominus$};
\node (s1) at  (.5,1) [xshift=.07cm, yshift=.07cm, blue] {\tiny $ \oplus$};
\node (s2) at  (.5,1) [xshift=-.07cm, yshift=-.07cm, red] {\tiny $ \ominus$};
\node (s1) at  (y3) [xshift=.07cm, yshift=.07cm, blue] {\tiny $ \oplus$};
\node (s2) at  (y3)  [xshift=-.07cm, yshift=-.07cm, red] {\tiny $ \ominus$};
\node (s1) at  (x3) [xshift=.07cm, yshift=.07cm, blue] {\tiny $ \oplus$};
\node (s1) at  (x3) [xshift=-.07cm, yshift=-.07cm, red] {\tiny $ \ominus$};
\node (l2) at (1,-.7)  [xshift=.00cm, yshift=.00cm] {\footnotesize  $H^3({\cal P}_3^{(\textrm{even})})$};

\coordinate (O) at (2.5,0);
\coordinate (y1) at (2.5,.5);
\coordinate (y2) at (2.5,1);
\coordinate (y3) at (2.5,1.5);
\coordinate (y4) at (2.5,2);
\coordinate (y5) at (2.5,2.5);
\coordinate (y6) at (2.5,3);
\coordinate (y7) at (2.5,3.5);
\coordinate (x1) at (3,0);
\coordinate (x2) at (3.5,0);
\coordinate (x3) at (4,0);
\coordinate (x4) at (4.5,0);
\coordinate (x5) at (5,0);
\coordinate (x6) at (5.5,0);
\coordinate (x7) at (6,0);
\draw[-, thick] (O) to (5.4,0);
\draw[->, thick] (5.6,0) to (x7);
\draw[-, thick] (O) to (2.5,2.9);
\draw[->, thick] (2.5,3.1) to (y7);
\draw[-,thick](3,-.1) to (3,.1); 
\node (x1) at (x1) [below=0 of x1]{\tiny $1$};
\draw[-,thick](3.5,-.1) to (3.5,.1); 
\node (x2) at (x2) [below=0 of x2]{\tiny $2$};   
\draw[-,thick](4,-.1) to (4,.1); 
\node (x3) at (x3) [below=0 of x3]{\tiny $3$};   
\draw[-,thick](4.5,-.1) to (4.5,.1); 
\node (x4) at (x4) [below=0 of x4]{\tiny $4$};   
\draw[-,thick](5,-.1) to (5,.1); 
\node (x5) at (x5) [below=0 of x5]{\tiny $5$};   
\draw[-,thick](2.4,.5) to (2.6,.5); 
\node (y1) at (y1) [left=0 of y1]{\tiny $1$};   
\draw[-,thick](2.4,1) to (2.6,1); 
\node (y2) at (y2) [left=0 of y2]{\tiny $2$};
\draw[-,thick](2.4,1.5) to (2.6,1.5); 
\node (y3) at (y3) [left=0 of y3]{\tiny $3$};   
\draw[-,thick](2.4,2) to (2.6,2); 
\node (y4) at (y4) [left=0 of y4]{\tiny $4$};
\draw[-,thick](2.4,2.5) to (2.6,2.5); 
\node (y5) at (y5) [left=0 of y5]{\tiny $5$};

\node (x7) at  (x7) [below=0 of x7] {$p$};
\node (y7) at  (y7) [left=0 of y7] {$q$};
\node (sm1) at  (y6) [xshift=.0cm, yshift=.0cm, red] {\tiny $ \ominus$};
\node (sm2) at  (3,2.5) [xshift=.0cm, yshift=.0cm, red] {\tiny $ \ominus$};
\node (sm3) at  (3.5,2) [xshift=.0cm, yshift=.0cm, red] {\tiny $ \ominus$};
\node (sm4) at  (4,1.5) [xshift=-.07cm, yshift=-.07cm, red] {\tiny $ \ominus$};
\node (sp1) at  (x6) [xshift=.0cm, yshift=.0cm, blue] {\tiny $ \oplus$};
\node (sp2) at  (5,.5) [xshift=.0cm, yshift=.0cm, blue] {\tiny $ \oplus$};
\node (sp3) at  (4.5,1) [xshift=.0cm, yshift=.0cm, blue] {\tiny $ \oplus$};
\node (sp4) at  (4,1.5) [xshift=.07cm, yshift=.07cm, blue] {\tiny $ \oplus$};
%\node (s1) at  (x3) [xshift=-.08cm, yshift=-.08cm, red] {\tiny $ \ominus$};
\node (l2) at (3.5,-.7)  [xshift=.00cm, yshift=.00cm] {\footnotesize  $H^6(M_{Z_6})$};

\coordinate (0) at (7,0);
\coordinate (y1) at (7,.5);
\coordinate (y2) at (7,1);
\coordinate (y3) at (7,1.5);
\coordinate (y4) at (7,2);
\coordinate (x1) at (7.5,0);
\coordinate (x2) at (8,0);
\coordinate (x3) at (8.5,0);
\coordinate (x4) at (9,0);
\draw[-, thick] (0) to (8.4,0);
\draw[->, thick] (8.6,0) to (x4);
\draw[-, thick] (0) to (7,1.4);
\draw[->, thick] (7,1.6) to (y4);
\draw[-,thick](7.5,-.1) to (7.5,.1); 
\node (x1) at (x1) [below=0 of x1]{\tiny $1$};   
\draw[-,thick](6.9,.5) to (7.1,.5); 
\node (y1) at (y1) [left=0 of y1]{\tiny $1$};   
\draw[-,thick](8,-.1) to (8,.1); 
\node (x2) at (x2) [below=0 of x2]{\tiny $2$};   
\draw[-,thick](6.9,1) to (7.1,1); 
\node (y2) at (y2) [left=0 of y2]{\tiny $2$};   
\node (x4) at  (x4) [below=0 of x4] {$p$};
\node (y4) at  (y4) [left=0 of y4] {$q$};
\node (s1) at  (8,.5) [xshift=.07cm, yshift=.07cm, blue] {\tiny $ \oplus$};
\node (s2) at  (8,.5) [xshift=-.07cm, yshift=-.07cm, red] {\tiny $\ominus$};
\node (s1) at  (7.5,1) [xshift=.07cm, yshift=.07cm, blue] {\tiny $ \oplus$};
\node (s2) at  (7.5,1) [xshift=-.07cm, yshift=-.07cm, red] {\tiny $ \ominus$};
\node (s1) at  (y3) [xshift=.07cm, yshift=.07cm, blue] {\tiny $ \oplus$};
\node (s2) at  (y3)  [xshift=-.07cm, yshift=-.07cm, red] {\tiny $ \ominus$};
\node (s1) at  (x3) [xshift=.07cm, yshift=.07cm, blue] {\tiny $ \oplus$};
\node (s1) at  (x3) [xshift=-.07cm, yshift=-.07cm, red] {\tiny $ \ominus$};
\node (l2) at (8,-.7)  [xshift=.00cm, yshift=.00cm] {\footnotesize  $H^3({\cal P}_3^{(\textrm{odd})})$};

\coordinate (O) at (9.5,0);
\coordinate (y1) at (9.5,.5);
\coordinate (y2) at (9.5,1);
\coordinate (y3) at (9.5,1.5);
\coordinate (y4) at (9.5,2);
\coordinate (y5) at (9.5,2.5);
\coordinate (y6) at (9.5,3);
\coordinate (y7) at (9.5,3.5);
\coordinate (y8) at (9.5,4);
\coordinate (x1) at (10,0);
\coordinate (x2) at (10.5,0);
\coordinate (x3) at (11,0);
\coordinate (x4) at (11.5,0);
\coordinate (x5) at (12,0);
\coordinate (x6) at (12.5,0);
\coordinate (x7) at (13,0);
\coordinate (x8) at (13.5,0);
\draw[-, thick] (O) to (12.9,0);
\draw[->, thick] (13.1,0) to (x8);
\draw[-, thick] (O) to (9.5,3.4);
\draw[->, thick] (9.5,3.6) to (y8);
\draw[-,thick](10,-.1) to (10,.1); 
\node (x1) at (x1) [below=0 of x1]{\tiny $1$};
\draw[-,thick](10.5,-.1) to (10.5,.1); 
\node (x2) at (x2) [below=0 of x2]{\tiny $2$};   
\draw[-,thick](11,-.1) to (11,.1); 
\node (x3) at (x3) [below=0 of x3]{\tiny $3$};   
\draw[-,thick](11.5,-.1) to (11.5,.1); 
\node (x4) at (x4) [below=0 of x4]{\tiny $4$};   
\draw[-,thick](12,-.1) to (12,.1); 
\node (x5) at (x5) [below=0 of x5]{\tiny $5$};   
\draw[-,thick](12.5,-.1) to (12.5,.1); 
\node (x6) at (x6) [below=0 of x6]{\tiny $6$};  
\draw[-,thick](9.4,.5) to (9.6,.5); 
\node (y1) at (y1) [left=0 of y1]{\tiny $1$};   
\draw[-,thick](9.4,1) to (9.6,1); 
\node (y2) at (y2) [left=0 of y2]{\tiny $2$};
\draw[-,thick](9.4,1.5) to (9.6,1.5); 
\node (y3) at (y3) [left=0 of y3]{\tiny $3$};   
\draw[-,thick](9.4,2) to (9.6,2); 
\node (y4) at (y4) [left=0 of y4]{\tiny $4$};
\draw[-,thick](9.4,2.5) to (9.6,2.5); 
\node (y5) at (y5) [left=0 of y5]{\tiny $5$};
\draw[-,thick](9.4,3) to (9.6,3); 
\node (y6) at (y6) [left=0 of y6]{\tiny $6$};
\node (x8) at  (x8) [below=0 of x8] {$p$};
\node (y8) at  (y8) [left=0 of y8] {$q$};
\node (sm1) at  (y7) [xshift=.0cm, yshift=.0cm, red] {\tiny $ \ominus$};
\node (sm2) at  (10,3) [xshift=.0cm, yshift=.0cm, red] {\tiny $ \ominus$};
\node (sm3) at  (10.5,2.5) [xshift=.0cm, yshift=.0cm, red] {\tiny $ \ominus$};
\node (sm4) at  (11,2) [xshift=-.00cm, yshift=-.00cm, red] {\tiny $ \ominus$};
\node (sp1) at  (x7) [xshift=.0cm, yshift=.0cm, blue] {\tiny $ \oplus$};
\node (sp2) at  (12.5,.5) [xshift=.0cm, yshift=.0cm, blue] {\tiny $ \oplus$};
\node (sp3) at  (12,1) [xshift=.0cm, yshift=.0cm, blue] {\tiny $ \oplus$};
\node (sp4) at  (11.5,1.5) [xshift=.00cm, yshift=.00cm, blue] {\tiny $ \oplus$};
%\node (s1) at  (x3) [xshift=-.08cm, yshift=-.08cm, red] {\tiny $ \ominus$};
\node (l2) at (11,-.7)  [xshift=.00cm, yshift=.00cm] {\footnotesize  $H^7(M_{Z_7})$};

\end{tikzpicture}
\end{center}
\caption{Hodge structure $H^{q,p}$ of the one-parameter Picard three-folds $\mathcal{P}_3^{(\textrm{even})}$ and $\mathcal{P}_3^{(\textrm{odd})}$ versus the one-parameter Calabi-Yau six- and seven-varieties for  triangle 
track graphs with $\ell=6$ (even) and $\ell=7$ (odd) . }
\label{rel:Hodgeoneparam}
\end{figure}
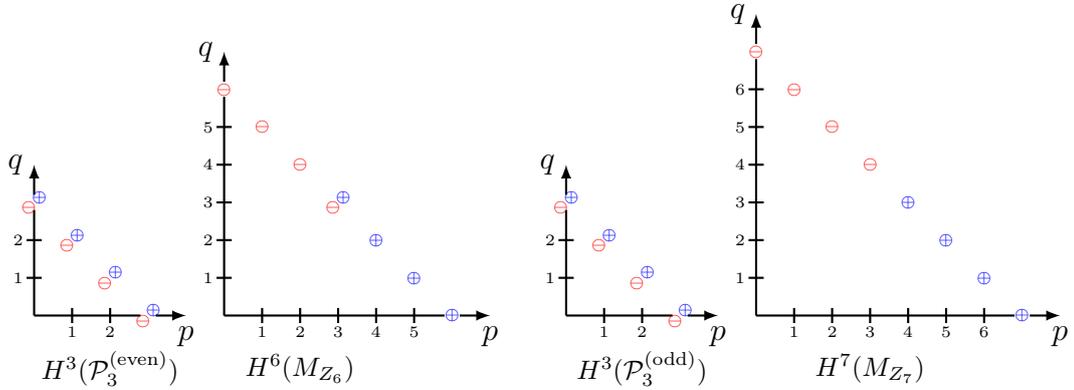

  More generally, on the  CY motives, there will be an order 3 Galois action that involves the $S_3$ action permuting the three sheets. 
  For these general cases, we discuss the one-parameter families in  section \ref{sec:fourponttriangle}. For the $\ell=7$ case, we find that the ring 
  ${\cal R}$ for the  graphs     from the hexagonal lattice can be $\mathbb{Z}[\alpha]$, while for the square lattice  
   we simply have $\mathcal{R}=\mathbb{Z}$. A triangle track example with ${\cal R}=\mathbb{Z}[\alpha]$ in the monodromy group 
  ${\cal O}(\Sigma,{\cal R})$ is discussed after eq.~\eqref{rho4F3}.  The relation of the Hodge structure between the Picard $m$-varieties 
  and  the Calabi-Yau $(2m)$ and $(2m+1)$ varieties is made explicit in figure    \ref{rel:Hodgeoneparam} (in  examples for $m=3$).

 Note that also in higher dimensions for the square case  all singularities are normal crossing and can be easily 
resolved, if and only if for co-dimension $k$ loci in $B$ the vanishing order is $s=k$ and local singularities are of type $y^2=\epsilon^s$. 
If there are $\mu$ meshes in $G$ this order gets as large as $s=k+\mu$. 
Hence, the train track CY varieties have a nested structure of normal 
crossing singularities. We expect that the resolution construction extends  
to all cases yielding smooth graph manifolds $M_{G}$'s or at 
least a consistent truncation of the subspace of the complex  moduli space of $M_G$, 
leading to a CY motive.  

\section{Low-loop examples and four-point limits}
\label{sec:examples}

In this section, we discuss several examples of fishnet integrals with a small number of loops as well as the four-point limit of triangle track graphs. 
These results do not only illustrate the abstract mathematical concepts of the previous sections, but they also support our conjecture from section~\ref{sec:fishnet_CYs} that the PFI is generated by the holomorphic symmetry algebra $\Perm\ltimes Y_G$.
We start by presenting some details about train track integrals omitted in ref.~\cite{Duhr:2022pch}, and we also give new results for low-loop triangle track integrals. Appendix~\ref{app:examples} discusses additional examples of PFIs of fishnet graphs with loops in position space, which gives additional support to our conjecture.

%==============================
\subsection{Square tilings and train track graphs}
\begin{figure}[t]
\centering
 % \resizebox{0.35\textwidth}{!}{
\begin{tikzpicture}[very thick]

  \draw [ultra thick, gray!50, rounded corners] (0.5,0.8) -- (4.5,0.8) -- (0.5,-0.8) -- (4.5,-0.8);
  % \draw [ultra thick, gray!50] (0.5,-0.8) -- (4.5,0.8);
  % \draw [ultra thick, gray!50] (0.5,-0.8) -- (4.5,-0.8);

  \node (G1) at (0,-1) [font=\small, text width=1 cm]{$G_{1,\ell}$};
  \draw (0,0) -- (5,0);
  \draw (1,1) -- (1,-1);
  \node [{below right}] at (1,0) {$x_1$};
  \draw[fill=black!100] (1,0)  circle (1.5pt);
  \draw (2,1) -- (2,-1);
  \node [{below right}] at (2,0) {$x_2$};
  \draw[fill=black!100] (2,0)  circle (1.5pt);
  \draw (4,1) -- (4,-1);
  \node [{below right}] at (4,0) {$x_\ell$};
  \draw[fill=black!100] (4,0)  circle (1.5pt);
  \draw[fill=black!100] (2.85,0.5)  circle (0.5pt);
  \draw[fill=black!100] (3,0.5)  circle (0.5pt);
  \draw[fill=black!100] (3.15,0.5)  circle (0.5pt);
  \draw[fill=black!100] (2.85,-0.5)  circle (0.5pt);
  \draw[fill=black!100] (3,-0.5)  circle (0.5pt);
  \draw[fill=black!100] (3.15,-0.5)  circle (0.5pt);
  \node [{left}] at (0,0) {$a_1$};
  \node [{above}] at (1,1) {$a_2$};
  \node [{above}] at (2,1) {$a_3$};
  \node [{above}] at (4,1) {$a_{\ell+1}$};
  \node [{right}] at (5,0) {$a_{\ell+2}$};
  \node [{below}] at (4,-1) {$a_{\ell+3}$};
  \node [{below}] at (2,-1) {$a_{2\ell+1}$};
  \node [{below}] at (1,-1) {$a_{2\ell+2}$};
  
\end{tikzpicture}
% }
\caption{The $\ell$-loop train track graph $G_{1,\ell}$ with Z-rule defining the MUM-point cross ratios.}
\label{fig:traintracks}
\end{figure}
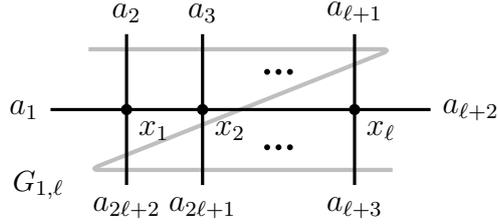

We start by discussing examples of train track graphs $G_{1,\ell}$ from \Figref{fig:traintracks} at low loop orders. These results were already presented in ref.~\cite{Duhr:2022pch}, but no details on the structure of the periods and the PFI were given.
Up to three loops, all fishnet graphs associated to a square tiling are train track graphs. We start by discussing some general properties of train track graphs. We then focus on the one-, two- and three-loop train track graphs $G_{1,1}, G_{1,2}$ and $G_{1,3}$. The PFI of the four-loop window graph $G_{2,2}$ is discussed in Appendix~\ref{app:window}.

For train track graphs there are no hidden permutation symmetries, and we have
\beq
\textrm{Perm}_{G_{1,\ell}} = \Aut(G_{1,\ell}) = \left\{\begin{array}{ll}
S_4\,,&\ell=1\\
S_3^2\times \mathbb{Z}_2^{\ell-2}\times\mathbb{Z}_2\,,\,  &\ell>1\,,
\end{array}\right.
\eeq
where for $\ell>1$ the $S_3$'s exchange the external legs at the end of the train track, and there is a $\mathbb{Z}_2$ that exchanges each pair of external legs attached to the interior of the train track. In addition, there is a $\mathbb{Z}_2$ symmetry that exchanges the two ends of the train track.

By conformal invariance, the only non-trivial functional dependence is through the cross ratios in eq.~\eqref{defzcrossratios}. There is some ambiguity in how we choose a set of independent cross ratios.
For train track graphs, we choose appropriate cross ratios by the Z-rule%
\footnote{The name comes from the fact that we choose the cross ratios following the gray Z-shaped line in \Figref{fig:traintracks}.} (cf. eq. \eqref{defcrossratios})
\begin{align}
z_k=\frac14\, \chi_{1,k+1,k+2,\ell+2}\, , \quad z_\ell =\frac1{4^{3-l}}\,\chi_{1,\ell+1,2\ell+2,\ell+2}\, ,
\quad z_{\ell+k} =\frac14\,\chi_{1,2\ell+3-k,2\ell+2-k,\ell+2}\, ,
\label{eq:traintrackcrossratios}
\end{align}
for $k=1,\hdots, \ell-1$. The advantage of working with these cross ratios comes from the fact that, for all examples we have studied, there is a point of maximal unipotent monodromy (MUM) at $\uz=(z_1,\ldots,z_{2\ell-1})=0$, and we expect this to hold in general. The factors of $1/4$ are included so that the holomorphic solution has an integer coefficient expansion around the MUM-point $\uz=0$.

It is sufficient to look at the conformally invariant function $\phi_{G_{1,\ell}}(\uz)$ defined through eq.~\eqref{eq:I-Fphi}.
%We write the 2$D$ traintrack graphs of figure \ref{fig:traintracks} as
 For the prefactor, we choose
\beq
    F_{G_{1,\ell}}^{(2)}(\ua)=\frac{|a_1-a_{\ell+2}|^{\ell-1}}{|a_{\ell+3}-a_1||a_{\ell+4}-a_1|\cdots|a_{2\ell+2}-a_1||a_2-a_{\ell+2}||a_3-a_{\ell+2}|\cdots|a_{\ell+1}-a_{\ell+2}|} \, .
\eeq

%==================================
\paragraph{The one-loop train track.}

\begin{figure}[t]
\centering
 % \resizebox{3.5cm}{!}{
\begin{tikzpicture}[very thick]
  \node (G1) at (0,-1) [font=\small, text width=1 cm]{$G_{1,1}$};
  \draw (0,0) -- (2,0);
  \draw (1,1) -- (1,-1);
  \node [{below right}] at (1,0) {$x_1$};
  \draw[fill=black!100] (1,0)  circle (2pt);
  \node [{left}] at (0,0) {$a_1$};
  \node [{above}] at (1,1) {$a_2$};
  \node [{right}] at (2,0) {$a_3$};
  \node [{below}] at (1,-1) {$a_4$};
\end{tikzpicture}
% }
\caption{One-loop train track graph also known as box graph.}
\label{fig:onetraintrack}
\end{figure}
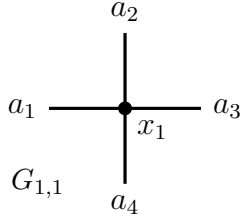

The simplest train track integral is given by the one-loop box graph shown in \Figref{fig:onetraintrack}. This integral only depends on the single cross ratio
\beq
\label{var1traintrack}
    z = \frac1{4^2}\chi_{1,2,4,3}\, .
\eeq
The Yangian differential ideal $\YDI(G_{1,1})$ is generated by the single operator
\begin{equation}
\begin{aligned}
    \mathcal D_{G_{1,1}} &= \theta ^2-4 z (1+2 \theta )^2\, ,
\end{aligned}
\end{equation}
which has a two-dimensional solution space given by the two periods of an elliptic curve
\beq
    \Sol(\PFI(M_{G_{1,1}}))  %=  \Sol(\PFI(\mathcal E))   
    =    \Sol(\mathcal D_{G_{1,1}}) = \Sol(\YDI(G_{1,1})) =  \big\langle \Phi_{G_{1,1},0}(z), \Phi_{G_{1,1},1}(z)\big\rangle_{\mathbb{C}}\, .
\eeq
These two functions are given by elliptic integrals
\begin{equation}
\begin{aligned}
    \Phi_{G_{1,1},0}(z) &= \frac2\pi \K(4^2z) = \sum_{n=0}^\infty \binom{2n}{n}^2z^n = 1+4z+36z^2+400z^3+\mathcal O(z^4) \, , \\
    \Phi_{G_{1,1},1}(z) &= -2\K(1-4^2z) = \Phi_{G_{1,1},0}(z)\log(z) + 8z+84z^2+\frac{2960}3z^3+\mathcal O(z^4) \, ,
\end{aligned}
\end{equation}
where we introduced the complete elliptic integral of the first kind, defined by
\beq
\K(\lambda) \coloneqq \int_0^1\frac{\rd x}{\sqrt{x(1-x)(1-\lambda x)}}\,.
\eeq
Note that the functions $ \Phi_{G_{1,1},i}(z)$ have singularities at $z\in\{0,1/16,\infty\}$. 
We can now change basis to an integral monodromy basis
\beq
% \label{rotmonl1}
    \underline{\Pi}_{G_{1,1}}(z) = \begin{pmatrix}
         1 & 0  \\
 0 & \frac{1}{2 \pi  i} 
    \end{pmatrix} \underline \Phi_{G_{1,1}}(z)\, ,
\eeq
such that the one-loop train track integral is given by
\beq
\phi_{G_{1,1}}(z) = -i\, \underline{ \Pi}_{G_{1,1}}(z)^\dagger \Sigma_{G_{1,1}} \underline{ \Pi}_{G_{1,1}}(z)
\eeq
with intersection form
\beq
    \Sigma_{G_{1,1}} = \begin{pmatrix}
         0 & 1  \\
 -1 & 0 
    \end{pmatrix} \, .
\eeq
This result agrees with refs.~\cite{Chicherin:2017frs,Corcoran:2021gda}.

%==================================
\paragraph{The two-loop train track.}

\begin{figure}[t]
\centering
 % \resizebox{3.5cm}{!}{
\begin{tikzpicture}[very thick]
  \node (G1) at (0,-1) [font=\small, text width=1 cm]{$G_{1,2}$};
  \draw (0,0) -- (3,0);
  \draw (1,1) -- (1,-1);
  \node [{below right}] at (1,0) {$x_1$};
  \draw[fill=black!100] (1,0)  circle (2pt);
  \draw (2,1) -- (2,-1);
  \node [{below right}] at (2,0) {$x_2$};
  \draw[fill=black!100] (2,0)  circle (2pt);
  \node [{left}] at (0,0) {$a_1$};
  \node [{above}] at (1,1) {$a_2$};
  \node [{above}] at (2,1) {$a_3$};
  \node [{right}] at (3,0) {$a_4$};
  \node [{below}] at (2,-1) {$a_5$};
  \node [{below}] at (1,-1) {$a_6$};
\end{tikzpicture}
% }
\caption{Two-loop train track graph.}
\label{fig:twotraintrack}
\end{figure}
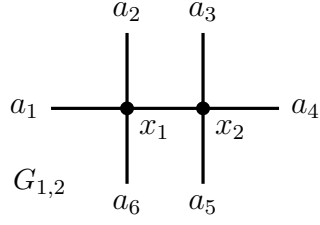

This graph is shown in \Figref{fig:twotraintrack}, where we also indicate the labelling of the external points. The necessary period integrals depend on the  three cross ratios
\beq
\label{var2traintrack}
    z_1 = \frac14\chi_{1,2,3,4}\,, \quad z_2 = \frac14\chi_{1,3,6,4}\,, \quad z_3 = \frac14\chi_{1,6,5,4}\, .
\eeq
At two loops, the Yangian (including permutations) furnishes the full differential operator ideal, which specifies the period integrals. In the variables from eq.~\eqref{var2traintrack} a generating set of the Yangian differential ideal can be chosen to be
\begin{align}
 \nonumber   \mathcal D_{G_{1,2},1} &= \theta_1^2 -2 z_1 \left(\theta _1-\theta _2\right) \left(1+2 \theta _1+2 \theta _2\right)-4 z_1 z_2 \left(1+2 \theta _2-2 \theta _3\right)^2 \\
    &\quad -32 z_1 z_2 z_3 \left(1+2 \theta _2-\theta _3\right) \left(1+2 \theta _3\right) \, , \\
 \nonumber   \mathcal D_{G_{1,2},2} &= \theta _1 \theta _2-\theta _3 \left(\theta _2-\theta _3\right) +2 z_3 \left(\theta _2-\theta _3\right) \left(1+2 \theta _3\right)-4 z_1 z_2 \left(1+2 \theta _1\right) \left(1+2 \theta _2-2 \theta _3\right)\\
 \nonumber   &\quad -4 z_1 z_2 z_3 \left(1+2 \theta _1\right) \left(4+8 \theta _3\right) \, , \\
\nonumber    \mathcal D_{G_{1,2},3} &= (\theta_1-\theta_2)\theta_3 + 4 z_3 \left(\theta _1-\theta _2\right) \left(\theta _2-\theta _3\right) \\
\nonumber    &\quad + 4 z_2 z_3 \left(-4 \theta _1 \left(1+\theta _2\right)+\left(1+2 \theta _2\right){}^2-4 \theta _2 \theta _3+4 \theta _3^2\right) +32 z_2 z_3^2 \left(\theta _2-\theta _3\right) \left(1+2 \theta _3\right) \, ,
\end{align}
where we used logarithmic derivatives $\theta_i=z_i\partial_i$ for $i=1,2,3$. Around the MUM-point $\uz=0$ for $i=1,2,3$, these operators have as solution space
\beq
\label{solspacel2}
    \Sol(\left\{ \mathcal D_{G_{1,2},k} \right\}_{k=1,2,3}) = \big\langle \Phi_{G_{1,2},0}(\uz),\Phi_{G_{1,2},1,1}(\uz),\Phi_{G_{1,2},1,2}(\uz),\Phi_{G_{1,2},1,3}(\uz),\Phi_{G_{1,2},2}(\uz)\big\rangle_{\mathbb{C}} \, ,
\eeq
which can be constructed in the following way
\begin{equation}
\begin{aligned}
\label{frobl2}
    \Phi_{G_{1,2},0}(\uz)   &= \varpi(\uz;0)  \\
                            &= 1+\left(4 z_1 z_2+4 z_2 z_3\right)+8 z_1 z_2 z_3+\left(36 z_1^2 z_2^2+16 z_1 z_2^2 z_3+36z_2^2 z_3^2\right) + \mathcal O(z_i^5)\,,\\
    \Phi_{G_{1,2},1,i}(\uz) &= \partial_{\rho_i}\varpi(\uz;\underline\rho)|_{\underline\rho=0} = \phi_{G_{1,2},0}(\uz) \log(z_i)+ \Sigma_{G_{1,2},1,i}\,,\quad \text{for } i=1,2,3 \, , \\
    \Phi_{G_{1,2},2}(\uz)   &= \left[  \partial_{\rho_2}^2+2\left(  \partial_{\rho_1}\partial_{\rho_2}+\partial_{\rho_1}\partial_{\rho_3}+\partial_{\rho_2}\partial_{\rho_3}\right)\right]\varpi(\uz;\underline\rho)|_{\underline\rho=0} \\
                            &= \Phi_{G_{1,2},0}(\uz)\left[\log ^2\left(z_2\right)+2 \left(\log \left(z_1\right) \log \left(z_2\right)+\log \left(z_1\right) \log \left(z_3\right)+\log\left(z_2\right) \log \left(z_3\right)\right)\right] \\
                            &\quad + 2\log(z_1)(\sigma_{G_{1,2},1,2}+\sigma_{G_{1,2},1,3}) +2\log(z_2)(\sigma_{G_{1,2},1,1}+\sigma_{G_{1,2},1,2}+\sigma_{G_{1,2},1,3}) \\
                            &\quad   +2\log(z_3)(\sigma_{G_{1,2},1,1}+\sigma_{G_{1,2},1,2})  + \sigma_{G_{1,2},2} - \pi^2\Phi_{G_{1,2},0}(\uz)\,,
\end{aligned}
\end{equation}
with coefficients
\begin{equation}
\begin{aligned}
     \varpi(\uz;\underline\rho) &= \sum_{\underline n=0}^\infty c(\underline n+\underline \rho) \uz^{\underline n+\underline \rho}\, , \quad c(\underline n) = (n_1)(n_3)(n_2-n_1)(n_2-n_3)(n_1-n_2+n_3) \, , \\
     \Sigma_{G_{1,2},1,1}       &=  2 z_1-2 z_2+\left(3 z_1^2+4 z_1 z_2-3 z_2^2-4 z_2 z_3\right)+ \mathcal O(z_i^3)    \, , \\
     \Sigma_{G_{1,2},1,2}       &=   -2 z_1+2 z_2-2 z_3+\left(-3 z_1^2+4 z_1 z_2+3 z_2^2+4 z_2 z_3-3 z_3^2\right)+ \mathcal O(z_i^3)   \, , \\
     \Sigma_{G_{1,2},1,3}       &=   -2 z_2+2 z_3+\left(-4 z_1 z_2-3 z_2^2+4 z_2 z_3+3 z_3^2\right) + \mathcal O(z_i^3)  \, , \\
     \Sigma_{G_{1,2},2}         &=     -4 z_1^2+8 z_1 z_2-4 z_2^2+8 z_2 z_3-4 z_3^2+\mathcal O(z_i^3)  \, ,
\end{aligned}
\end{equation}
where we used the shorthand notation for the binomials $(n) = \binom{2n}{n}$. This solution space was constructed as the solution space to the Yangian differential ideal $\YDI(G_{1,2})$. We find that the solution space in eq.~\eqref{solspacel2} is exactly the space of K3 periods, in agreement with our conjecture:
\beq
    \Sol(\PFI(M_{G_{1,2}}))  =  \Sol(\left\{ \mathcal D_{G_{1,2},k} \right\}_{k=1,2,3}) =\Sol(\YDI(G_{1,2}))\, .
\eeq

The solutions in eq.~\eqref{frobl2} are given as locally convergent series around the MUM-point $\uz=0$. By analytic continuation, one can extend them to global solutions. For this it is essential to understand their singularity structure which is determined by the discriminant locus given by
\beq
    \Delta_{G_{1,2}} =  (1-4z_1)(1-4z_2)(1-4z_3)(1-16z_1z_2)(1-16z_2z_3)(1-64z_1z_2z_3) \, .
\eeq
To obtain the monodromy-invariant bilinear in the periods, we have to choose a monodromy basis which has at least real monodromies, better integral monodromies as expected in our geometric setting, around all singular divisors. From our construction of the solution space by taking derivatives of the $\rho$-deformed series $\varpi(\uz;\underline\rho)$, we get directly a rational monodromy basis if we normalize the logarithmic solutions by an appropriate power of $2\pi i$. As it is argued in~\cite{kerr2020unipotent}, one always obtains at least a rational monodromy basis by constructing the solution space from the $\rho$-deformed series $\varpi(\uz;\underline\rho)$ for hypergeometric systems at a MUM point. To simplify the intersection form later, we also add with a factor of $-1/4$ the holomorphic solution to the double logarithmic one. Since this is a rational contribution, the monodromy matrices are still at least rational. In total, we have the following rotation
\beq
\label{rotmonl2}
    \underline{\Pi}_{G_{1,2}}(\uz) = \begin{pmatrix}
         1 & 0 & 0 & 0 & 0 \\
 0 & \frac{1}{2 \pi  i} & 0 & 0 & 0 \\
 0 & 0 & \frac{1}{2 \pi  i} & 0 & 0 \\
 0 & 0 & 0 & \frac{1}{2 \pi  i} & 0 \\
 -\frac{1}{4} & 0 & 0 & 0 & \frac{1}{(2 \pi  i)^2}
    \end{pmatrix} \underline \Phi_{G_{1,2}}(\uz)\, .
\eeq
Then the two-loop train track integral is given by
\beq
  \phi_{G_{1,2}}(\uz) = - \underline{ \Pi}_{G_{1,2}}(\uz)^\dagger \Sigma_{G_{1,2}} \underline{ \Pi}_{G_{1,2}}(\uz)\,,
\eeq
with the intersection form
\beq
    \Sigma_{G_{1,2}} = \begin{pmatrix}
                       0 & 0 & 0 & 0 & 1 \\
 0 & 0 & -2 & -2 & 0 \\
 0 & -2 & -2 & -2 & 0 \\
 0 & -2 & -2 & 0 & 0 \\
 1 & 0 & 0 & 0 & 0   \end{pmatrix} \, .
\eeq
We have checked that our result agrees with a direct numerical evaluation of the integral.
%

%==================================
%==================================
\paragraph{The three-loop train track.}

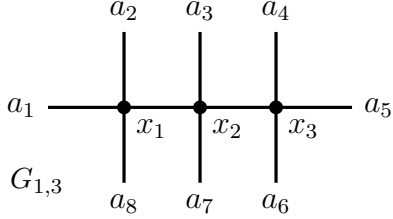
\begin{figure}[t]
\centering
 % \resizebox{4.5cm}{!}{
\begin{tikzpicture}[very thick]
  \node (G1) at (0,-1) [font=\small, text width=1 cm]{$G_{1,3}$};
  \draw (0,0) -- (4,0);
  \draw (1,1) -- (1,-1);
  \node [{below right}] at (1,0) {$x_1$};
  \draw[fill=black!100] (1,0)  circle (2pt);
  \draw (2,1) -- (2,-1);
  \node [{below right}] at (2,0) {$x_2$};
  \draw[fill=black!100] (2,0)  circle (2pt);
  \draw (3,1) -- (3,-1);
  \node [{below right}] at (3,0) {$x_3$};
  \draw[fill=black!100] (3,0)  circle (2pt);
  \node [{left}] at (0,0) {$a_1$};
  \node [{above}] at (1,1) {$a_2$};
  \node [{above}] at (2,1) {$a_3$};
  \node [{above}] at (3,1) {$a_4$};
  \node [{right}] at (4,0) {$a_5$};
  \node [{below}] at (3,-1) {$a_6$};
  \node [{below}] at (2,-1) {$a_7$};
  \node [{below}] at (1,-1) {$a_8$};
\end{tikzpicture}
% }
\caption{Three-loop train track graph.}
\label{fig:threetraintrack}
\end{figure}
From the Z-rule we obtain the following five cross ratios
\beq
\label{var3traintrack}
    z_1 = \frac14\chi_{1,2,3,5}\,, \quad z_2 = \frac14\chi_{1,3,4,5}\,, \quad z_3 = \chi_{1,4,8,5}\,, \quad z_4 = \frac14\chi_{1,8,7,5}\,, \quad z_5 = \frac14\chi_{1,7,6,5}\, .
\eeq
At three loops, the Yangian (including permutation and two-site symmetries) furnishes the full differential operator ideal, which specifies the period integrals. 
%We need at least one of the two-site symmetries described above, see \eqref{eq:twositesyms}. 
In the variables of eq.~\eqref{var3traintrack}, a generating set of the Yangian differential ideal is given by five operators $\mathcal D_{G_{1,3},k}$ for $k=1,\hdots,5$, where their explicit expressions are shown in Appendix~\ref{app:examples3t}. At $\uz=0$ we have a MUM-point, and the solution space of the Yangian differential operators is eight-dimensional
\begin{align}
    &\Sol(\PFI(M_{G_{1,3}}))  =     \Sol(\left\{ \mathcal D_{G_{1,3},k} \right\}_{k=1,\hdots,5}) =\Sol(\YDI(G_{1,3})) = \\
\nonumber     &\quad= \big\langle \Phi_{G_{1,3},0}(\uz),\Phi_{G_{1,3},1,1}(\uz),\hdots,\Phi_{G_{1,3},1,5}(\uz),\Phi_{G_{1,3},2,1}(\uz),\hdots,\Phi_{G_{1,3},2,5}(\uz),\Phi_{G_{1,3},3}(\uz)\big\rangle_{\mathbb{C}}\, ,
\end{align}
and was constructed from a $\rho$-deformed series $\varpi(\uz;\underline\rho)$ to obtain a rational mondromy basis. Moreover, these solutions form the periods a CY three-fold and their explicit expressions are also given in Appendix~\ref{app:examples3t}. They exhibit singularities on the discriminant locus
\begin{equation}
\begin{aligned}
    \Delta_{G_{1,3}} =&  (1-4z_1)(1-4z_2)(1-4z_3)(1-4z_4)(1-4z_5)(1-16z_1z_2)(1-16z_2z_3)(1-16z_3z_4)\\
    &\quad (1-16z_4z_5)(1-64z_1z_2z_3)(1-64z_2z_3z_4)(1-64z_3z_4z_5)(1-256z_1z_2z_3z_4) \\
    &\quad (1-256z_2z_3z_4z_5)(1-1024z_1z_2z_3z_4z_5) \, .
\end{aligned}
\end{equation}

With these periods, the three-loop train track integral is given by
\beq
\label{resultl3}
\phi_{G_{1,3}}(\uz) = i\, \underline{ \Pi}_{G_{1,3}}(\uz)^\dagger \Sigma_{G_{1,3}} \underline{ \Pi}_{G_{1,3}}(\uz)\,,
\eeq
with intersection form
\beq
    \Sigma_{G_{1,3}} = \left(
\begin{array}{c c}
    0 & \mathbb J_{6\times6} \\
    -\mathbb J_{6\times6} & 0
\end{array}
\right) \, \quad\text{with}\quad(\mathbb J_{6\times6})_{ij} = \delta_{6-i-j}
\eeq
and basis $\underline{ \Pi}_{G_{1,3}} = T_{G_{1,3}} \underline{ \Phi}_{G_{1,3}}$. The rotation $T_{G_{1,3}}$ is shown in Appendix~\ref{app:examples3t}. We have checked that our result in eq.~\eqref{resultl3} coincides with a direct numerical evaluation of the integral.

%==================================

%%%%%%%%%%%%%%%%%%%%%%%%%%%%%%%%%%%%%%%%
%%%%%%%%%%%%%%%%%%%%%%%%%%%%%%%%%%%%%%%%
\subsection{Hexagonal tilings and triangle track graphs}
\label{subsec:zigzags}

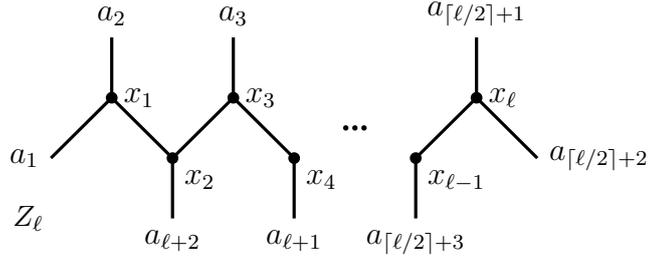
\begin{figure}[t]
\centering
 % \resizebox{6cm}{!}{
\begin{tikzpicture}[very thick,scale=0.8]
  \node (G1) at (0,-1) [font=\small, text width=1 cm]{$Z_\ell$};
  \draw (0,0) -- (1,1);
  \draw (1,1) -- (2,0);
  \draw (2,0) -- (3,1);
  \draw (3,1) -- (4,0);
  \draw (6,0) -- (7,1);
  \draw (7,1) -- (8,0);
  \draw (1,1) -- (1,2);
  \draw (3,1) -- (3,2);
  \draw (7,1) -- (7,2);
  \draw (2,0) -- (2,-1);
  \draw (4,0) -- (4,-1);
  \draw (6,0) -- (6,-1);

  \draw[fill=black!100] (1,1)  circle (2pt);
  \node [{right}] at (1,1) {$x_1$};
  \draw[fill=black!100] (2,0)  circle (2pt);
  \node [{below right}] at (2,0) {$x_2$};
  \draw[fill=black!100] (3,1)  circle (2pt);
  \node [{right}] at (3,1) {$x_3$};
  \draw[fill=black!100] (4,0)  circle (2pt);
  \node [{below right}] at (4,0) {$x_4$};
  \draw[fill=black!100] (6,0)  circle (2pt);
  \node [{below right}] at (6,0) {$x_{\ell-1}$};
  \draw[fill=black!100] (7,1)  circle (2pt);
  \node [{right}] at (7,1) {$x_\ell$};

  \draw[fill=black!100] (4.85,0.5)  circle (0.5pt);
  \draw[fill=black!100] (5,0.5)  circle (0.5pt);
  \draw[fill=black!100] (5.15,0.5)  circle (0.5pt);

  \node [{left}] at (0,0) {$a_1$};
  \node [{above}] at (1,2) {$a_2$};
  \node [{above}] at (3,2) {$a_3$};
  \node [{above}] at (7,2) {$a_{\lceil\ell/2\rceil+1}$};
  \node [{right}] at (8,0) {$a_{\lceil\ell/2\rceil+2}$};
  \node [{below}] at (6,-1) {$a_{\lceil\ell/2\rceil+3}$};
  \node [{below}] at (4,-1) {$a_{\ell+1}$};
  \node [{below}] at (2,-1) {$a_{\ell+2}$};
\end{tikzpicture}
% }
\caption{$\ell$-loop triangle track graph.}
\label{fig:zigzag}
\end{figure}

We now discuss some examples of low-loop fishnet integrals for a hexagonal tiling. Up to three loops, all such graphs are instances of 
{triangle track graphs} $Z_{\ell}$ (see~\Figref{fig:zigzag}). In the following, we present explicit results for triangle track graphs up to eight external points. Results for the triangle wheel graph can be found in appendix~\ref{app:T6}. We do not discuss the one-loop case, because it is entirely fixed by the star-triangle relation.

Our goal is to determine the periods for the CY varieties $M_{Z_\ell}$ attached to triangle track graphs, for which we need to know the PFI. Our conjecture implies that the PFI is generated by the Yangian generators and two-side densities (and its permutations). The automorphisms of $Z_{\ell}$ are
\beq
\Aut(Z_\ell) = \left\{\begin{array}{ll}
S_3\,,\, & \ell=1\,,\\
\mathbb{Z}_2^3\,,\,  & \ell>1\,.
\end{array}\right.
\eeq

Unlike for the square tilings, in this case there are hidden permutation symmetries which are not part of $\Aut(Z_\ell)$. In order to expose these hidden symmetries, 
we bring the graph into a canonical form by employing the admissible star-triangle relation to the left- and right-most vertices (there are no other admissible star-triangle relations). However, since the powers of the propagators change under the application of the star-triangle relation, this does not increase the permtuation symmetries of the fishnet integral for $\ell\neq 3$, and so $\textrm{Perm}_{Z_{\ell}} = \Aut(Z_\ell)$ for $\ell\neq 3$. For $\ell=3$, we have (see the discussion in section~\ref{sec:hidden})
\beq
\textrm{Perm}_{Z_3} = S_4\,.
\eeq
Following the discussion in section~\ref{sec:hidden}, we do find additional hidden symmetries from two-site densities for $\ell>1$.

An alternative calculation for the below examples of two- and three-loop triangle tracks in the framework of twisted cohomology was given in ref.~\cite{Duhr:2023bku}.

%%%%%%%%%%%%%%%%%%%%%%%%%%%%%%%%%%%%%%%%
\paragraph{The two-loop triangle track graph $Z_2$.}
The simplest non-trivial triangle track graph is given by the two-loop one shown in \Figref{fig:2zigzag}. The corresponding integral only depends on a single cross ratio
\beq
\label{var2zigzag}
    z = \frac1{3^3}\chi_{1,4,2,3} \, ,
\eeq
where the additional factor of $1/3^3$ ensures an integer series expansion later.

The Yangian differential ideal $\YDI(Z_2)$ is generated by a single differential operator
\begin{equation}
\begin{aligned}
    \mathcal D_{Z_2}  &=   \theta ^2-3 z (1+3 \theta ) (2+3 \theta )    \, ,                      
\end{aligned}
\end{equation}
having a two-dimensional solution space
\beq
    \Sol(\PFI(M_{Z_2}))  =    \Sol(\mathcal D_{Z_2}) =\Sol(\YDI(Z_2)) =   \big\langle \Phi_{Z_2,0}(z), \Phi_{Z_2,1}(z)\big\rangle_{\mathbb{C}}\, .
\eeq
These two solutions can be related to ${_2F_1}$ hypergeometric functions
\begin{equation}
\begin{aligned}
    \Phi_{Z_2,0}(z) &\,= {_2F_1}(1/3,2/3,1;3^3z)\\
    &\, = 1+6 z+90 z^2+1680 z^3+\mathcal O(z^4) \, , \\
    \Phi_{Z_2,1}(z) &\,= -\frac{2 \pi }{\sqrt{3}}\,{_2F_1}(1/3,2/3,1;1-3^3z)\\
    &\, = \Phi_{Z_2,0}(z)\log(z) + 15 z+\frac{513 }{2}z^2+5018 z^3+\mathcal O(z^4) \, .
\end{aligned}
\end{equation}
The functions have singularities for $z\in\{0,1/27,\infty \}$.

We can now change basis to an integral monodromy basis
\beq
\label{rotmonl1}
    \underline{\Pi}_{Z_2}(z) = \begin{pmatrix}
         1 & 0  \\
 0 & \frac{1}{2 \pi  i} 
    \end{pmatrix} \underline \Phi_{Z_2}(z)\, ,
\eeq
such that the integral is given by
\beq
\label{resultz2}
    I_{Z_2}(z) = i\frac1{|a_{12}|^{4/3}|a_{34}|^{4/3}}\, \underline{ \Pi}_{Z_2}(z)^\dagger \Sigma_{Z_2} \underline{ \Pi}_{Z_2}(z)\,,
\eeq
with intersection form
\beq
    \Sigma_{Z_2} = \begin{pmatrix}
         0 & 1  \\
 -1 & 0 
    \end{pmatrix} \, .
\eeq
We have compared our result to a direct numerical evaluation of the integral and found very good agreement.

%%%%%%%%%%%%%%%%%%%%%%%%%%%%%%%%%%%%%%%%
\paragraph{The three-loop triangle track graph $Z_3$.}

The three-loop triangle track graph shown in \Figref{fig:3zigzag} is associated to a family of Calabi-Yau three-folds which depends on two parameters given by the following cross ratios
\beq
\label{var3zigzag}
    z_1 = \frac1{3^3}\chi_{1,5,3,4}\,, \quad z_2 = \chi_{1,3,2,4}\, .
\eeq

After applying the star-triangle identity as shown in \Figref{fig:3zigzag}, we can derive the Yangian differential ideal $\YDI(Z_3)$, which is generated by
\begin{equation}
\begin{aligned}
    \mathcal D_{Z_3,1}  &=  \theta _1 \theta _2-9 z_1 \left(1+3 \theta _1-3 \theta _2\right) \theta _2-3 z_1 z_2 \left[2+9 \theta _2 \left(1+\theta _2\right)\right]        \, , \\
    \mathcal D_{Z_3,2}  &=   \theta _2 \left(-1+3 \theta _2\right)+z_2 \left[-3 \theta _1^2+\theta _1 \left(1+3 \theta _2\right)-\theta _2 \left(1+3 \theta
   _2\right)\right] \\
                        &\quad +27 z_1 z_2 \left[6 \theta _1^2+\theta _1 \left(2-6 \theta _2\right)-3 \theta _2 \left(1+\theta _2\right)\right]-9 z_1
   z_2 \left[27 z_1 \left(2+3 \theta _1\right) \left(1+3 \theta _1-3 \theta _2\right)\right. \\
                        &\quad \left. -z_2 \left(2+9 \theta _2 \left(1+\theta
   _2\right)\right)\right]       \, .                      
\end{aligned}
\end{equation}
The solution space is three-dimensional
\begin{equation}
\begin{aligned}
    \Sol(\PFI(M_{Z_3}))  =    \Sol(\left\{ \mathcal D_{Z_3,1}, \mathcal D_{Z_3,2} \right\}) =\Sol(\YDI(Z_3)) = \big\langle  \Phi_0(\uz), \Phi_1(\uz),\, \varphi_0(\uz) \big\rangle_{\mathbb{C}}\, ,
\end{aligned}
\end{equation}
and can be related to the Appell hypergeometric function $F_1$:
\begin{equation}
\begin{aligned}
\label{frob3zigzag}
    \Phi_{Z_3,0}(\uz)   &=   F_1(2/3,1/3,1/3,1;3^3z_1z_2,3^3z_1) \\
                    &= 1+6 z_1+\left(90 z_1^2+6 z_1 z_2\right)+\left(1680 z_1^3+45 z_1^2 z_2\right) + \mathcal O(z_i^4)     \, , \\
    \Phi_{Z_3,1}(\uz)   &= \Phi_{0}(\uz)\log(z_1)  +  \left(15 z_1-\frac{z_2}{2}\right)+\left(\frac{513 z_1^2}{2}+3 z_1 z_2-\frac{z_2^2}{5}\right)+ \mathcal O(z_i^3)     \, , \\
    \varphi_{Z_3,0}(\uz)   &= z_2^{1/3} \left[ 1+\frac{\lambda  z_2}{6}+\lambda ^2 \left(9 z_1 z_2+\frac{5 z_2^2}{63}\right)+\lambda ^3 \left(\frac{15}{7} z_1 z_2^2+\frac{4
   z_2^3}{81}\right) + \mathcal O(z_i^4) \right]     \, .
\end{aligned}
\end{equation}
These solutions exhibit singularities on the discriminant locus given by
\beq
\label{deltaz3}
    \Delta_{Z_3} =  \left(1-27 z_1\right) \left(1-z_2\right) \left(1-27 z_1 z_2\right) \, .
\eeq

To obtain the monodromy-invariant bilinear for the three-loop triangle track graph, we have to analyse all monodromies of the solutions around the singular divisors determined by eq.~\eqref{deltaz3}. For this, we have to perform analytic continuations around all singular divisors such that we obtain a global basis of solutions. We found that the monodromy-invariant bilinear can be constructed from the basis
\beq
    \underline{\Pi}_{Z_3}(\uz) = \left(
\begin{array}{ccc}
 1 & 0 & 0 \\
 0 & \frac{1}{2 \pi i } & 0 \\
 0 & 0 & \frac{2 i \pi }{\Gamma \left(\frac{1}{3}\right)^3} \\
\end{array}
\right) \underline \Phi_{Z_3}(\uz)\, ,
\eeq
where $\underline \Phi_{Z_3}(\uz) = \big( \Phi_{Z_3,0}(\uz),\Phi_{Z_3,1}(\uz),\varphi_{Z_3,0}(\uz)\big)^T$,
and we find
\beq
\label{resultz3}
    I_{Z_3}(\ua) =  i\frac{|a_{14}|^{2/3}}{|a_{12}|^{4/3}|a_{13}|^{2/3}|a_{45}|^{4/3}|a_{34}|^{2/3}}\, \underline{ \Pi}_{Z_3}(\uz)^\dagger \Sigma_{Z_3} \underline{ \Pi}_{Z_3}(\uz)\,,
\eeq
with the intersection form
\beq
    \Sigma_{Z_3} = \left(
\begin{array}{ccc}
 0 & 1 & 0 \\
 -1 & 0 & 0 \\
 0 & 0 & -i \sqrt{3} \\
\end{array}
\right) \, .
\eeq
The intersection form $\Sigma_{Z_3}$ is constructed by the requirement that it is invariant under conjugation of all monodromy matrices, i.e. $M^T\Sigma_{Z_3}M=\Sigma_{Z_3}$ for any monodromy matrix $M$. We have checked numerically by a direct evaluation of the integral our result in eq. \eqref{resultz3}.

\subsection{Four-point limit of triangle track graphs}
\label{sec:fourponttriangle}
In this subsection, we want to consider a specific four-point limit of the triangle track graphs, similar to the four-point limit of the fishnet graphs (also called ladder graphs) discussed in ref.~\cite{Duhr:2023eld}. We show that these integrals are related to one-parameter hypergeometric period motives. This means that the periods are given by hypergeometric functions, where their arguments are related to the propagator powers, as it is also the case for ladder graphs.

\begin{figure}[t]
\begin{center}

\begin{tikzpicture}[very thick, scale=0.75]
\node (G1) at (0,3) [font=\small, text width=3 cm]{$\text{Odd: } \ell=2m+1$};

  \draw (0,1) -- (1,1);
  \draw (1,1) -- (2,0);
  \draw (2,0) -- (3,1);
  \draw (3,1) -- (4,0);
  \draw (4,0) -- (5,1);
  \draw (5,1) -- (6,0);
  \draw (8,0) -- (9,1);
  \draw (9,1) -- (11,1);
  
  \draw (1,1) -- (5.5,2.5);
  \draw (3,1) -- (5.5,2.5);
  \draw (5,1) -- (5.5,2.5);
  \draw (9,1) -- (5.5,2.5);
  \draw (2,0) -- (5.5,-1.5);
  \draw (4,0) -- (5.5,-1.5);
  \draw (6,0) -- (5.5,-1.5);
  \draw (8,0) -- (5.5,-1.5);

  \draw[fill=black!100] (1,1)  circle (2pt);
  \node [{right}] at (1,1) {$x_1$};
  \draw[fill=black!100] (2,0)  circle (2pt);
  \node [{right}] at (2,0) {$x_2$};
  \draw[fill=black!100] (3,1)  circle (2pt);
  \node [{right}] at (3,1) {$x_3$};
  \draw[fill=black!100] (4,0)  circle (2pt);
  \node [{right}] at (4,0) {$x_4$};
  \draw[fill=black!100] (5,1)  circle (2pt);
  \node [{right}] at (5,1) {$x_5$};
  \draw[fill=black!100] (6,0)  circle (2pt);
  \node [{right}] at (6,0) {$x_6$};
  \draw[fill=black!100] (8,0)  circle (2pt);
  \node [{right}] at (8,0) {$x_{2m}$};
  \draw[fill=black!100] (9,1)  circle (2pt);
  \node [{below right}] at (9,1) {$x_{2m+1}$};

  \draw[fill=black!100] (6.85,0.5)  circle (0.5pt);
  \draw[fill=black!100] (7,0.5)  circle (0.5pt);
  \draw[fill=black!100] (7.15,0.5)  circle (0.5pt);

  \node [{left}] at (0,1) {$z$};
  \node [{above}] at (5.5,2.5) {$\infty$};
  \node [{right}] at (11,1) {$1$};
  \node [{below}] at (5.5,-1.5) {$0$};

\node (G2) at (13.5,0.5) [font=\small, text width=2 cm]{$\rightarrow$};

  \draw [darkred] (14,0.5) -- (19,0.5);
  \draw (15,0.5) -- (16.5,2.5);
  \draw (16,0.5) -- (16.5,2.5);
  \draw (17,0.5) -- (16.5,2.5);
  \draw (18,0.5) -- (16.5,2.5);
  \draw (15,0.5) -- (16.5,-1.5);
  \draw (16,0.5) -- (16.5,-1.5);
  \draw (17,0.5) -- (16.5,-1.5);
  \draw (18,0.5) -- (16.5,-1.5);

  \draw[fill=black!100] (16.35,0.75)  circle (0.5pt);
  \draw[fill=black!100] (16.5,0.75)  circle (0.5pt);
  \draw[fill=black!100] (16.65,0.75)  circle (0.5pt);

  \draw[fill=black!100] (15,0.5)  circle (2pt);
  \node [{below }] at (15,0.5) {$x_1$};
  \draw[fill=black!100] (16,0.5)  circle (2pt);
  \node [{below right}] at (16,0.5) {$x_2$};
  \draw[fill=black!100] (17,0.5)  circle (2pt);
  \node [{below right}] at (16.8,0.5) {$x_{m-1}$};
  \draw[fill=black!100] (18,0.5)  circle (2pt);
  \node [{below right}] at (18,0.5) {$x_m$};

  \node [{left}] at (14,0.5) {$z$};
  \node [{above}] at (16.5,2.5) {$\infty$};
  \node [{right}] at (19,0.5) {$1$};
  \node [{below}] at (16.5,-1.5) {$0$};
  
\end{tikzpicture}
\\
\begin{tikzpicture}[very thick, scale=0.75]
\node (G1) at (0,3) [font=\small, text width=3 cm]{$\text{Even: } \ell=2m$};

  \draw (0,1) -- (1,1);
  \draw (1,1) -- (2,0);
  \draw (2,0) -- (3,1);
  \draw (3,1) -- (4,0);
  \draw (4,0) -- (5,1);
  \draw (5,1) -- (6,0);
  \draw (8,0) -- (9,1);
  \draw (9,1) -- (10,0);
  \draw (10,0) -- (11,0);
  
  \draw (1,1) -- (5.5,2.5);
  \draw (3,1) -- (5.5,2.5);
  \draw (5,1) -- (5.5,2.5);
  \draw (9,1) -- (5.5,2.5);
  \draw (2,0) -- (5.5,-1.5);
  \draw (4,0) -- (5.5,-1.5);
  \draw (6,0) -- (5.5,-1.5);
  \draw (8,0) -- (5.5,-1.5);
  \draw (10,0) -- (5.5,-1.5);

  \draw[fill=black!100] (1,1)  circle (2pt);
  \node [{right}] at (1,1) {$x_1$};
  \draw[fill=black!100] (2,0)  circle (2pt);
  \node [{right}] at (2,0) {$x_2$};
  \draw[fill=black!100] (3,1)  circle (2pt);
  \node [{right}] at (3,1) {$x_3$};
  \draw[fill=black!100] (4,0)  circle (2pt);
  \node [{right}] at (4,0) {$x_4$};
  \draw[fill=black!100] (5,1)  circle (2pt);
  \node [{right}] at (5,1) {$x_5$};
  \draw[fill=black!100] (6,0)  circle (2pt);
  \node [{right}] at (6,0) {$x_6$};
  \draw[fill=black!100] (8,0)  circle (2pt);
  \node [{right}] at (8,0) {$x_{2m-2}$};
  \draw[fill=black!100] (9,1)  circle (2pt);
  \node [{right}] at (9,1) {$x_{2m-1}$};
  \draw[fill=black!100] (10,0)  circle (2pt);
  \node [{below}] at (10,0) {$x_{2m}$};

  \draw[fill=black!100] (6.85,0.5)  circle (0.5pt);
  \draw[fill=black!100] (7,0.5)  circle (0.5pt);
  \draw[fill=black!100] (7.15,0.5)  circle (0.5pt);

  \node [{left}] at (0,1) {$z$};
  \node [{above}] at (5.5,2.5) {$\infty$};
  \node [{right}] at (11,0) {$1$};
  \node [{below}] at (5.5,-1.5) {$0$};

\node (G2) at (13.5,0.5) [font=\small, text width=2 cm]{$\rightarrow$};

  \draw [darkred] (14,0.5) -- (18,0.5);
  \draw (18,0.5) -- (19,0.5);  
  \draw (15,0.5) -- (16.5,2.5);
  \draw (16,0.5) -- (16.5,2.5);
  \draw (17,0.5) -- (16.5,2.5);
  \draw [darkred] (18,0.5) -- (16.5,2.5);
  \draw (15,0.5) -- (16.5,-1.5);
  \draw (16,0.5) -- (16.5,-1.5);
  \draw (17,0.5) -- (16.5,-1.5);
  \draw (18,0.5) -- (16.5,-1.5);

  \draw[fill=black!100] (16.35,0.75)  circle (0.5pt);
  \draw[fill=black!100] (16.5,0.75)  circle (0.5pt);
  \draw[fill=black!100] (16.65,0.75)  circle (0.5pt);

  \draw[fill=black!100] (15,0.5)  circle (2pt);
  \node [{below }] at (15,0.5) {$x_1$};
  \draw[fill=black!100] (16,0.5)  circle (2pt);
  \node [{below right}] at (16,0.5) {$x_2$};
  \draw[fill=black!100] (17,0.5)  circle (2pt);
  \node [{below right}] at (16.8,0.5) {$x_{m-1}$};
  \draw[fill=black!100] (18,0.5)  circle (2pt);
  \node [{below right}] at (18,0.5) {$x_m$};

  \node [{left}] at (14,0.5) {$z$};
  \node [{above}] at (16.5,2.5) {$\infty$};
  \node [{right}] at (19,0.5) {$1$};
  \node [{below}] at (16.5,-1.5) {$0$};
  
\end{tikzpicture}

\end{center}
\caption{Four-point limit of triangle track graphs together with non-isotropic ladder graphs obtained after using the star-triangle identity. Black lines represent propagators with power $2/3$ whereas red lines have propagator power $1/3$.}
\label{fig:4pttriangle}
\end{figure}

We can distinguish two different cases, where the original triangle track graph has either odd $\ell=(2m+1)$ or even $\ell=(2m)$ loops. After using the star-triangle identity, the two types are given by $m$-loop non-isotropic ladder graphs as shown in figure~\ref{fig:4pttriangle}, where the different propagator powers are $2/3$ (black) or $1/3$ (red). The corresponding integrals depend only on a single cross ratio, for which we take the following choice
\beq
    z = \frac1{(3\sqrt3)^{m+1}}\chi_{1,4,2,3} \, .
\label{4ptcross}
\eeq
The additional factor of $1/(3\sqrt3)^{m+1}$ makes the later series expansions $\mathbb Z[\sqrt3]$-integral (i.e., we allow for integer coefficients and the appearance of $\sqrt3$ contributions) and simplifies their monodromies.

The PFIs of triangle track graphs in the four-point limit can be derived from a limit of the results from section \ref{subsec:zigzags}, or even simpler from an expansion of the torus integral over the holomorphic $(m,0)$-form $\Omega$ in eq.~\eqref{eq:Omega} as explained in section 5 of ref.~\cite{Duhr:2023eld}. For the two different cases, we find\footnote{Note that despite the fact that they come from restriction 
of a Calabi-Yau Picard Fuchs system and application of the star-triangle relations, these 
operators are not Calabi-Yau operators in the sense of~\cite{BognerCY,MR3822913}, as they lack self-adjointness 
as well as the integrality properties.} 
\begin{equation}
\begin{aligned}
\text{Odd:} \hspace{1cm} {\mathcal L}_m^\text{Odd} &= \theta^{m+1} - (\sqrt3)^{m+1}z(1+3\theta)^{m+1} \, , \\
\Phi_{m,0}^\text{Odd}(z) &= {_{m+1}F_m}(1/3,\hdots,1/3;1,\hdots,1;(3\sqrt3)^{m+1}z) \, , \\ 
\text{Even:} \hspace{1cm} {\mathcal L}_m^\text{Even} &= \theta^{m+1} - (\sqrt3)^{m+1}z(2+3\theta)(1+3\theta)^{m} \, , \\
\Phi_{m,0}^\text{Even}(z) &= {_{m+1}F_m}(2/3,1/3,\hdots,1/3;1,\hdots,1;(3\sqrt3)^{m+1}z) \, .
\label{opsmotivetri}
\end{aligned}
\end{equation}
The singularity structure of both hypergeometric systems is rather simple. Their singularities are located at $0,\mu\coloneqq1/(3\sqrt3)^{m+1}$ and $\infty$. In more detail, their Riemann $\mathcal P$-symbols are
\begin{equation}
\begin{aligned}
\mathcal P^\text{Odd}\left\{ \begin{matrix}  0 & \mu & \infty \\ \hline
                                            0 & 0                    & 1/3 \\
                                            0 & 1                    & 1/3 \\
                                            0 & 2                    & 1/3 \\
                                            \vdots & \vdots & \vdots       \\
                                            0 & m-1               & 1/3 \\
                                            0 & \frac{2m-1}3      & 1/3
                            \end{matrix} \right\} \quad \text{and} \quad
\mathcal P^\text{Even}\left\{ \begin{matrix}  0 & \mu & \infty \\ \hline
                                            0 & 0                    & 1/3 \\
                                            0 & 1                    & 1/3 \\
                                            0 & 2                    & 1/3 \\
                                            \vdots & \vdots & \vdots       \\
                                            0 & m-1               & 1/3 \\
                                            0 & \frac{2(m-1)}3      & 2/3
                            \end{matrix} \right\} \, .
\label{RPmotivetri}
\end{aligned}
\end{equation}
These local exponents of general hypergeometric systems have been stated  in ref.~\cite{MR0974906} together with a 
complete analysis of the properties of their  global monodromy groups. In particular, the monodromy group\footnote{Notice, that $\mu$ 
is scaled to $1$ by a redefinition of $z$ and $m+1$ is denoted $n$ in ref.~\cite{MR0974906}.} 
is generated by $M_0,M_\mu,M_\infty$ with the relation $M_\infty=M_\mu^{-1}M_0^{-1}$, and generate for 
our cases an irreducible subgroup of GL$(m+1,\mathbb{C})$. According to ref.~\cite{MR0974906},  one can choose a basis such that these monodromies are defined over an algebraic extension in the integers $\mathbb{Z}[\underline \alpha ]$.  The tuple $\underline \alpha$ is determined in great generality by the $(m+1)$-tuple $\underline a$ and the $m$-tuple $\underline b$ appearing in the definition of  
$_{m+1}F_m({\underline a},{\underline b},z)$ in ref.~\cite{MR0974906}. In our cases, this extension of the integers  
can be identified  simply with $\mathbb{Z}[\alpha]$ with $\alpha=\exp(\pi i /3)$. Moreover, quite generally  
and in particular in our cases, there exists a monodromy invariant hermitian form $F({\underline X},{\underline Y})=i^{m^2} {\underline X}^\dagger\Sigma {\underline Y}$ on $\mathbb{C}^{m+1}$ with signature $(p,q)$, where  $|p-q|=0$ for $m$ odd and $|p-q|=1$ for $m$ even and $F({\underline\Pi}^\dagger, {\underline \Pi})\in \mathbb{R}_{\ge 0}$. The latter quantity can be seen as the volume of the singular Calabi-Yau 
geometry and is up to a normalisation, the Feynman integral. 
The actual calculation of a $\mathbb{Z}[\alpha]$ basis and the corresponding intersection 
form $\Sigma$ is not explicitly performed in ref.~\cite{MR0974906}, even though many  useful intermediate statements  
are made.        

Since a basis and the corresponding intersection form $\Sigma$ is essential to evaluate the Feynman integral, we have constructed the latter information  from a numerical analysis of transition matrices yielding  the global monodromies for various examples of the corresponding hypergeometric systems in eq.~\eqref{opsmotivetri}. For our two specific hypergeometric systems in eq.~\eqref{opsmotivetri} we find in this way the  bases $\underline \Pi_m^\text{Odd}$ and $\underline \Pi_m^\text{Even}$, respectively, such that all monodromy matrices have entries\footnote{This is also related to the $\mathbb Z[\sqrt3]$-integrality  of the holomorphic period in eq.~\eqref{4ptcross}.} in $\mathbb{Z}[\alpha]$.
The corresponding intersection forms $\Sigma$ for the $\mathbb Z[\alpha]$-integral monodromy bases are determined by the requirement that the form $F$ is globally monodromy invariant, i.e. $F((M_*\underline \Pi)^\dagger, M_* {\underline \Pi})=F({\underline\Pi}^\dagger, {\underline \Pi})$ for $*=0,\mu,\infty$.

To give an explicit example, we can consider the odd $m=3$ case.\footnote{We drop here the superscript `$\text{Odd}$'.} We can construct a $\mathbb Z[\alpha]$-integral monodromy basis $\underline \Pi_3$ using again the $\rho$-deformed series
\begin{equation}
    \varpi(z;\rho) = \sum_{n=0}^\infty c(n+\rho)(3^6z)^{n+\rho}\, , \quad c(n) = \left(\frac{\Gamma(1/3+n)}{\Gamma(1/3)\Gamma(1+n)}\right)^4
\label{rho4F3}    
\end{equation}
to obtain directly a $\mathbb Z[\alpha]$-rational monodromy basis. Explicitly, we find
\begin{equation}
\begin{aligned}
    \Phi_{3,0}(z) &\,= \varpi(z;0) = {_4F_3}(1/3,1/3,1/3,1/3;1,1,1;3^6z)\\
    &\, = 1+9 z+1296 z^2+345744 z^3+\mathcal O(z^4) \, , \\
    \Phi_{3,1}(z) &\,= \partial_{\rho}\varpi(z;\rho)|_{\underline\rho=0} = \Phi_{3,0}(z)\log(z) + f_1-\frac{2 \pi }{\sqrt{3}}\Phi_{3,0}(z) \, , \\
    \Phi_{3,2}(z) &\,= \partial_{\rho}^2\varpi(z;\rho)|_{\underline\rho=0} = \Phi_{3,0}(z)\log^2(z) + 2f_1 \log(z) + 2f_2 \\
    &\quad -\frac{4 \pi }{\sqrt{3}}\Phi_{3,1}(z)+\frac{2}{3} \left(\pi ^2+6 \zeta(2,1/3)\right)\Phi_{3,0}(z)\, , \\
    \Phi_{3,3}(z) &\,= \partial_{\rho}^3\varpi(z;\rho)|_{\underline\rho=0} = \Phi_{3,0}(z)\log^3(z) + 3f_1 \log^2(z) + 6f_2\log(z) +6f_3 -4 \sqrt{3} \pi  \Phi_{3,2}(z)\\
    &\quad +2 \left(\pi ^2+6 \zeta(2,1/3)\right) \Phi_{3,1}(z) -4 \left(24 \zeta (3)+2 \sqrt{3} \pi  \zeta(2,1/3)+\frac{\pi ^3}{\sqrt{3}}\right)\Phi_{3,0}(z)            \, , \\
    f_1 &= 72 z+11664 z^2+3243408 z^3 +\mathcal O(z^4) \, , \\
    f_2 &= 144 z+30942 z^2+9414958 z^3 +\mathcal O(z^4) \, , \\
    f_3 &= -72 z+8991 z^2+\frac{15962149 }{3}z^3 +\mathcal O(z^4) \, , \\
\end{aligned}
\end{equation}
including the new transcendental number $\zeta(2,1/3)$ given by the Hurwitz $\zeta$-function, defined by $\zeta(s,a)\coloneqq \sum_{k=0}^\infty 1/(k+a)^s$. After normalizing the logarithms with powers of $2\pi i$ $\underline \Phi_3$ is indeed a $\mathbb Z[\alpha]$-rational monodromy basis. For a $\mathbb Z[\alpha]$-integer monodromy basis we use the additional rotation $T_3$ such that $\underline \Pi_3 = T_3 \underline \Phi_3$ with 
\beq
T_3 = 
\left(
\begin{array}{cccc}
 1 & 0 & 0 & 0 \\
 \frac{1}{3} (1-2 \alpha ) & 1 & 0 & 0 \\
 \frac{1}{3}(2-\alpha) & \frac{1}{2} (3-4 \alpha ) & \frac{3}{2} & 0 \\
 0 & 0 & \frac{1}{2} (1-2 \alpha ) & \frac{1}{2} \\
\end{array}
\right)
\left(
\begin{array}{cccc}
 1 & 0 & 0 & 0 \\
 0 & \frac{1}{2 \pi  i} & 0 & 0 \\
 0 & 0 & \frac{1}{(2 \pi  i)^2} & 0 \\
 0 & 0 & 0 & \frac{1}{(2 \pi  i)^3} \\
\end{array}
\right) \, ,
\eeq
In this basis, the global monodromies are given by
\begin{equation}
\begin{aligned}
    M_0 &= \left( \begin{array}{cccc}
 1 & 0 & 0 & 0 \\
 1 & 1 & 0 & 0 \\
 2 & 3 & 1 & 0 \\
 0 & 1 & 1 & 1 \\
\end{array} \right)\, ,\, \, M_\frac1{3^6} = \left( \begin{array}{cccc}
 1 & 2-\alpha  & -(1+\alpha ) & 3 (1-\alpha ) \\
 0 & 1-\alpha  & -1+\alpha  & -(1+\alpha)  \\
 0 & 1-\alpha  & 0 & 1-2 \alpha  \\
 0 & 0 & 0 & 1 \\
\end{array} \right) \quad\text{and}\\
M_\infty &= \left( \begin{array}{cccc}
-2+3 \alpha  & 7-11 \alpha  & -2 (1-2 \alpha ) & -3 \alpha  \\
 \alpha  & -(4+\alpha ) & 2 & -2+\alpha  \\
 2 \alpha  & -2 (1+3 \alpha ) & 1+2 \alpha  & -1-\alpha  \\
 0 & 2 & -1 & 1 \\
\end{array} \right) ,
\end{aligned}
\end{equation}
with $\alpha= e^{i\pi/3}$. We checked  that these  matrices satisfy $M_0 M_\frac1{3^6} M_\infty = \mathbb 1$. The intersection form for this basis $\underline \Pi_3$ is given by
\begin{equation}
\begin{aligned}
    \Sigma_3 = \left(
\begin{array}{cccc}
 0 & 0 & 0 & 1 \\
 0 & 0 & -1 & 0 \\
 0 & 1 & 0 & 0 \\
 -1 & 0 & 0 & 0 \\
\end{array}
\right) \, . 
\end{aligned}
\end{equation}

Using the $\rho$-deformed series $\varpi(z;\rho)$ one can easily construct for all hypergeometric systems \eqref{opsmotivetri} a $\mathbb Z[\alpha]$-rational monodromy basis. From the $\rho$-derivatives one obtains new transcendental numbers besides factors of $\pi, \sqrt3$ and normal $\zeta$-values which are again given by the Hurwitz $\zeta$-function, more precisely by $\zeta(m,1/3)$. These numbers can also be related to $L$-function values of Dirichlet characters and the methods of their derivation from \eqref{rho4F3} is an extension of the 
$\widehat \Gamma$-class method used in \cite{Bonisch:2020qmm}. The extension is 
also described  in \cite{kerr2020unipotent}.

%% file: AppendixExamples.tex
% !TEX root = 2DFishnetsLong.tex

\section{Low-loop examples: Additional material}
\label{app:examples}

In this appendix, we give additional material necessary to compute low loop fishnet integrals.

\subsection{Three-loop train track}
\label{app:examples3t}

The Yangian differential ideal $\YDI(G_{1,3})$ associated to the three-loop train track graph is generated by five differential operators:
\begin{equation}
\begin{aligned}
    \mathcal D_{G_{1,3},1}  &= 4\theta_5^2 -2 z_5 \left(2 \theta _4-2 \theta _5-1\right) \left(\theta _4-\theta _5\right)-4 z_4 z_5 \left(\theta _3-\theta _4\right) \left(\theta _3+\theta _4-2 \theta _5\right)  \\
                            &\quad +z_3 z_4 z_5 \left(2 \theta _2-2 \theta _3-1\right) \left(2 \theta _2-2 \theta _3+4 \theta _5+1\right)+2 z_2 z_3 z_4 z_5 \left(2 \theta _1-2 \theta _2-1\right) \\
                            &\quad \times\left(\theta _1+\theta _2-2 \theta _3+2 \theta _5\right)-z_1 z_2 z_3 z_4 z_5 \left(2 \theta _1+1\right) \left(2 \theta _1-4 \theta _3+4 \theta _5-1\right)      \, , \\
    \mathcal D_{G_{1,3},2}  &=  4 \theta _5 \left(2 \theta _2-2 \theta _3+\theta _5\right)+2 z_5 \left(\theta _4-\theta _5\right) \left(4 \theta _2-4 \theta _3+2 \theta _4+2 \theta _5+1\right)   \\
                            & \quad  -4 z_4 z_5 \left(\theta _3-\theta _4\right) \left(-2 \theta _2+\theta _3-\theta _4-1\right)+z_3 z_4 z_5 \left(-2 \theta _2+2 \theta _3+1\right){}^2 \\
                            &\quad -2 z_2 z_3 z_4 z_5 \left(2 \theta _1-2 \theta _2-1\right) \left(\theta _1-\theta _2-1\right)+z_1 z_2 z_3 z_4 z_5 \left(2 \theta _1+1\right) \left(2 \theta _1-4 \theta _2-3\right)\, , \\
    \mathcal D_{G_{1,3},3}  &= 4\theta_5^2+2 z_5 \left(\theta _4-\theta _5\right) \left(2 \theta _4+2 \theta _5+1\right) \\ 
                            &\quad-4 z_4 z_5 \left(\theta _1 \theta _2-\theta _2^2-\theta _3+\theta _2 \theta _3-\theta _3^2+\theta _4-\theta _3 \theta _4+2
   \theta _4^2-\theta _4 \theta _5+\theta _5^2\right)   \\
                            & \quad  -z_4 z_5 \left(z_3 \left(-2 \theta _2+2 \theta _3+1\right){}^2+2 \left(z_2 \left(-2 \theta _1+2 \theta _2+1\right) \left(\theta_2-\theta _3\right)+z_4 \left(\theta _3-\theta _4\right) \right.\right. \\ 
                            &\qquad \times \left.\left.\left(2 \theta _4-2 \theta _5+1\right)+z_5 \left(\theta _4-\theta_5\right) \left(2 \theta _5+1\right)\right)\right)       \\ 
                            &\quad -2 z_4 z_5 \left(z_2 z_3 \left(2 \theta _1-2 \theta _2-1\right) \left(\theta _1+\theta _2-2 \theta _3-1\right)+z_4 z_5\left(\theta _3-\theta _4\right) \left(2 \theta _5+1\right)\right) \\
                            &\quad +2 z_1 z_2 z_3 z_4 z_5 \left(2 \theta _1+1\right) \left(\theta _1-\theta _3-\theta _4-1\right)+z_1 z_2 z_3 z_4^2 z_5 \left(2 \theta _1+1\right) \left(2 \theta _4-2 \theta _5+1\right)\\ 
                            &\quad+z_1 z_2 z_3 z_4^2 z_5^2 \left(2 \theta _1+1\right) \left(2 \theta _5+1\right) \, , \\
    \mathcal D_{G_{1,3},4}  &= 4 \left(z_5 \theta _1 \left(\theta _1+2 \theta _3-2 \theta _4\right)+z_1 \theta _5 \left(-2 \theta _2+2 \theta _3+\theta
   _5\right)\right) \\
                            &\quad +2 z_1 z_5 \left(-2 \theta _1^2+\theta _2 \left(2 \theta _3-2 \theta _4+1\right)+\theta _1 \left(2 \theta _2-2 \theta _3+2
   \theta _5-1\right) \right. \\
                            &\qquad \left.+\left(\theta _4-\theta _5\right) \left(2 \theta _3+2 \theta _5+1\right)\right)\\
                            & \quad  +2 z_1 z_5 \left(-z_2 \left(\theta _2-\theta _3\right) \left(2 \theta _1+2 \theta _3-2 \theta _4+1\right)+z_4 \left(\theta
   _3-\theta _4\right) \left(-2 \theta _2+2 \theta _3+2 \theta _5+1\right)\right) \\
                            &\quad -z_1 z_3 z_5 \left(z_2 \left(2 \theta _3-2 \theta _4+1\right) \left(4 \theta _1+2 \theta _3-2 \theta _4+3\right) \right. \\
                            &\qquad \left.+z_4 \left(2
   \theta _2-2 \theta _3-1\right) \left(2 \theta _2-2 \theta _3-4 \theta _5-3\right)\right) \\
                            &\quad  -z_1 z_2 z_3 z_4 z_5 \left(3-4 \theta _2^2+4 \theta _3+2 \theta _4+\theta _2 \left(4 \theta _3+4 \theta _4+2\right)+\theta _1
   \left(-4 \theta _2-4 \theta _3-4 \theta _5 2\right) \right. \\
                            &\qquad \left.+4 \left(\theta _3-\theta _4\right) \left(\theta _4-\theta _5\right)-2
   \theta _5\right) \\
                            &\quad   +z_1 z_2 z_3 z_4 z_5 \left(z_1 \left(2 \theta _1+1\right) \left(2 \theta _2-2 \theta _3-2 \theta _5-1\right) \right. \\
                            &\qquad \left.-z_5 \left(2
   \theta _1+2 \theta _3-2 \theta _4+1\right) \left(2 \theta _5+1\right)\right) \, , \\ 
    \mathcal D_{G_{1,3},5}  &= 4 z_4 z_5 \theta _1^2+8 z_1 z_2 \left(\theta _2-\theta _3\right) \theta _5 \\
                            &\quad -2 z_1 \left(z_4 z_5 \left(\theta _1-\theta _2\right) \left(2 \theta _1+2 \theta _2+1\right)+2 z_2 \left(-2 z_5 \left(\theta
   _2-\theta _3\right) \left(\theta _4-\theta _5\right) \right.\right. \\
                            &\qquad \left.\left.+z_3 \theta _5 \left(2 \theta _2-2 \theta _3+\theta
   _5\right)\right)\right) \\
                            &\quad -2 z_1 z_2 z_5 \left(2 z_4 \left(\theta _2-\theta _3\right) \left(\theta _2-\theta _3+2 \theta _4+1\right) \right. \\
                            &\qquad \left.+z_3 \left(\theta
   _4-\theta _5\right) \left(4 \theta _2-4 \theta _3+2 \theta _4+2 \theta _5+1\right)\right)\\
                            & \quad +z_1 z_2 z_3 z_4 z_5 \left(1-8 \theta _1^2+8 \theta _1 \theta _2+8 \theta _4+8 \left(\theta _3^2-\theta _3 \theta _4+\theta
   _4^2 \right.\right. \\
                            &\qquad \left.\left.+\theta _2 \left(-2 \theta _3+\theta _4-1\right)-\theta _4 \theta _5+\theta _5^2\right)\right) \\
                            &\quad  +z_1 z_2 z_3 z_4 z_5 \left(4 z_1 \left(2 \theta _1+1\right) \left(\theta _1-\theta _2\right)-z_3 \left(-2 \theta _2+2 \theta
   _3+1\right){}^2 \right. \\
                            &\qquad \left.-2 z_4 \left(2 \theta _4-2 \theta _5+1\right) \left(\theta _4-\theta _5+1\right)+4 z_5 \left(\theta
   _4-\theta _5\right) \left(2 \theta _5+1\right)\right)  \\
                            &\quad  +z_1 z_2 z_3 z_4 z_5 \left(2 z_2 z_3 \left(2 \theta _1-2 \theta _2-1\right) \left(\theta _1-\theta _2-1\right)-z_4 z_5 \left(4
   \theta _4-2 \theta _5+3\right) \left(2 \theta _5+1\right)\right) \\
                            &\quad -z_1^2 z_2^2 z_3^2 z_4 z_5 \left(2 \theta _1+1\right) \left(2 \theta _1-4 \theta _2-3\right) \, . 
\end{aligned}
\end{equation}
The solutions to these operators can be computed using the Frobenius method
\begin{equation}
\vspace{-2ex}
\begin{aligned}
\label{frobl3}
    \phi_{0}(\uz)   &= \varpi(\uz;0) \,, \\
                            &= 1+\left(4 z_1 z_2 z_3+4 z_2 z_3 z_4+4 z_3 z_4 z_5\right)+\left(8 z_1 z_2 z_3 z_4+8 z_2 z_3 z_4 z_5\right)+ \mathcal O(z_i^5)\\
    \phi_{1,i}(\uz) &= \partial_{\rho_i}\varpi(\uz;\underline\rho)|_{\underline\rho=0} \, , \quad \text{for } i=1,\hdots,5 \, , \\
    \phi_{1,1}(\uz) &=   \phi_{0}(\uz)\log \left(z_1\right)+2 z_1+\left(3 z_1^2-2 z_2 z_3\right)    + \mathcal O(z_i^3) \, , \\
    \phi_{1,2}(\uz) &=   \phi_{0}(\uz)\log \left(z_2\right)-\left(2 z_1-2 z_2\right)-\left(3 z_1^2-4 z_1 z_2-3 z_2^2+2 z_3 z_4\right)    + \mathcal O(z_i^3) \, , \\
    \phi_{1,3}(\uz) &=   \phi_{0}(\uz)\log \left(z_3\right)-\left(2 z_2+2 z_4\right) \\
                    &\quad -\left(4 z_1 z_2+3 z_2^2-2 z_2 z_3-2 z_3 z_4+3 z_4^2+4 z_4 z_5\right)    + \mathcal O(z_i^3) \, , \\
    \phi_{1,4}(\uz) &=   \phi_{0}(\uz)\log \left(z_4\right)+\left(2 z_4-2 z_5\right)-\left(2 z_2 z_3-3 z_4^2-4 z_4 z_5+3 z_5^2\right)    + \mathcal O(z_i^3) \, ,  \\
    \phi_{1,5}(\uz) &=   \phi_{0}(\uz)\log \left(z_5\right)+2 z_5-\left(2 z_3 z_4-3 z_5^2\right) + \mathcal O(z_i^3) \, ,            \\
    \phi_{2,1}(\uz)   &= \left(2 \partial_{\rho_2} \left(\partial_{\rho_3}+\partial_{\rho_4}+\partial_{\rho_5}\right)+\left(\partial_{\rho_3}+\partial_{\rho_4}\right)  \left(\partial_{\rho_3}+\partial_{\rho_4}+2 \partial_{\rho_5}\right)\right)\varpi(\uz;\underline\rho)|_{\underline\rho=0} \, , \\
                    &= 2 \log \left(z_2\right) \left(\log \left(z_3\right)+\log \left(z_4\right)+\log \left(z_5\right)\right)+\left(\log \left(z_3\right)+\log \left(z_4\right)\right) \\
                    &\quad \times \left(\log\left(z_3\right)+\log \left(z_4\right)+2 \log \left(z_5\right)\right)+ \mathcal O(z_i^1)-\pi^2\phi_{0}(\uz) \, , \\
    \phi_{2,2}(\uz)   &= \left(\left(\partial_{\rho_2}+\partial_{\rho_3}\right) \left(\partial_{\rho_2}+\partial_{\rho_3}+2 \partial_{\rho_4}+\partial_{\rho_5}\right)+\partial_{\rho_1} \left(2 \partial_{\rho_2}+2 \partial_{\rho_3}+2 \partial_{\rho_4}+\partial_{\rho_5}\right) \right)\varpi(\uz;\underline\rho)|_{\underline\rho=0} \, , \\
                    &= \left(\log \left(z_2\right)+\log \left(z_3\right)\right) \left(\log \left(z_2\right)+\log \left(z_3\right)+2 \log \left(z_4\right)+\log \left(z_5\right)\right) \\
                    &\quad +\log
   \left(z_1\right) \left(2 \log \left(z_2\right)+2 \log \left(z_3\right)+2 \log \left(z_4\right)+\log \left(z_5\right)\right)+ \mathcal O(z_i^1)-\pi^2\phi_{0}(\uz) \, , \\
    \phi_{2,3}(\uz)   &= \left(   2 \partial_{\rho_1} \left(\partial_{\rho_2}+\partial_{\rho_3}+\partial_{\rho_4}\right)+\left(\partial_{\rho_2}+\partial_{\rho_3}\right) \left(\partial_{\rho_2}+\partial_{\rho_3}+2 \partial_{\rho_4}\right)   \right)\varpi(\uz;\underline\rho)|_{\underline\rho=0} \, , \\
                    &= 2 \log \left(z_1\right) \left(\log \left(z_2\right)+\log \left(z_3\right)+\log \left(z_4\right)\right) \\
                    &\quad +\left(\log \left(z_2\right)+\log \left(z_3\right)\right) \left(\log
   \left(z_2\right)+\log \left(z_3\right)+2 \log \left(z_4\right)\right)+ \mathcal O(z_i^1)-\pi^2\phi_{0}(\uz) \, , \\
    \phi_{2,4}(\uz)   &= \left(  \partial_{\rho_2}^2+2 \partial_{\rho_1} \left(\partial_{\rho_2}+\partial_{\rho_3}+\partial_{\rho_4}+\partial_{\rho_5}\right)    \right)\varpi(\uz;\underline\rho)|_{\underline\rho=0} \, , \\
                    &= \log ^2\left(z_2\right)+2 \log \left(z_1\right) \left(\log \left(z_2\right)+\log \left(z_3\right)+\log \left(z_4\right)+\log \left(z_5\right)\right)+ \mathcal O(z_i^1) +\frac13\pi^2\phi_{0}(\uz)\, , \\
     \phi_{2,5}(\uz)   &= \left(  \partial_{\rho_1} \left(\partial_{\rho_3}+\partial_{\rho_4}+\partial_{\rho_5}\right)    \right)\varpi(\uz;\underline\rho)|_{\underline\rho=0} \, , \\
                    &= \log \left(z_1\right) \left(\log \left(z_3\right)+\log \left(z_4\right)+\log \left(z_5\right)\right)+ \mathcal O(z_i^1) \, , \\
    \phi_{3}(\uz)   &= \left(   3 \partial_{\rho_2}^2 \left(\partial_{\rho_3}+\partial_{\rho_4}+\partial_{\rho_5}\right)+3 \partial_{\rho_2} \left(\partial_{\rho_3}+\partial_{\rho_4}\right) \left(\partial_{\rho_3}+\partial_{\rho_4}+2
   \partial_{\rho_5}\right) \right. \\
                    &\quad \left. +\partial_{\rho_3} \left(\partial_{\rho_3}^2+3 \partial_{\rho_3} \left(\partial_{\rho_4}+\partial_{\rho_5}\right)+3 \partial_{\rho_4} \left(\partial_{\rho_4}+2
   \partial_{\rho_5}\right)\right)+3 \partial_{\rho_1} \left(2 \partial_{\rho_2} \left(\partial_{\rho_3}+\partial_{\rho_4}+\partial_{\rho_5}\right) \right. \right. \\
                    &\quad \left. \left. +\left(\partial_{\rho_3}+\partial_{\rho_4}\right)
   \left(\partial_{\rho_3}+\partial_{\rho_4}+2 \partial_{\rho_5}\right)\right)   \right)\varpi(\uz;\underline\rho)|_{\underline\rho=0} \\
                    &= 3 \log ^2\left(z_2\right) \left(\log \left(z_3\right)+\log \left(z_4\right)+\log \left(z_5\right)\right)+3 \log \left(z_2\right) \left(\log \left(z_3\right)+\log
   \left(z_4\right)\right) \\
                    &\quad \times\left(\log \left(z_3\right)+\log \left(z_4\right)+2 \log \left(z_5\right)\right)+\log \left(z_3\right) \left(\log ^2\left(z_3\right)+3 \log
   \left(z_3\right) \right. \\
                    &\quad \times \left. \left(\log \left(z_4\right)+\log \left(z_5\right)\right)+3 \log \left(z_4\right) \left(\log \left(z_4\right)+2 \log \left(z_5\right)\right)\right)+3 \log
   \left(z_1\right) \\
                    &\quad \times\left(2 \log \left(z_2\right) \left(\log \left(z_3\right)+\log \left(z_4\right)+\log \left(z_5\right)\right)+\left(\log \left(z_3\right)+\log
   \left(z_4\right)\right)\right. \\
                    &\quad \times\left.\left(\log \left(z_3\right)+\log \left(z_4\right)+2 \log \left(z_5\right)\right)\right) + \mathcal O(z_i^1) \\
                    &\quad -3\pi^2(\phi_{1,1}(\uz)+\phi_{1,2}(\uz)+\phi_{1,4}(\uz)+\phi_{1,5}(\uz))-2\pi^2\phi_{1,3}(\uz)
\end{aligned}
\end{equation}
with coefficients
\begin{equation}
\begin{aligned}
     \varpi(\uz;\underline\rho) &= \sum_{\underline n=0}^\infty c(\underline n+\underline \rho) \uz^{\underline n+\underline \rho} \, , \\ 
     c(\underline n) &= (n_1)(n_5)(n_2-n_1)(n_3-n_2)(n_3-n_4)(n_4-n_5)(n_1-n_3+n_4)(n_2-n_3+n_5) \, .
\end{aligned}
\end{equation}
Here, we have computed the basis using the $\rho$-deformed series $\varpi(\uz;\underline\rho)$ to obtain at least a rational monodromy basis after normalizing the logarithms. Moreover, we have modified the basis without loosing rational monodromies to simplify the intersection form. This rotation $T_{G_{1,3}}$ is given by
\beq
    T_{G_{1,3}} = \left(
\begin{array}{cccccccccccc}
 1 & 0 & 0 & 0 & 0 & 0 & 0 & 0 & 0 & 0 & 0 & 0 \\
 0 & -\frac{3 i}{\pi } & 0 & 0 & 0 & 0 & 0 & 0 & 0 & 0 & 0 & 0 \\
 0 & -\frac{3 i}{2 \pi } & -\frac{3 i}{2 \pi } & 0 & 0 & 0 & 0 & 0 & 0 & 0 & 0 & 0 \\
 0 & 0 & 0 & 0 & -\frac{3 i}{2 \pi } & \frac{3 i}{2 \pi } & 0 & 0 & 0 & 0 & 0 & 0 \\
 0 & 0 & 0 & 0 & \frac{3 i}{\pi } & 0 & 0 & 0 & 0 & 0 & 0 & 0 \\
 0 & \frac{3 i}{2 \pi } & \frac{3 i}{2 \pi } & \frac{3 i}{2 \pi } & 0 & 0 & 0 & 0 & 0 & 0 & 0 & 0 \\
 -\frac{1}{3} & 0 & 0 & 0 & 0 & 0 & -\frac{1}{4 \pi ^2} & 0 & 0 & -\frac{1}{4 \pi ^2} & 0 & 0 \\
 -\frac{1}{2} & 0 & 0 & 0 & 0 & 0 & 0 & -\frac{1}{4 \pi ^2} & 0 & 0 & 0 & 0 \\
 -\frac{1}{2} & 0 & 0 & 0 & 0 & 0 & 0 & 0 & -\frac{1}{4 \pi ^2} & 0 & 0 & 0 \\
 \frac{1}{6} & 0 & 0 & 0 & 0 & 0 & 0 & 0 & 0 & -\frac{1}{4 \pi ^2} & \frac{1}{2 \pi ^2} & 0 \\
 0 & 0 & 0 & 0 & 0 & 0 & 0 & 0 & 0 & 0 & -\frac{1}{4 \pi ^2} & 0 \\
 0 & 0 & 0 & 0 & 0 & 0 & 0 & 0 & 0 & 0 & 0 & \frac{i}{8 \pi ^3} \\
\end{array}
\right) \, .
\eeq

\subsection{The four-loop window graph}
\label{app:window}

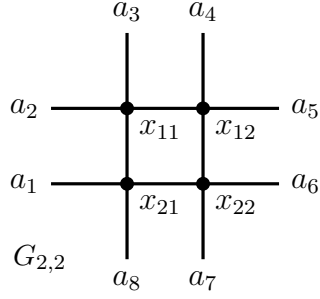
\begin{figure}[t]
\centering
 % \resizebox{3.5cm}{!}{
\begin{tikzpicture}[very thick]
  \node (G1) at (0,-2) [font=\small, text width=1 cm]{$G_{2,2}$};
  \draw (0,0) -- (3,0);
  \draw (0,-1) -- (3,-1);
  \draw (1,1) -- (1,-2);
  \node [{below right}] at (1,0) {$x_{11}$};
  \draw[fill=black!100] (1,0)  circle (2pt);
  \draw (2,1) -- (2,-2);
  \node [{below right}] at (2,0) {$x_{12}$};
  \draw[fill=black!100] (2,0)  circle (2pt);
  \node [{below right}] at (1,-1) {$x_{21}$};
  \draw[fill=black!100] (1,-1)  circle (2pt);
  \node [{below right}] at (2,-1) {$x_{22}$};
  \draw[fill=black!100] (2,-1)  circle (2pt);
  \node [{left}] at (0,-1) {$a_1$};
  \node [{left}] at (0,0) {$a_2$};
  \node [{above}] at (1,1) {$a_3$};
  \node [{above}] at (2,1) {$a_4$};
  \node [{right}] at (3,0) {$a_5$};
  \node [{right}] at (3,-1) {$a_6$};
  \node [{below}] at (2,-2) {$a_7$};
  \node [{below}] at (1,-2) {$a_8$};
\end{tikzpicture}
% }
\caption{Four-loop window graph.}
\label{fig:window}
\end{figure}

In this appendix, we consider the four-loop window graph shown in \Figref{fig:window} which is the first fishnet graph that is not a train track graph. The automorphism and permutation symmetries are
\beq
\textrm{Perm}_{G_{2,2}} = \Aut(G_{2,2}) = D_4\times \mathbb{Z}_2^4\,.
\eeq
The necessary period integrals again depend on five cross ratios
\beq
\label{varwindow}
    z_1 = \frac14\chi_{2,3,4,6}\,, \quad z_2 = \frac1{4^2}\chi_{2,4,5,6}\,, \quad z_3 = \frac1{4^2}\chi_{2,5,1,6}\,, \quad z_4 = \frac1{4^2}\chi_{2,1,8,6}\,, \quad z_5 = \frac14\chi_{2,8,7,6}\, .
\eeq
At four loops, the Yangian (including permutation and two-site symmetries) furnishes the full differential operator ideal. In the variables from eq.~\eqref{varwindow} a generating set of the Yangian differential ideal $\YDI(G_{2,2})$ can be chosen to be 
\begin{equation}
\begin{aligned}
\label{windowops}
    \mathcal D_{G_{2,2},1}  &=  \left(\theta _1-\theta _2\right) \left(\theta _2-\theta _3\right)-4 z_2 \left(-1+2 \theta _1-2 \theta _2\right) \left(1+2 \theta _2-2 \theta _3\right)        \, , \\
    \mathcal D_{G_{2,2},2}  &=   \left(\theta _3-\theta _4\right) \left(\theta _4-\theta _5\right)-4 z_4 \left(-1+2 \theta _3-2 \theta _4\right) \left(1+2 \theta _4-2 \theta _5\right)       \, , \\
    \mathcal D_{G_{2,2},3}  &=   \theta_1^2-2 z_1 \left(1+2 \theta _1\right) \left(\theta _1-\theta _2\right) -16 z_1 z_2 \left(1+2 \theta _1\right) \left(1+2 \theta _2-2 \theta _3\right) \\
                            &\quad -32 z_1 z_2 z_3 \left(1+2 \theta _1\right) \left(\theta _3-\theta _4\right)  -256 z_1 z_2 z_3 z_4 \left(1+2 \theta _1\right) \left(1+2 \theta _4-2 \theta _5\right)\\
                            &\quad -1024 z_1 z_2 z_3 z_4 z_5 \left(1+2 \theta _1\right) \left(1+2 \theta _5\right)   \, , \\
    \mathcal D_{G_{2,2},4}  &=  \theta _5^2 + 2 z_5 \left(\theta _4-\theta _5\right) \left(1+2 \theta _5\right)+16 z_4 z_5 \left(-1+2 \theta _3-2 \theta _4\right) \left(1+2 \theta
   _5\right) \\
                            &\quad +32 z_3 z_4 z_5 \left(\theta _2-\theta _3\right) \left(1+2 \theta _5\right)+256 z_2 z_3 z_4 z_5 \left(-1+2 \theta _1-2 \theta _2\right)
   \left(1+2 \theta _5\right) \\
                            &\quad -1024 z_1 z_2 z_3 z_4 z_5 \left(1+2 \theta _1\right) \left(1+2 \theta _5\right)\, , \\ 
    \mathcal D_{G_{2,2},5}  &=  \theta _5^2+2 z_5 \left[\theta _4-\theta _5+2 \left(\theta _1^2-\theta _1 \theta _2+\theta _2^2-\theta _2 \theta _3+\theta _3 \theta
   _4+\theta _4 \theta _5-2 \theta _5^2\right)\right] \\
                            &\quad -8 z_5 \left[z_1 \left(1+2 \theta _1\right) \left(\theta _1-\theta _2\right)-2 z_2 \left(-1+2 \theta _1-2 \theta _2\right) \left(1+2
   \theta _2-2 \theta _3\right)\right. \\
                            &\quad \left.+z_5 \left(\theta _4-\theta _5\right) \left(1+2 \theta _5\right)-2 z_4 \left(-1+2 \theta _3-2 \theta
   _4\right) \left(1+2 \theta _3+2 \theta _5\right)\right] \\
                            &\quad -32 z_5 \left\{2 z_1 z_2 \left(1+2 \theta _1\right) \left(1+2 \theta _2-2 \theta _3\right)+z_4 \left[z_3 \left(-1+2 \theta _2-2 \theta
   _3\right) \left(\theta _2-\theta _3\right) \right. \right. \\
                            &\quad \left. \left.+2 z_5 \left(-1+2 \theta _3-2 \theta _4\right) \left(1+2 \theta _5\right)\right]\right\}  -512 z_2 z_3 z_4 z_5 \left(-1+2 \theta _1-2 \theta _2\right) \left(\theta _1+\theta _2-2 \theta _3\right) \\
                            &\quad +1024 z_1 z_2 z_3 z_4 z_5
   \left(1+2 \theta _1\right) \left(-1+2 \theta _1-4 \theta _3\right)\, .                      
\end{aligned}
\end{equation}
Unlike for train track graphs, $\uz=0$ is not a MUM-point. Nevertheless, we can compute the solution space of the Yangian which is twenty-dimensional\footnote{For simplicity, we have removed the label $G_{2,2}$ and the dependence on the  variables $\uz$ to lighten the notation.}
\begin{equation}
\label{solg22}
\begin{aligned}
    &\Sol(\PFI(M_{G_{2,2}}))  =    \Sol(\left\{ \mathcal D_{G_{2,2},k} \right\}_{k=1,\hdots,5}) =\Sol(YDI(G_{2,2})) = \\
     &\quad= \big\langle \Phi_0,\Phi_{1,1},\hdots, \Phi_{1,4}, \Phi_{2,1},\hdots, \Phi_{2,6},\Phi_{3,1},\hdots, \Phi_{3,4},\Phi_4,\, \varphi_0,\varphi_{1,1},\varphi_{1,2},\varphi_2       \big\rangle_{\mathbb{C}}\, .
\end{aligned}
\end{equation}
The explicit solutions for the four-loop window graph are given by
\begin{equation}
\begin{aligned}
\label{frobwindow}
    \Phi_{0}(\uz)   &= 1+4z_2+4z_4 +\mathcal O(z_i^2) \, , \\
    \Phi_{1,1}(\uz) &= \Phi_{0}(\uz) \log(z_1) + 2 z_1-4 z_2 +\mathcal O(z_i^2) \, , \\
    \Phi_{1,2}(\uz) &= \Phi_{0}(\uz) \log(z_2) -2 z_1+8 z_2  +\mathcal O(z_i^2) \, , \\
    \Phi_{1,3}(\uz) &= \Phi_{0}(\uz) \log(z_4) +8 z_4-2 z_5  +\mathcal O(z_i^2) \, , \\
    \Phi_{1,4}(\uz) &= \Phi_{0}(\uz) \log(z_5) -4 z_4+2 z_5  +\mathcal O(z_i^2) \, , \\
    \Phi_{2,1}(\uz) &= \log(z_1)\log(z_4)  +\mathcal O(z_i^1) \, , \\
    \Phi_{2,2}(\uz) &= \log(z_1)\log(z_5)  +\mathcal O(z_i^1) \, , \\
    \Phi_{2,3}(\uz) &= \log(z_2)\log(z_4)  +\mathcal O(z_i^1) \, , \\
    \Phi_{2,4}(\uz) &= \log(z_2)\log(z_5)  +\mathcal O(z_i^1) \, , \\
    \Phi_{2,5}(\uz) &= \frac12\log(z_2)^2+\log(z_1)\log(z_2)  +\mathcal O(z_i^1) \, , \\
    \Phi_{2,6}(\uz) &= \frac12\log(z_4)^2+\log(z_4)\log(z_5)  +\mathcal O(z_i^1) \, , \\
    \Phi_{3,1}(\uz) &= \log \left(z_2\right) \left(2 \log \left(z_1\right)+\log \left(z_2\right)\right) \left(4 \log \left(z_4\right)-\log
   \left(z_5\right)\right) + \mathcal O(z_i^1) \, , \\
    \Phi_{3,2}(\uz) &= \log \left(z_2\right) \left(2 \log \left(z_1\right)+\log \left(z_2\right)-\log \left(z_4\right)\right)
   \left(\log \left(z_4\right)+2 \log \left(z_5\right)\right) + \mathcal O(z_i^1) \, , \\
    \Phi_{3,3}(\uz) &= \log \left(z_4\right) \left(\log \left(z_1\right)-4 \log
   \left(z_2\right)\right) \left(\log \left(z_4\right)+2 \log \left(z_5\right)\right) + \mathcal O(z_i^1) \, , \\
    \Phi_{3,4}(\uz) &= \log \left(z_4\right) \left(2 \log
   \left(z_1\right)+\log \left(z_2\right)\right) \left(\log \left(z_2\right)-\log \left(z_4\right)-2 \log \left(z_5\right)\right) + \mathcal O(z_i^1) \, , \\
    \Phi_4(\uz)     &= \log(z_2)\log(z_4) \left(2 \log \left(z_1\right)+\log \left(z_2\right)\right)  \left(\log \left(z_4\right)+2 \log
   \left(z_5\right)\right)     +\mathcal O(z_i^1) \, , \\
   \varphi_0(\uz)         &= \sqrt{z_3} \left(  1+\frac{z_3}{12}   +\mathcal O(z_i^2)   \right)     \, , \\
   \varphi_{1,1}(\uz)     &= \varphi_0(\uz)\log(z_3) + \sqrt{z_3} \left(  -16 z_2+\frac{5 z_3}{18}-16 z_4   +\mathcal O(z_i^2)   \right)     \, , \\
   \varphi_{1,2}(\uz)     &= \varphi_0(\uz)\log(z_1)\log(z_2)\log(z_4)\log(z_5) + \sqrt{z_3} \left(  16 z_2-\frac{z_3}{2}+16 z_4   +\mathcal O(z_i^2)   \right)     \, , \\
   \varphi_{2}(\uz)       &= \sqrt{z_3}\log \left(z_3\right) \left(\log \left(z_1\right)+\log \left(z_2\right)+\log \left(z_3\right)+\log \left(z_4\right)+\log
   \left(z_5\right)\right)      +\sqrt{z_3}\mathcal O(z_i^1)        \, . \\
\end{aligned}
\end{equation}
 The solution space in eq.~\eqref{solg22} is the space of CY four-fold periods. The number of power series solutions ($\Phi_0$ and $\varphi_0$) also indicates that we are not at a MUM-point of this four-dimensional CY variety, where we would expect only a single solution of this type and exactly five single logarithmic ones. We still expect that the corresponding CY four-fold $M_{G_{2,2}}$ has a MUM-point, although we have not found it. A full analysis of this situation would require a full understanding of the moduli space, which is far beyond the scope of our paper. However, we verified that the PFI is generated by the Yangian and the automorphisms of the graph $G_{2,2}$.

\subsection{Triangle track graphs up to six loops}

%%%%%%%%%%%%%%%%%%%%%%%%%%%%%%%%%%%%%%%%

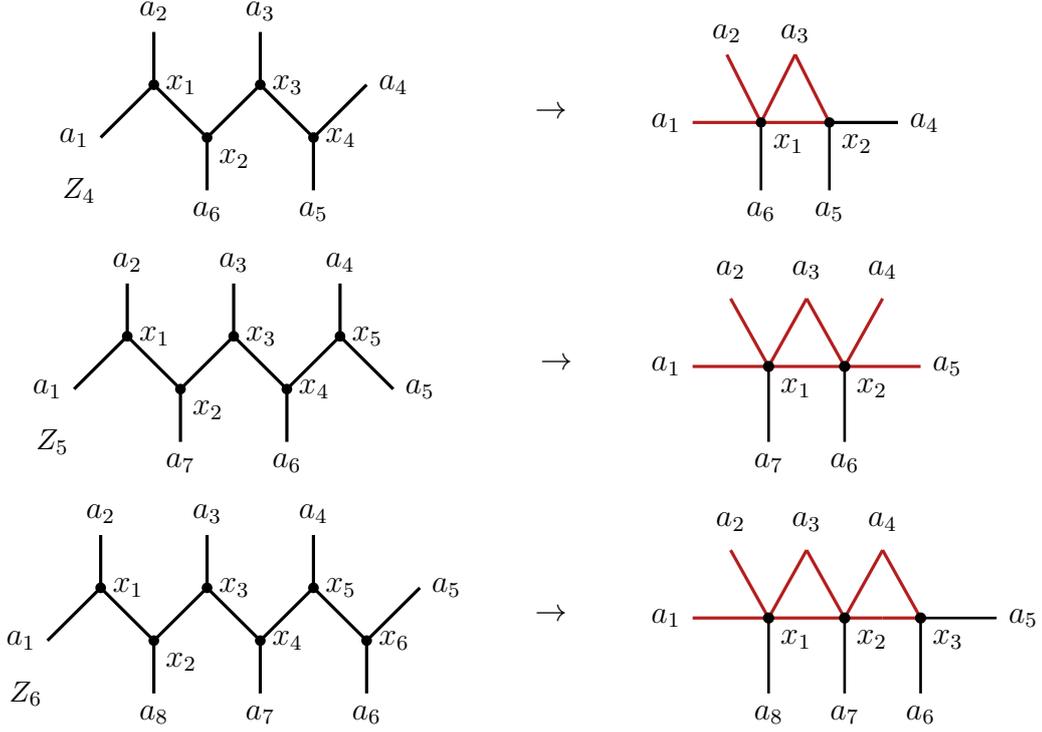
\begin{figure}[t]
% \centering
\begin{center}
\begin{tabular}{c c l}
 % \resizebox{7cm}{!}{
\begin{tikzpicture}[very thick, scale=.7]
  \node (G1) at (0,-1) [font=\small, text width=1 cm]{$Z_4$};
  \draw (0,0) -- (1,1);
  \draw (1,1) -- (2,0);
  \draw (2,0) -- (3,1);
  \draw (3,1) -- (4,0);
  \draw (4,0) -- (5,1);
  \draw (1,1) -- (1,2);
  \draw (3,1) -- (3,2);
  \draw (2,0) -- (2,-1);
  \draw (4,0) -- (4,-1);

  \draw[fill=black!100] (1,1)  circle (2pt);
  \node [{right}] at (1,1) {$x_1$};
  \draw[fill=black!100] (2,0)  circle (2pt);
  \node [{below right}] at (2,0) {$x_2$};
  \draw[fill=black!100] (3,1)  circle (2pt);
  \node [{right}] at (3,1) {$x_3$};
  \draw[fill=black!100] (4,0)  circle (2pt);
  \node [{right}] at (4,0) {$x_4$};

  \node [{left}] at (0,0) {$a_1$};
  \node [{above}] at (1,2) {$a_2$};
  \node [{above}] at (3,2) {$a_3$};
  \node [{right}] at (5,1) {$a_4$};
  \node [{below}] at (4,-1) {$a_5$};
  \node [{below}] at (2,-1) {$a_6$};
\end{tikzpicture}
&
\raisebox{1.5cm}{\quad $\to$ \quad}
 &
  % \draw[-stealth,line width=0.75pt] (7,1) -- (8,1);
  %
\begin{tikzpicture}[very thick, scale=.9]
  \draw [darkred] (10,1) -- (12,1);
  \draw (12,1) -- (13,1);
  \draw [darkred] (11,1) -- (10.5,2);
  \draw [darkred] (12,1) -- (11.5,2);
  \draw [darkred] (11,1) -- (11.5,2);
  \draw [line width=1pt] (11,1) -- (11,0);
  \draw [line width=1pt] (12,1) -- (12,0);
  \draw [line width=1pt] (12,1) -- (13,1);
  % \draw [darkred, dashed] (10,1) -- (11,2);

  \draw[fill=black!100] (11,1)  circle (1.5pt);
  \draw[fill=black!100] (12,1)  circle (1.5pt);
  \node [{below right}] at (11,1) {$x_1$};
  \node [{below right}] at (12,1) {$x_2$};

  \node [{left}] at (10,1) {$a_1$};
  \node [{above}] at (10.5,2) {$a_2$};
  \node [{above}] at (11.5,2) {$a_3$};
  \node [{right}] at (13,1) {$a_4$};
  \node [{below}] at (12,0) {$a_5$};
  \node [{below}] at (11,0) {$a_6$};
\end{tikzpicture}
% }
\\
% \resizebox{0.75cm}{!}{\textcolor{white}{$\phantom{hi}$}}

% \resizebox{7cm}{!}{
\begin{tikzpicture}[very thick, scale=.7]
  \node (G1) at (0,-1) [font=\small, text width=1 cm]{$Z_5$};
  \draw (0,0) -- (1,1);
  \draw (1,1) -- (2,0);
  \draw (2,0) -- (3,1);
  \draw (3,1) -- (4,0);
  \draw (4,0) -- (5,1);
  \draw (5,1) -- (6,0);
  \draw (1,1) -- (1,2);
  \draw (3,1) -- (3,2);
  \draw (5,1) -- (5,2);
  \draw (2,0) -- (2,-1);
  \draw (4,0) -- (4,-1);

  \draw[fill=black!100] (1,1)  circle (2pt);
  \node [{right}] at (1,1) {$x_1$};
  \draw[fill=black!100] (2,0)  circle (2pt);
  \node [{below right}] at (2,0) {$x_2$};
  \draw[fill=black!100] (3,1)  circle (2pt);
  \node [{right}] at (3,1) {$x_3$};
  \draw[fill=black!100] (4,0)  circle (2pt);
  \node [{right}] at (4,0) {$x_4$};
  \draw[fill=black!100] (5,1)  circle (2pt);
  \node [{right}] at (5,1) {$x_5$};

  \node [{left}] at (0,0) {$a_1$};
  \node [{above}] at (1,2) {$a_2$};
  \node [{above}] at (3,2) {$a_3$};
  \node [{above}] at (5,2) {$a_4$};
  \node [{right}] at (6,0) {$a_5$};
  \node [{below}] at (4,-1) {$a_6$};
  \node [{below}] at (2,-1) {$a_7$};

\end{tikzpicture}
&
\raisebox{1.5cm}{\quad$\to$\quad}
  % \draw[-stealth,line width=0.75pt] (7,1) -- (8,1);
  &
\begin{tikzpicture}[very thick, scale=1]
  \draw [darkred] (11,1) -- (14,1);
  \draw [darkred] (11.5,1.9) -- (12,1);
  \draw [darkred] (12.5,1.9) -- (12,1);
  \draw [darkred] (12.5,1.9) -- (13,1);
  \draw [darkred] (13.5,1.9) -- (13,1);
  \draw [line width=1pt] (12,1) -- (12,0);
  \draw [line width=1pt] (13,1) -- (13,0);
  % \draw [darkred, dashed] (11.5,1.9) -- (11,1);
  % \draw [darkred, dashed] (13.5,1.9) -- (14,1);

  \draw[fill=black!100] (12,1)  circle (1.5pt);
  \draw[fill=black!100] (13,1)  circle (1.5pt);
  \node [{below right}] at (12,1) {$x_1$};
  \node [{below right}] at (13,1) {$x_2$};

  \node [{left}] at (11,1) {$a_1$};
  \node [{above}] at (11.5,2) {$a_2$};
  \node [{above}] at (12.5,2) {$a_3$};
  \node [{above}] at (13.5,2) {$a_4$};
  \node [{right}] at (14,1) {$a_5$};
  \node [{below}] at (13,0) {$a_6$};
  \node [{below}] at (12,0) {$a_7$};
\end{tikzpicture}
\\
% }

% \resizebox{0.75cm}{!}{\textcolor{white}{$\phantom{hi}$}}

 % \resizebox{7cm}{!}{
\begin{tikzpicture}[very thick, scale=.7]
  \node (G1) at (0,-1) [font=\small, text width=1 cm]{$Z_6$};
  \draw (0,0) -- (1,1);
  \draw (1,1) -- (2,0);
  \draw (2,0) -- (3,1);
  \draw (3,1) -- (4,0);
  \draw (4,0) -- (5,1);
  \draw (5,1) -- (6,0);
  \draw (6,0) -- (7,1);
  \draw (1,1) -- (1,2);
  \draw (3,1) -- (3,2);
  \draw (5,1) -- (5,2);
  \draw (2,0) -- (2,-1);
  \draw (4,0) -- (4,-1);
  \draw (6,0) -- (6,-1);

  \draw[fill=black!100] (1,1)  circle (2pt);
  \node [{right}] at (1,1) {$x_1$};
  \draw[fill=black!100] (2,0)  circle (2pt);
  \node [{below right}] at (2,0) {$x_2$};
  \draw[fill=black!100] (3,1)  circle (2pt);
  \node [{right}] at (3,1) {$x_3$};
  \draw[fill=black!100] (4,0)  circle (2pt);
  \node [{right}] at (4,0) {$x_4$};
  \draw[fill=black!100] (5,1)  circle (2pt);
  \node [{right}] at (5,1) {$x_5$};
  \draw[fill=black!100] (6,0)  circle (2pt);
  \node [{right}] at (6,0) {$x_6$};

  \node [{left}] at (0,0) {$a_1$};
  \node [{above}] at (1,2) {$a_2$};
  \node [{above}] at (3,2) {$a_3$};
  \node [{above}] at (5,2) {$a_4$};
  \node [{right}] at (7,1) {$a_5$};
  \node [{below}] at (6,-1) {$a_6$};
  \node [{below}] at (4,-1) {$a_7$};
  \node [{below}] at (2,-1) {$a_8$};
\end{tikzpicture}
  % \draw[-stealth,line width=0.75pt] (9,1) -- (10,1);
  &
\raisebox{1.5cm}{\quad $\to$ \quad}
&
\begin{tikzpicture}[very thick, scale=1]
  \draw [darkred] (12,1) -- (14.5,1);
  \draw [darkred] (14.5,1) -- (15,1);
  \draw [darkred] (12.5,1.9) -- (13,1);
  \draw [darkred] (13.5,1.9) -- (13,1);
  \draw [darkred] (13.5,1.9) -- (14,1);
  \draw [darkred] (14.5,1.9) -- (14,1);
  \draw [darkred] (14.5,1.9) -- (15,1);
  \draw [line width=1pt] (15,1) -- (16,1);
  \draw [line width=1pt] (13,1) -- (13,0);
  \draw [line width=1pt] (14,1) -- (14,0);
  \draw [line width=1pt] (15,1) -- (15,0);
  % \draw [darkred, dashed] (12.5,1.9) -- (12,1);

  \draw[fill=black!100] (13,1)  circle (1.5pt);
  \draw[fill=black!100] (14,1)  circle (1.5pt);
  \draw[fill=black!100] (15,1)  circle (1.5pt);
  \node [{below right}] at (13,1) {$x_1$};
  \node [{below right}] at (14,1) {$x_2$};
  \node [{below right}] at (15,1) {$x_3$};

  \node [{left}] at (12,1) {$a_1$};
  \node [{above}] at (12.5,2) {$a_2$};
  \node [{above}] at (13.5,2) {$a_3$};
  \node [{above}] at (14.5,2) {$a_4$};
  \node [{right}] at (16,1) {$a_5$};
  \node [{below}] at (15,0) {$a_6$};
  \node [{below}] at (14,0) {$a_7$};
  \node [{below}] at (13,0) {$a_8$};
\end{tikzpicture}
\end{tabular}
% }
\end{center}
\caption{The four-, five- and six-loop triangle track graphs together with diagrams that are associated via the star-triangle identity.}
\label{fig:zigzags}
\end{figure}

In this appendix, we give collective results for triangle track graphs up to six loops (see \Figref{fig:zigzags}). The corresponding $\ell$-loop integrals depend on $\ell-1$ cross ratios, which we have chosen as:
\begin{equation}
\begin{aligned}
    &Z_4: \qquad z_1 = \chi_{1,5,6,4}\,,  &z_2 &= \frac1{3^4\sqrt3}\chi_{1,6,3,4}\, ,    
    &z_3 &= \chi_{1,3,2,4}\, , &&\\
    &Z_5: \qquad z_1 = \chi_{1,6,7,5}\,,  &z_2 &= \frac1{3^4\sqrt3}\chi_{1,7,4,5}\, ,    
    &z_3 &= \chi_{1,4,3,5}\, , &z_4 &= \chi_{1,3,2,5}\, , \\
    &Z_6: \qquad z_1 = \chi_{1,6,7,5}\,,  &z_2 &= \chi_{1,7,8,5}\, ,
    &z_3 &= \frac1{3^6}\chi_{1,8,4,5}\, , &z_4 &= \chi_{1,4,3,5}\, ,\quad    z_5 = \chi_{1,3,2,5}\, .
\end{aligned}
\end{equation}

\paragraph{The four-loop triangle track graph.}
At four loops, we find
\begin{equation}
\begin{aligned}
    \mathcal D_{Z_4,1}  &=  \left(\theta _1-\theta _2\right) \theta _3-9 \sqrt{3} z_2 z_3 \left(-2+3 \theta _1-3 \theta _2\right) \left(1+3 \theta _3\right)        \, , \\
    \mathcal D_{Z_4,2}  &=   \left(2+6 \theta _2-3 \theta _3\right) \theta _3+3 z_3 \left(\theta _2-\theta _3\right){}^2+81 \sqrt{3} z_2 z_3 \left(-2+3 \theta _1-3
   \theta _2\right) \left(1+\theta _1+\theta _2\right) \\
                        &\quad -27 \sqrt{3} z_1 z_2 z_3 \left(2+9 \theta _1 \left(1+\theta _1\right)\right)       \, , \\
    \mathcal D_{Z_4,3}  &=   \theta _1 \left(2-3 \theta _1+3 \theta _2\right)+3 z_1 \left(\theta _1-\theta _2\right){}^2+27 \sqrt{3} z_1 z_2 \left(-2+3 \theta _1-3
   \theta _2\right) \left(2+3 \theta _2-3 \theta _3\right) \\
                        &\quad +27 \sqrt{3} z_1 z_2 z_3 \left(-2+3 \theta _1-3 \theta _2\right) \left(1+3
   \theta _3\right)       \,               
\end{aligned}
\end{equation}
with solutions
\beq
     \Sol(\PFI(M_{Z_4}))  =  \Sol(\left\{ \mathcal D_{Z_4,k} \right\}_{k=1,2,3}) = \Sol(\YDI(Z_4)) = \big\langle \Phi_0,\Phi_1,\Phi_2,\,\varphi_1, \varphi_2   \big\rangle_{\mathbb{C}}\, , 
     \eeq
     with
\begin{equation}
\begin{aligned}
% \label{frob3zigzag}
    \Phi_{0}(\uz)   &=  1+24 \sqrt{3} z_2+\left(18 \sqrt{3} z_1 z_2+3375 z_2^2+18 \sqrt{3} z_2 z_3\right)  + \mathcal O(z_i^3)     \, , \\
    \Phi_{1}(\uz)   &= \Phi_{1}(\uz)\log(z_2)  + 36 \sqrt{3}   z_2+ \left(9 \sqrt{3} z_1 z_2+6075 z_2^2+9 \sqrt{3} z_2 z_3\right) + \mathcal O(z_i^3)     \, , \\
    \Phi_{2}(\uz)   &= \frac{1}{2} \log ^2\left(z_2\right)+\left(3 z_1-18 \sqrt{3} z_2+36 \sqrt{3} \log \left(z_2\right) z_2+12 \sqrt{3} \log ^2\left(z_2\right)
   z_2+3 z_3\right) + \mathcal O(z_i^2)      \, , \\
    \varphi_{1}(\uz)   &= z_1^{2/3}\left[1+\frac{4 z_1}{15}+\left(\frac{5 z_1^2}{36}+\frac{162}{5} \sqrt{3} z_1 z_2\right)+\left(\frac{80 z_1^3}{891}+\frac{54}{5} \sqrt{3} z_1^2
   z_2 \right. \right. \\
                    &\quad \left. \left.+\frac{54}{5} \sqrt{3} z_1 z_2 z_3\right) + \mathcal O(z_i^4) \right]     \, , \\
    \varphi_{2}(\uz)   &=  z_3^{2/3}\left[ 1+\frac{4 z_3}{15}+\left(\frac{2 z_2 z_3}{5}+\frac{5 z_3^2}{36}\right)+\left(\frac{2}{15} z_1 z_2 z_3+\frac{2}{15} z_2 z_3^2+\frac{80
   z_3^3}{891}\right)+ \mathcal O(z_i^4) \right]     \, .
\end{aligned}
\end{equation}

\paragraph{The five-loop triangle track graph.}
At five loops, we find
\begin{align}
  \nonumber  \mathcal D_{Z_5,1}  &=   9 \theta _1 \left(\theta _2-\theta _3\right)-3^4\sqrt{3}z_1 z_2 \left(2+3 \theta _1\right) \left(1+3 \theta _2-3 \theta _3\right)       \, , \\
  \nonumber  \mathcal D_{Z_5,2}  &=   3 \theta _1 \left(2-3 \theta _1+3 \theta _2\right)+9 z_1 \left(\theta _1-\theta _2\right){}^2 +3^4\sqrt{3}z_1 z_2 \left(-2+3 \theta _1-3 \theta _2\right) \left(1+3 \theta _2-3 \theta
   _3\right)        \\
    \nonumber                    &\quad +3^4\sqrt{3}z_1 z_2 z_3 \left(-2+3 \theta _1-3 \theta _2\right) \left(2+3 \theta _3-3 \theta _4\right) \\
                     \nonumber   &\quad +3^4\sqrt{3}z_1 z_2 z_3 z_4 \left(-2+3 \theta _1-3 \theta _2\right) \left(1+3 \theta _4\right)  \, , \\
    \mathcal D_{Z_5,3}  &=    3 \left(1+3 \theta _2-3 \theta _3\right) \theta _4+3 z_4 \left(1+3 \theta _2-3 \theta _3\right) \left(\theta _3-\theta _4\right)\\
\nonumber    &\quad+3 z_3 z_4 \left(-1+3 \theta _2-3 \theta
   _3\right) \left(\theta _2-\theta _3\right)       \\
                 \nonumber       &\quad  +3^4\sqrt{3}z_2 z_3 z_4 \left(-2+3 \theta _1-3 \theta _2\right) \left(1+3 \theta _2-3 \theta _3\right) \\
                      \nonumber  &\quad -3^4\sqrt{3}z_1 z_2 z_3 z_4 \left(2+3 \theta _1\right) \left(1+3 \theta _2-3 \theta _3\right)        \, , \\
\nonumber    \mathcal D_{Z_5,4}  &=    3 \left(1+6 \theta _3-3 \theta _4\right) \theta _4+3 z_4 \left(-1+3 \theta _3-3 \theta _4\right) \left(\theta _3-\theta _4\right) \\
                 \nonumber       &\quad -3 z_3 z_4 \left(-1+3 \theta _2-3 \theta
   _3\right) \left(\theta _2-\theta _3\right)   \\
\nonumber                        &\quad   +3^4\sqrt{3}z_2 z_3 z_4 \left(-2+3 \theta _1-3 \theta _2\right) \left(2+3 \theta _1-3 \theta _2+6 \theta _3\right) \\
         \nonumber               &\quad -3^5\sqrt{3} z_1 z_2 z_3 z_4 \left(2+3 \theta _1\right) \left(\theta _1-2 \theta
   _2+2 \theta _3\right)       \,  
\end{align}
with solutions
\beq\bsp
     \Sol(\PFI(M_{Z_5}))  &\,=  \Sol(\left\{ \mathcal D_{Z_5,k} \right\}_{k=1,\hdots,4}) = \Sol(\YDI(Z_4)) =  \\
     &\,=\big\langle \Phi_0,\Phi_1,\Phi_2,\,\varphi_0,\tilde\varphi_0,\varphi_{1,1},\varphi_{1,2},\, \psi_1   \big\rangle_{\mathbb{C}}\, , 
   \esp  \eeq
     with
\begin{equation}
\begin{aligned}
% \label{frob3zigzag}
    \Phi_{0}(\uz)   &=  1+6 \sqrt{3} z_2+\left(18 \sqrt{3} z_1 z_2+540 z_2^2+24 \sqrt{3} z_2 z_3\right)+ \mathcal O(z_i^3)     \, , \\
    \Phi_{1}(\uz)   &= \Phi_{1}(\uz)\log(z_2)+\left(27 \sqrt{3} z_2-\frac{z_3}{2}\right)+\left(36 \sqrt{3} z_1 z_2+2754 z_2^2 \right. \\
                    &\quad \left.+24 \sqrt{3} z_2 z_3-\frac{z_3^2}{5}\right)   + \mathcal O(z_i^3)     \, , \\
    \Phi_{2}(\uz)   &= \frac{1}{2} \Phi_{1}(\uz)\log(z_2)^2+\log \left(z_2\right) \left(27 \sqrt{3} z_2-\frac{z_3}{2}\right)+3 z_1+9 \sqrt{3} z_2+ \mathcal O(z_i^2)      \, , \\
    \varphi_{0}(\uz)    &= z_3^{1/3}\left[1+\frac{z_3}{6}+\left(27 \sqrt{3} z_2 z_3+\frac{5 z_3^2}{63}\right) + \mathcal O(z_i^3)\right]    \, , \\
    \tilde{\varphi}_{0}(\uz)    &=  z_3^{1/3}z_4\left[ 1+\frac{z_4}{3}+\left(\frac{z_3 z_4}{9}+\frac{5 z_4^2}{27}\right)      + \mathcal O(z_i^4) \right] \, , \\
    \varphi_{1,1}(\uz)   &= \varphi_{0}(\uz)\log(z_2)+\frac13\left(\log(z_3)+2\log(z_4)\right) \tilde{\varphi}_{0}(\uz) -\frac{5}{12} z_3^{4/3}   + \mathcal O(z_i^{7/3})     \, , \\
    \varphi_{1,2}(\uz)   &= \varphi_{0}(\uz)\log(z_3)-\frac13\left(\log(z_3)+\log(z_4)\right) \tilde{\varphi}_{0}(\uz)+\frac{11}{24} z_3^{4/3}   + \mathcal O(z_i^{7/3})     \, , \\
    \psi_{1}(\uz)   &= z_1^{2/3}\left[1+\frac{4 z_1}{15}+\left(\frac{5 z_1^2}{36}+\frac{108}{5} \sqrt{3} z_1 z_2\right)          + \mathcal O(z_i^4) \right]      \, .
\end{aligned}
\end{equation}

\paragraph{The six-loop triangle track graph.}
At six loops, we find
\begin{equation}
\begin{aligned}
    \mathcal D_{Z_6,1}  &=   9 \theta _1 \left(\theta _3-\theta _4\right)-3^6z_1 z_2 z_3 \left(1+3 \theta _1\right) \left(2+3 \theta _3-3 \theta _4\right)       \, , \\
    \mathcal D_{Z_6,2}  &=   9 \left(\theta _2-\theta _3\right) \theta _5+3^6z_3 z_4 z_5 \left(2-3 \theta _2+3 \theta _3\right) \left(1+3 \theta _5\right)       \, , \\
    \mathcal D_{Z_6,3}  &=  3 \left(2+3 \theta _3-3 \theta _4\right) \theta _5+3 z_5 \left(2+3 \theta _3-3 \theta _4\right) \left(\theta _4-\theta _5\right)+9 z_4 z_5 \left(\theta _3-\theta _4\right){}^2 \\
                        &\quad   -3^6z_3 z_4 z_5 \left(2-3 \theta _2+3 \theta _3\right) \left(2+3 \theta _3-3 \theta _4\right) \\
                        &\quad +3^6z_2 z_3 z_4 z_5 \left(-2+3 \theta _1-3 \theta _2\right) \left(2+3 \theta _3-3 \theta
   _4\right) \\
                        &\quad -3^6z_1 z_2 z_3 z_4 z_5 \left(1+3 \theta _1\right) \left(2+3 \theta _3-3 \theta _4\right)          \, , \\
    \mathcal D_{Z_6,4}  &=  3 \theta _1 \left(2-3 \theta _2+3 \theta _3\right)-3 z_1 \left(\theta _1-\theta _2\right) \left(2-3 \theta _2+3 \theta _3\right)+9 z_1 z_2 \left(\theta _2-\theta _3\right){}^2  \\
                        &\quad  -3^6z_1 z_2 z_3 \left(2-3 \theta _2+3 \theta _3\right) \left(2+3 \theta _3-3 \theta _4\right)\\
                        &\quad -3^6z_1 z_2 z_3 z_4 \left(2-3 \theta _2+3 \theta _3\right) \left(2+3 \theta _4-3 \theta
   _5\right)     \\
                        &\quad  -3^6z_1 z_2 z_3 z_4 z_5 \left(2-3 \theta _2+3 \theta _3\right) \left(1+3 \theta _5\right)     \, , \\
    \mathcal D_{Z_6,5}  &=   3 \left(4+6 \theta _3-3 \theta _5\right) \theta _5+3 z_5 \left(\theta _4-\theta _5\right) \left(2+6 \theta _3-3 \theta _4-3 \theta _5\right)+9 z_4 z_5 \left(\theta _3-\theta
   _4\right){}^2      \\
                        &\quad  -3^6z_3 z_4 z_5 \left(2-3 \theta _2+3 \theta _3\right) \left(5+3 \theta _2+3 \theta _3\right)-3^7 z_2 z_3 z_4 z_5 \left(2-3 \theta _1+3 \theta _2\right) \left(1+\theta _1+\theta
   _2\right)     \\
                        &\quad  -3^6z_1 z_2 z_3 z_4 z_5 \left(2+9 \theta _1 \left(1+\theta _1\right)\right)       \,                 
\end{aligned}
\end{equation}
with solutions
\beq\bsp
     \Sol(\PFI(M_{Z_6}))  &\,=  \Sol(\left\{ \mathcal D_{Z_6,k} \right\}_{k=1,\hdots,5}) = \Sol(\YDI(Z_6)) = \\
     &\, =\big\langle \Phi_0,\Phi_1,\Phi_2,\Phi_3\,\varphi_0,\varphi_1,\,\tilde\varphi_0,\tilde\varphi_1,\, \psi_1,\hdots, \psi_5   \big\rangle_{\mathbb{C}}\, .
     \esp\eeq
     with
\begin{equation}
\begin{aligned}
% \label{frob3zigzag}
    \Phi_{0}(\uz)   &= 1+72 z_3+\left(216 z_2 z_3+20250 z_3^2+216 z_3 z_4\right) + \mathcal O(z_i^3)     \, , \\
    \Phi_{1}(\uz)   &= \Phi_{1}(\uz)\log(z_3)+252 z_3+\left(216 z_2 z_3+\frac{164025 z_3^2}{2}+216 z_3 z_4\right)   + \mathcal O(z_i^3)     \, , \\
    \Phi_{2}(\uz)   &= \frac{1}{2}\Phi_{1}(\uz) \log ^2(z_3)+252z_3\log (z_3)+3 z_2+18 z_3+3 z_4 +\mathcal O(z_i^2)      \, , \\
    \Phi_{3}(\uz)   &= \frac{1}{6}\Phi_{1}(\uz) \log ^3(z_3)+126 \log ^2\left(z_3\right) z_3+\log \left(z_3\right) \left(3 z_2+18 z_3+3 z_4\right)   \\
                    &\quad   + \frac{27 z_2}{2}-171 z_3+\frac{27 z_4}{2}+\mathcal O(z_i^2)      \, , \\
    \varphi_{0}(\uz)    &= z_2^{2/3}\left(1+\frac{4 z_2}{15}+\left(\frac{5 z_2^2}{36}+\frac{1296 z_2 z_3}{5}\right) + \mathcal O(z_i^3)\right)     \, , \\
    \varphi_{1}(\uz)   &= \varphi_{0}(\uz)(\log(z_2)+\log(z_3))-2 z_1 z_2^{2/3}+\frac{z_2^{5/3}}{25}   + \mathcal O(z_i^{7/3})     \, , \\
    \tilde\varphi_{0}(\uz)    &=z_4^{2/3}\left(1+\frac{4 z_4}{15}+\left(\frac{1296 z_3 z_4}{5}+\frac{5 z_4^2}{36}\right)\right)  + \mathcal O(z_i^3)     \, , \\
    \tilde\varphi_{1}(\uz)   &= \tilde\varphi_{0}(\uz)(\log(z_3)+\log(z_4))+\frac{z_4^{5/3}}{25}-2 z_4^{2/3} z_5   + \mathcal O(z_i^{7/3})     \, , \\
    \psi_{1}(\uz)   &= z_2^{1/3}\left(1+\left(-\frac{z_1}{3}+\frac{z_2}{8}\right)+\left(-\frac{z_1^2}{9}+\frac{2 z_2^2}{35}+243 z_2 z_3+\frac{243 z_3 z_4}{2}\right)          + \mathcal O(z_i^3) \right)     \, , \\
    \psi_{2}(\uz)   &= z_4^{1/3}\left(1+\left(\frac{z_4}{8}-\frac{z_5}{3}\right)+\left(\frac{243 z_2 z_3}{2}+243 z_3 z_4+\frac{2 z_4^2}{35}-\frac{z_5^2}{9}\right)         + \mathcal O(z_i^3) \right)     \, , \\
    \psi_{3}(\uz)   &= z_1^{4/3}z_2^{2/3}\left(1+\frac{8 z_1}{21}+\left(\frac{2 z_1^2}{9}+\frac{4 z_1 z_2}{21}\right)          + \mathcal O(z_i^3) \right)     \, , \\
    \psi_{4}(\uz)   &= z_4^{2/3}z_5^{4/3}\left(1+\frac{8 z_5}{21}+\left(\frac{4 z_4 z_5}{21}+\frac{2 z_5^2}{9}\right)          + \mathcal O(z_i^3) \right)     \, , \\
    \psi_{5}(\uz)   &=  (z_2z_3z_4)^{1/3}\left( 1-\left(\frac{z_1}{3}+\frac{z_5}{3}\right)+\left(-\frac{z_1^2}{9}+243 z_2 z_3+243 z_3 z_4+\frac{z_1 z_5}{9}-\frac{z_5^2}{9}\right) + \mathcal O(z_i^3) \right)     \, .
\end{aligned}
\end{equation}

\subsection{Solution basis for the six-loop triangle wheel graph}
\label{app:T6}

%%%%%%%%%%%%%%%%%%%%%%%%%%%%%%%%%%%%%%%%

\begin{figure}[t]
\centering
\includegraphicsbox{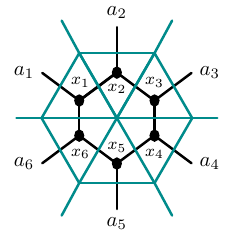}
\caption{Six-loop triangle wheel graph named after its momentum space representation (green lines).}
\label{fig:hexagon}
\end{figure}

As our last example we want to consider the six-loop graph cut from a hexagon lattice, which we denote as triangle wheel graph $T_6$ (see \Figref{fig:hexagon}). This is the first fishnet integral for a hexagonal lattice that cannot be written as a triangle track graph. 
There are no admissible star-triangle relations, and so
\beq
\textrm{Perm}_{T_6} = \Aut(T_6) = D_6\,.
\eeq
The graph $T_6$ depends on the following three cross ratios
\beq
\label{varH6}
    z_1 = \chi_{1,5,6,4}\,, \quad z_2 = \chi_{1,6,3,4}\, ,\quad    z_3 = \chi_{1,3,2,4}  \, .
\eeq
The Yangian differential operator ideal is generated by
\begin{equation}
\begin{aligned}
    \mathcal D_{T_6,1}  &=   9 \left(1+2 \theta _2-\theta _3\right) \theta _3+9 z_3 \left(\theta _2-\theta _3\right){}^2+3 z_2 z_3 \left(-2+3 \theta _1-3 \theta _2\right) \left(1+\theta _1+\theta _2\right)   \\
                        &\quad -z_1 z_2 z_3 \left(2+3 \theta _1\right){}^2  \, , \\
    \mathcal D_{T_6,2}  &=  9 \left(1+2 \theta _2-\theta _3\right) \theta _3+9 z_3 \left(\theta _2-\theta _3\right){}^2+3 z_2 z_3 \left(-2+3 \theta _1-3 \theta _2\right) \left(1+\theta _1+\theta _2\right)    \\
                        &\quad  -z_1 z_2 z_3 \left(2+3 \theta _1\right){}^2    \, , \\
    \mathcal D_{T_6,3}  &=  9 z_1 \theta _3 \left(1+2 \theta _2-\theta _3\right)+\left(9 z_1 z_3 \left(\theta _2-\theta _3\right){}^2+3 z_2 \left(z_3 \theta _1 \left(2-3 \theta _1+3 \theta _2\right)\right.\right. \\
                        &\quad \left. \left.-z_1
   \left(2+3 \theta _2-3 \theta _3\right) \theta _3\right)\right)  -3 z_1 z_2 z_3 \left(2-6 \theta _1^2+3 \theta _2^2-\theta _1 \left(2+3 \theta _2\right) \right. \\
                        &\quad \left.+\theta _2 \left(5-3 \theta _3\right)+\theta _3+3 \theta _3^2\right)-z_1^2 z_2 z_3
   \left(2+3 \theta _1\right){}^2    \, .                
\end{aligned}
\end{equation}
The corresponding solution space is spanned by
\beq\bsp
\label{solspacelH6}
    \Sol(\PFI(M_{T_6})) &\, =  \Sol(\left\{ \mathcal D_{T_6,k} \right\}_{k=1,2,3}) =\Sol(\YDI(T_6)) = \\
    &\,=\big\langle \Phi_{0},\Phi_{1},\Phi_{2},\, \varphi_{1},\hdots, \varphi_{5}\big\rangle_{\mathbb{C}} \, .
\esp\eeq
The explicit series for the basis of solutions for the triangle wheel graph $T_6$ are:
\begin{equation}
\begin{aligned}
% \label{frob3zigzag}
    \Phi_{0}(\uz)   &= 1+\frac{2 z_2}{9}+\left(\frac{2 z_1 z_2}{9}+\frac{20 z_2^2}{189}+\frac{2 z_2 z_3}{9}\right)  + \mathcal O(z_i^3)     \, , \\
    \Phi_{1}(\uz)   &= \phi_{0}(\uz)\log(z_2)+\frac{4 z_2}{9}+\left(\frac{z_1 z_2}{6}+\frac{4243 z_2^2}{15876}+\frac{z_2 z_3}{6}\right)    + \mathcal O(z_i^3)     \, , \\
    \Phi_{2}(\uz)   &= \frac{1}{2}\phi_{0}(\uz) \log ^2(z_2)       -\log \left(z_2\right) \left(\frac{z_1}{2}-\frac{4 z_2}{9}+\frac{z_3}{2}\right)+\frac{z_2}{12} +\mathcal O(z_i^2)      \, , \\
    \varphi_{1}(\uz)   &= \left(z_1+z_3\right)+\left(\frac{z_1^2}{2}+\frac{z_3^2}{2}\right)+\left(\frac{z_1^3}{3}+\frac{1}{3} z_1^2 z_2-\frac{1}{3} z_1 z_2 z_3+\frac{1}{3} z_2
   z_3^2+\frac{z_3^3}{3}\right)  + \mathcal O(z_i^4)      \, , \\
    \varphi_{2}(\uz)   &= z_2^{-1/3}\left[ 1+\left(\frac{z_1}{6}+\frac{z_2}{12}+\frac{z_3}{6}\right)+\left(\frac{4 z_1^2}{45}+\frac{5 z_1 z_2}{36}+\frac{37 z_2^2}{1125}+\frac{z_1 z_3}{36}\right.\right. \\
                    &\quad \left.\left.+\frac{5 z_2 z_3}{36} +\frac{4
   z_3^2}{45}\right)    + \mathcal O(z_i^3) \right]     \, , \\
    \varphi_{3}(\uz)   &= z_1^{1/3}z_2^{-1/3}\left[1+\lambda  \left(\frac{z_1}{3}+\frac{z_3}{6}\right)+\lambda ^2 \left(\frac{25 z_1^2}{126}+\frac{z_1 z_2}{6}+\frac{z_1 z_3}{18}+\frac{4 z_3^2}{45}\right)  + \mathcal O(z_i^3) \right]     \, , \\
    \varphi_{4}(\uz)   &= z_2^{-1/3}z_3^{1/3}\left[1+\left(\frac{z_1}{6}+\frac{z_3}{3}\right)+\left(\frac{4 z_1^2}{45}+\frac{z_1 z_3}{18}+\frac{z_2 z_3}{6}+\frac{25 z_3^2}{126}\right)    + \mathcal O(z_i^3) \right]    \, , \\
    \varphi_{5}(\uz)   &=  z_1^{1/3}z_2^{-1/3}z_3^{1/3}\left[ 1+\left(\frac{z_1}{3}+\frac{z_3}{3}\right)+\left(\frac{25 z_1^2}{126}+\frac{z_1 z_3}{9}+\frac{25 z_3^2}{126}\right)       + \mathcal O(z_i^3) \right]     \, .
\end{aligned}
\end{equation}

In this case, we checked the completeness of the differential operator ideal, i.e., that we have not found too many solutions, with the following analysis: First, we tried to find additional operators $\tilde{\mathcal D}_{T_6,k}$ annihilating the holomorphic solution $\Phi_1(\uz)$. Then we checked that these additional operators do not restrict the solution space further, compared to $\text{Sol}(\{  \mathcal D_{T_6,1}, \mathcal D_{T_6,2}, \mathcal D_{T_6,3}\})$. We have done this for second order operators up to multi-degree four in $\uz$.